\documentclass[onecolumn,showpacs,preprintnumbers]{revtex4}
\usepackage{graphicx}
\usepackage{bm}
\usepackage[english]{babel}

\begin{document}

\title{Microscopic theory of phase transitions and nonlocal corrections for free energy of a superconductor.}
\author{K.V. Grigorishin}
\email{gkonst@ukr.net}
\author{B.I. Lev}
\email{bohdan.lev@gmail.com} \affiliation{Boholyubov Institute for
Theoretical Physics of the Ukrainian National Academy of Sciences,
14-b Metrolohichna str. Kiev-03680, Ukraine.}
\date{\today}
\begin{abstract}
The new approach to the microscopic description of the phase
transitions starting from the only first principles was developed
on an example of the transition normal metal-superconductor. This
means mathematically, that the free energy is calculated in the
range of temperatures, which includes a point of pase transition,
without introducing any artificial parameters similar to an order
parameter, but only starting from microscopic parameters of
Hamiltonian. Moreover the theorems about connection of a vacuum
amplitude with thermodynamics potentials are realized. The
functional of a superconductor's free energy in a magnetic field
was obtained with help the developed method. The obtained
functional is generalization of Ginzburg-Landau functional for the
case of arbitrary value of a gap, arbitrary spatial
inhomogeneities and nonlocal magnetic response. The explicit
expressions for the extremals of this functional were obtained in
the low-temperature limit and the high-temperature limit at the
condition of slowness of gap's changes.
\end{abstract}

 \maketitle

\section{Formulation of the problem}\label{formulation}

Exact microscopic description of a phase transition is opened
problem of modern theoretical physics. The problem can be solved
for several simplest models only, but different phenomenological
approaches exist for the rest cases. The essence of the matter
lies in the following. Basic problem of statistical mechanics is
calculation of a partition function (of grand canonical ensemble
in a total case) $Z$ or calculation of a density matrix
$\widehat{\rho}$:
\begin{equation}\label{1.1}
Z=Sp\left(e^{-\beta(\widehat{H}-\mu \widehat{N})}\right)\equiv
e^{-\beta\Omega}, \qquad
\widehat{\rho}=\frac{1}{Z}e^{-\beta(\widehat{H}-\mu \widehat{N})},
\end{equation}
where $\widehat{H}=\widehat{H}_{0}+\widehat{V}$ is full
Hamiltonian of a system, $\widehat{N}$ is the particle operator,
$\mu$ is the chemical potential, $\Omega$ is the grand
thermodynamics potential. Replacement of Hamiltonian $\widehat{H}$
in canonical ensemble by Hamiltonian $\widehat{H}-\mu \widehat{N}$
in grand canonical ensemble leads to the shift of reference of
particle's energy from zero to Fermi surface:
$\varepsilon(k_{F})=0$. Hence, the potential $\Omega$ plays a part
of Helmholtz free energy with a reference of particle's energy
from Fermi surface. Therefore we shall call the grand
thermodynamics potential by free energy for brevity. Let's
transform the partition function (\ref{1.1}) to a form:
\begin{eqnarray}\label{1.2}
 Z=Sp\left(e^{-\beta(\widehat{H_{0}}-\mu
\widehat{N})}e^{\beta(\widehat{H_{0}}-\mu
\widehat{N})}e^{-\beta(\widehat{H}-\mu
\hat{N})}\right)=Sp\left(Z_{0}\widehat{\rho}_{0}\widetilde{U}(\beta)\right)=Z_{0}R(\beta),
\end{eqnarray}
where $Z_{0}$ is the partition function for a system of
noninteracting particles,
\begin{equation}\label{1.3}
    \widetilde{U}(\beta)=e^{+\beta(\hat{H_{0}}-\mu
\widehat{N})}e^{-\beta(\widehat{H}-\mu
\widehat{N})}=\sum_{n=0}^{\infty}\frac{(-1)^{n}}{n!}\int_{0}^{\beta}d\tau_{1}\ldots\int_{0}^{\beta}d\tau_{n}
\widehat{T}\left\{\widehat{H}_{I}(\tau_{1})\ldots\widehat{H}_{I}(\tau_{n})\right\}
\end{equation}
is the evolution operator in the interaction representation (it
describes evolution of the system in imaginary time
$it\rightarrow\tau$, $\widehat{T}$ is the ordering operator in
time), $\widehat{H}_{I}(\beta)=e^{+\beta(\hat{H_{0}}-\mu
\hat{N})}\widehat{V}e^{-\beta(\hat{H_{0}}-\mu \hat{N})}$ is the
interaction operator of particles in interaction representation,
\begin{eqnarray}\label{1.4}
R(\beta)=\langle
\widehat{U}(\beta)\rangle_{0}=Sp\left(\widehat{\rho}_{0}\widetilde{U}(\beta)\right)=\sum_{n=0}^{\infty}\frac{(-1)^{n}}{n!}\int_{0}^{\beta}d\tau_{1}\ldots\int_{0}^{\beta}d\tau_{n}
Sp\left(\widehat{\rho}_{0}
\widehat{T}\left\{\widehat{H}_{1}(\tau_{I})\ldots\widehat{H}_{I}(\tau_{n})\right\}\right)
\end{eqnarray}
is the vacuum amplitude of the system. The averaging
$\langle\rangle_{0}\equiv Sp(\widehat{\rho}_{0}\ldots)$ is done
over ensemble of \emph{noninteracting} particles. It is necessary
to note, that the vacuum amplitude describes dynamics of the
system in the multiparticle  state $\Phi_{0}$ under the influence
of the internal interaction, unlike Green function describing
dynamics of a particle in one-particle state $\phi_{k}$ under the
influence of its interaction with other particles.

The partition function $Z_{0}$ can be found exactly for any
system. If particles interact, then the situation becomes
complicated essentially. The only several models can be solved
exactly \citep{hua,bax}, but in the rest cases the values
(\ref{1.1}) can not be found exactly and it is necessary to
develop a perturbation theory. Solution of the basic problem of
the statistical mechanics reduces to the calculation of a
transition amplitude "vacuum-vacuum" (\ref{1.4}). In order to
formulate our problem let's consider the case of zero temperature
$T=0$. In this case the time is real, and the vacuum amplitude is
determined by the following:
\begin{eqnarray}\label{1.6}
R(t)&=&\langle\Phi_{0}|
\widetilde{U}(t-t_{0})|\Phi_{0}\rangle_{t_{0}=0}=\langle\Phi_{0}|
U(t)|\Phi_{0}\rangle e^{iW_{0}t}\nonumber\\
&=&\sum_{n=0}^{\infty}\frac{(-i)^{n}}{n!}\int_{0}^{t}dt_{1}\ldots\int_{0}^{t}dt_{n}
\left\langle\Phi_{0}|
\widehat{T}\left\{\widehat{H}_{I}(t_{1})\ldots\widehat{H}_{I}(t_{n})\right\}|\Phi_{0}\right\rangle,
\end{eqnarray}
where $\widehat{H}_{I}(t)=e^{+it(\hat{H_{0}}-\mu
\hat{N})}\widehat{V}e^{-it(\hat{H_{0}}-\mu \hat{N})}$ is operator
of particles' interaction in the interaction representation,
$W_{0}$ is ground state energy of a system without interaction.
The averaging $\langle\rangle_{0}\equiv
\langle\Phi_{0}|\ldots|\Phi_{0}\rangle$ is done over the ensemble
of \emph{noninteracting} particles. Knowing a vacuum amplitude we
can calculate the ground state energy of the system of the
interacting particles $E_{0}$ using the theorem \citep{matt}:
\begin{equation}\label{1.7}
    E_{0}=W_{0}+\lim_{t\rightarrow\infty(1-i\delta)}i\frac{d}{dt}\ln
    R(t),
\end{equation}
where $\delta$ is infinitely small value, but
$\infty\cdot\delta\rightarrow\infty$. The theorem is correct if
the limit transition
\begin{eqnarray}\label{1.8}
  [\frac{d}{dt}\ln R(t)]_{\infty(1-i\delta)}=iW_{0}
  +\frac{|\left\langle\Phi_{0}|\Psi_{0}\right\rangle|^{2}(-iE_{0})e^{-iE_{0}\infty}e^{-E_{0}\delta\cdot\infty}}{|\left\langle\Phi_{0}|\Psi_{0}\right\rangle|^{2}e^{-iE_{0}\infty}e^{-E_{0}\delta\cdot\infty}}
  =iW_{0}-iE_{0}
\end{eqnarray}
is possible. This transition is possible at the condition when the
symmetries of ground state of the system with interaction
$|\Psi_{0}\rangle$ and without interaction $|\Phi_{0}\rangle$ are
identical:
\begin{equation}\label{1.9}
    \langle\Phi_{0}|\Psi_{0}\rangle\neq0.
\end{equation}
The expressions (\ref{1.8}) and (\ref{1.9}) mean: 1)potential of
interaction is being switched slowly in the system in the ground
state without interaction, 2) the ground state of the system with
interaction is being obtained by continuous way from the ground
state without interaction while the switching of the interaction
(\textit{adiabatic hypothesis}). For nonzero temperature the
analog of the theorem (\ref{1.7}) has a form:
\begin{equation}\label{1.11}
    U=-\frac{\partial}{\partial\beta}\ln Z_{0}-\frac{\partial}{\partial\beta}\ln
    R(\beta),
\end{equation}
where $U$ is the internal energy of a system. Moreover, we can
calculate the free energy:
\begin{equation}\label{1.12}
    \Omega=-\frac{1}{\beta}\ln Z_{0}-\frac{1}{\beta}\ln
    R(\beta).
\end{equation}

If the wave functions are orthogonal
$\langle\Phi_{0}|\Psi_{0}\rangle=0$ - the adiabatic hypothesis is
not valid, then the symmetries of the system with interaction and
without one is different. This means, that an initial system
without interaction suffers phase transition stipulated by the
interaction. A nonfulfilment of the adiabatic hypothesis means
nonfulfilment of the theorem (\ref{1.7}). For the system with the
broken symmetry we can calculate a vacuum amplitude on the free
propagators $G_{0}(\textbf{k},t)$ and use the formula (\ref{1.7}).
However we shall find a wrong ground state energy $E_{0}$. This
means that a state exists with more low energy than the found
value. Moreover, in consequence of the breakdown of the condition
(\ref{1.9}) a system becomes unstable: $\Gamma$-matrix, which
determines vacuum amplitude $R(t)$, one-particle propagator
$G(\textbf{k},t)$ and two-particle propagator
$K(\textbf{k}_{1},\textbf{k}_{2},t)$ in stair approximation, has
the form:
\begin{equation}\label{1.10}
    \Gamma(t)=ce^{-\alpha t}+c'e^{+\alpha t}.
\end{equation}
The value of $\Gamma$ is increasing infinitely at
$t\rightarrow\infty$ \citep{matt,shri}, it means an instability of
the system. Similar instability was observed experimentally as the
process of formation of a charge density wave in
$\texttt{TbTe}_{3}$ \citep{yusup}.

From the aforesaid we can see, that perturbation theory making
possible to calculate the vacuum amplitude (thermodynamics
function $U$, $\Omega$ and so on) in ranges including a point of
phase transition (for example, temperature $T_{C}$) doesn't exist.
In other words, it is not possible to find $R(\beta)$ in the
system with condensed phase starting from the first principles.
However the phenomenological approach exists for calculation of
$\Omega$. It assumes, that at the temperature $T<T_{C}$ the
condensed phase is characterized by some order parameter $\eta$
exists. Near a transition point for type II phase transition or
near a point of overcooling for type I phase transition the
parameter $\eta$ is small, and the free energy can be represented
in a form of Landau expansion:
\begin{equation}\label{1.13}
    \Omega(T,V,\mu,h)=\Omega_{0}+\int d^{3}r (a\eta^{2}+b\eta^{4}+g(\nabla\eta)^{2}-\eta
    h),
\end{equation}
where $h$ is external field. The equilibrium value of order
parameter $\overline{\eta}$ is determined by an extremal of the
functional (\ref{1.13}):
$\frac{\delta\Omega}{\delta\eta}|_{\eta=\overline{\eta}}=0$.
However, the expansion (\ref{1.13}) is correct in the range
$(T_{C}-T)/T_{C}\ll 1$ only, hence $\overline{\eta}$ can be found
in this range only.

In order to obtain $\overline{\eta}$ at any temperature the
concept of \textit{quasi-averages} (\textit{anomalous averages})
is introduced. In the paper \citep{matt2} generalization of the
method of quasi-averages has been represented in terms of Green
function - Nambu-Gor'kov formalism \citep{gork,nambu}. It is
\textit{postulated}, that in the condensed phase, in addition to
normal propagators
$G(\textbf{k},\sigma,t'-t)=-i\langle\Psi_{0}|T\{C_{\textbf{k}\sigma}(t')C_{\textbf{k}\sigma}^{+}(t)\}|\Psi_{0}\rangle$,
anomalous propagators $F$ exist, for example:
\begin{eqnarray}\label{1.14}
  \texttt{Ferromagnetic} &:& F_{fer}=-i\langle\Psi_{0}|T\{C_{\textbf{k}\uparrow}(t')C_{\textbf{k}\uparrow}^{+}(t)\}|\Psi_{0}\rangle \nonumber\\
  \texttt{Solid, liquid} &:& F_{sol}=-i\langle\Psi_{0}|T\{C_{\textbf{k}+\textbf{q}\sigma}(t')C_{\textbf{k}\sigma}^{+}(t)\}|\Psi_{0}\rangle\\
  \texttt{Superconductor with singlet pairing} &:& F_{sup}=-i\langle\Psi_{0}|T\{C_{-\textbf{k}\downarrow}(t')C_{\textbf{k}\uparrow}(t)\}|\Psi_{0}\rangle \nonumber
  \end{eqnarray}
They are proportional to the according order parameters. The
anomalous propagators $F$ can not be obtained by summation of
diagrams consisting of normal propagators $G$ only. This fact is
result of different symmetries of a perturbed state and an
unperturbed state.

For construction of the self-consistent perturbation theory the
"sourse term" $\widehat{H}_{S}$ is introduced in Hamiltonian
instead of the interaction operator $\widehat{V}$. The sourse term
is induced by the order parameter (for ferromagnetic - magnetic
field orientating spins, for superconductor - a sourse of Cooper
pairs). The sourse term inducts specification structure and
changes symmetry of a system: $\Phi_{0}\rightarrow\Phi_{0}'$, such
that $\langle\Phi_{0}'|\Psi_{0}\rangle\neq0$. This means that a
system has an internal long range field $H$, generated by the
sourse $\widehat{H}_{S}$, and according order parameter $\eta$,
which is function of $H$ and depends on the interaction constant
$\lambda$ and temperature $T$: $\eta=\eta_{\lambda T}(H)$. In
turn, $H$ is function of $\eta$: $H=H_{\lambda T}(\eta)$. Hence,
these two equations can be combined:
\begin{equation}\label{1.15a}
    \eta=\eta_{\lambda T}(H_{\lambda T}(\eta))
\end{equation}
and they can be solved in $\eta$. This means, that order parameter
is determined in self consistent way. In Green function formalism
this fact has the following form: the free matrix propagator
$\textbf{G}_{0}$ (anomalous part is zero $F=0$) and the dressed
propagator $\textbf{G}$ (with normal $G$ part and anomalous $F$
parts) obey Dyson equation \citep{shri,matt2}:
\begin{equation}\label{1.15}
    \textbf{G}^{-1}=\textbf{G}^{-1}_{0}-\bf{\Sigma}(\textbf{G}),
\end{equation}
and what's more the mass operator $\bf{\Sigma}(\textbf{G})$ is
determined by self consistent way. This means, that elements of
diagrams for the mass operator are dressed propagators
$\textbf{G}$, which contain the sought mass operator.
Artificiality of the introduction of the sourse $\widehat{H}_{S}$
lies in the fact that the expression for a mass operator is
\emph{postulated}. Moreover, $\bf{\Sigma}$ can not be obtained by
summation of the diagram consisting of free propagators
$\textbf{G}_{0}$ only, that is $\bf{\Sigma}(\textbf{G}_{0})=0$.
For example, $\bf{\Sigma}$ has the form for a superconductor:
\begin{equation}\label{1.16}
    \bf{\Sigma}(\omega,p)=[1-Z(\omega,p)]\omega
    \textbf{1}+\chi(\omega,p)\tau_{3}+\varphi(\omega,p)\tau_{1}+\widetilde{\varphi}(\omega,p)\tau_{2},
\end{equation}
where $Z$ is some coefficient, the field $\chi$ determines a shift
of chemical potential $\mu$ at the transition to the
superconductive state, the fields $\varphi$ and
$\widetilde{\varphi}$ play a part of order parameter in
superconductivity - a gap, and specter of excitations is
represented by them
$E^{2}(p)=\varepsilon^{2}(p)+\varphi^{2}(p)+\widetilde{\varphi}^{2}(p)$;
$\textbf{1}$, $\tau_{i}$ are unit matrix and Pauli matrixes. The
dependence of all fields on frequency $\omega$ takes into account
a delay and a damping of quasi-particles. In Hartree-Fock
approximation we have $Z=1$ and $\bf{\Sigma}$ doesn't depend on
$\omega$. We can suppose $\widetilde{\varphi}=0$, $\chi=0$ and
denote $\varphi(p)\equiv\Delta(p)$, then
\begin{equation}\label{1.17}
    \Delta(T=0)=\frac{\lambda}{V}\int_{-\infty}^{+\infty}\frac{d\omega}{2\pi}\int
    \frac{d^{3}p}{(2\pi)^{3}}iF(\omega,\textbf{p},\Delta), \qquad \Delta(T)=\frac{\lambda}{V} T\sum_{n=-\infty}^{n=+\infty}\int
    \frac{d^{3}p}{(2\pi)^{3}}iF(\omega_{n},\textbf{p},\Delta),
\end{equation}
where $\lambda<0$ is the interaction constant, and the summation
is done over Fermy frequencies $\omega_{n}=(2n+1)\pi T$. We can
see, that the order parameter is determined by the anomalous
propagator $F$, and the equations (\ref{1.17}) are equations of
self consistency for order parameter $\Delta$ as specific case of
the total equation (\ref{1.15a}).

The described approach is phenomenological too, because the
anomalous propagators and corresponding order parameters are
introduced to the theory from elsewhere. This means that we select
the required states from all possible state artificially. In this
sense this approach likes Landau approach (\ref{1.13}). The sourse
of Cooper pairs for superconductor or the magnetic field
$H=\frac{J_{0}}{g\mu_{B}}\langle S_{z}\rangle$ for ferromagnetic
($J_{0}$ is the exchange integral, $g$ is the gyromagnetic
relation, $\mu_{B}$ is Bohr magneton, $\langle S_{z}\rangle$ is
the average projection of spin onto axis $z$) can not be
interpreted as real field. So, for iron the real internal magnetic
field is $\sim 10^{3}$ oersted, but for the ordering of spins it
is necessary the effective magnetic field $H\sim 10^{6}$ oersted.
Thus, self magnetization has nonmagnetic nature evidently.

As it has been noted in \citep{yukh}, \emph{in spite of all
progresses reached in description of phase transitions and in
calculation of main characteristics of a system in critical
region, the basic problem of phase transitions is not solved:
calculation explicit expressions for thermodynamical functions of
a system in ranges, which include a point of phase transition as
function of temperature, external fields and microscopic
parameters of Hamiltonian}. It is necessary to have a description
of phase transitions on microscopic level. This presupposes a
direct calculation of free energy $\Omega(T,V,h)$ from first
principles, but we must not construct it.

At the present moment the two approaches can be separated for
solution of the formulated problem. In the papers
\citep{yukh,yukh2,yukh3} a partition function of Ising model is
calculated by the method of collective variables, which are
oscillation modes of spin moment. Then a partition function can be
written via these variables as a functional. Investigation of
Euler-Lagrange equations shows, that among a set of collective
variables the variable connected with order parameter exists. In
the paper \citep{gran} a partition function is calculated by the
saddle point method for classical system with short range
attraction and repulsion between particles. It has been shown,
that the free energy can be represented by the form which is
analogous to Landau expansion where the saddle point plays a role
the order parameter. The saddle point method has the sense as
method of separation of states giving largest contribution in a
partition function and it is equivalent to the mean field method.
With the help of the saddle point method the processes of cluster
formation can be investigated. The cluster formation can be
considered as a phase transition from spatially homogeneous
distribution to spatially inhomogeneous distribution. In the
papers \citep{bel,zhu,grig1,grig2} both short range potentials and
long range potentials were considered. Critical temperature and
critical concentration, when clusters form in a system, dependence
of size of a cluster on temperature have been obtained.

One of the important applications of the microscopic theory of
phase transitions is description of thermodynamics and
electrodynamics of superconductors. It is necessary to know the
functional of free energy $\Omega(\beta,\Delta,\textbf{A})$, where
$\textbf{A}$ is potential of magnetic field. Then the equations
$\frac{\delta\Omega}{\delta\Delta}=0$,
$\frac{\delta\Omega}{\delta\textbf{A}}=0$ will describe
equilibrium states of the condensed phase and normal phase. Two
basic methods for obtaining of the sought equations exist. First
of them is joint solution of Gor'kov equations and the equation of
self consistency (\ref{1.17}). At the temperature $T\rightarrow
T_{C}$, when $\Delta/T_{C}\rightarrow0$, the solution of these
equation can be represented in the form of series in degrees of
$\Delta$. Moreover, magnetic penetration depth is bigger then
Pippard coherent length $l_{0}$, hence the potential $\textbf{A}$
changes a little on a coherent length. As a result we have the
well-known Ginzburg-Landau equation.

Another method, proposed in \citep{sad}, is the direct calculation
of a vacuum amplitude $R(\beta)$. The concept lies in the fact
that we consider electrons in a \emph{normal} metal propagating in
random "field" of thermodynamic fluctuations of order parameter
$\Delta_{\textbf{q}}$, where $\textbf{q}$ is small wave-vector.
The operator of the interaction of electrons with the fluctuations
can be written as:
\begin{equation}\label{1.27}
    \widehat{H}_{\texttt{int}}=\sum_{\textbf{p}}\left[\Delta_{\textbf{q}}\widehat{C}_{\textbf{p}_{+}}^{+}\widehat{C}_{-\textbf{p}_{-}}^{+}+
    \Delta_{\textbf{q}}^{\ast}\widehat{C}_{-\textbf{p}_{-}}\widehat{C}_{\textbf{p}_{+}}\right],
\end{equation}
where $\textbf{p}_{\pm}=\textbf{p}\pm \textbf{q}/2$. A correction
to the thermodynamics potential from any interaction is
represented via the vacuum amplitude $R$:
\begin{equation}\label{1.28}
    \Delta\Omega=-T\ln R(\beta)\approx -T[R(\beta)-1].
\end{equation}
Then using Wick theorem we can represent (\ref{1.4}) via the
propagators. For the correction of second order we have:
\begin{eqnarray}\label{1.29}
    \Delta\Omega&=&
    -\frac{T}{2!}\int_{0}^{1/T}d\tau_{1}\int_{0}^{1/T}d\tau_{2}\langle
    \widehat{T}_{\tau}(\widehat{H}_{int}(\tau_{1})\widehat{H}_{int}(\tau_{2}))\rangle\nonumber\\
    &=&-T\int_{0}^{1/T}d\tau_{1}\int_{0}^{1/T}d\tau_{2}|\Delta_{\textbf{q}}|^{2}\sum_{\textbf{p}}G_{0}(\textbf{p}_{+},\tau_{1}-\tau_{2})G_{0}(-\textbf{p}_{-},\tau_{1}-\tau_{2}).
\end{eqnarray}
The correction $\Delta\Omega$ is represented via the free
propagators $G_{0}$ of normal state only - we consider normal
metal at $T>T_{c}$, where the fluctuation sourse of Cooper pair
(\ref{1.27}) acts. As a result we have Landau expansion:
\begin{eqnarray}\label{1.30}
    \Omega_{s}-\Omega_{n}=\sum_{\textbf{q}}\left[\alpha(T)|\Delta_{\textbf{q}}|^{2}+\frac{b}{2}|\Delta_{\textbf{q}}|^{4}+\gamma q^{2}|\Delta_{\textbf{q}}|^{2}\right],
\end{eqnarray}
where $\alpha(T)\propto(T-T_{Ñ}),b,\gamma$ are expansion
coefficients.

In our opinion, this approach is not successively microscopic,
because the artificial element is used - the external sourse of
Cooper pairs (\ref{1.27}), that implies some seed order parameter.
As for calculation of the correction $\Delta\Omega$ the free
propagators $G_{0}$ of normal phase is used only, the condensed
phase is considered as fluctuations against the background of
normal phase. This means, that we can obtain the limit $\Omega
(T\rightarrow T_{Ñ})$ only.

Ginzburg-Landau equations are correct for description of
thermodynamics and electrodynamics of a superconductor at the
following restrictions:
\begin{enumerate}
    \item The gap is much less than critical temperature. Then the parameter $\Delta(\textbf{r},T)/T_{C}\ll
    1$ can be expansion parameter. This means, that the equations are correct in the range $T\rightarrow T_{C}$
    or in the range $H\rightarrow H_{C2}$ (intensity of magnetic field is near the second critical magnetic field $H_{C2}$).
    \item $\Delta(\textbf{r},T)$ changes slowly on the coherent length $l(T)$, which is size
    of a Cooper pair.
    \item Magnetic field $\textbf{H}(\textbf{r})=\texttt{rot}\textbf{A}(r)$
     changes slowly on the coherent length, that is the magnetic penetration depth is $\lambda(T)\gg l(0)$.
    This means, that electrodynamics of a superconductor is local.
\end{enumerate}
In the papers \citep{tew,werth} the equations has been proposed,
where the first restriction is absent. These equations were
obtained from Gor'kov equations and they are the generalization of
Ginzburg-Landau equations for the case of arbitrary value of
$\Delta(\textbf{r},T)/T$. However spatial inhomogeneities are slow
and electrodynamics is local.

Our aim is the description of phase transitions on microscopic
level starting from the first principles only. Mathematically this
means to develop a method of calculation of the partition function
(\ref{1.1}) (the free energy $\Omega$) in ranges of temperatures
and parameters of interaction, which include a point of pase
transition, without introducing any artificial parameters of type
of order parameter $\eta$ and sourses of ordering
$\widehat{H}_{S}$, but starting from microscopic parameters of
Hamiltonian $\hat{H}=\hat{H}_{0}+\hat{V}$ only. The theorems
(\ref{1.7},\ref{1.11},\ref{1.12}) about connection of a vacuum
amplitude with thermodynamics potentials must be realized. Thus,
in microscopic theory of phase transitions the equation of
self-consistency (\ref{1.15a}), the anomalous propagators
$F(\textbf{p},\omega)$ (\ref{1.14}) and Landau functional
(\ref{1.13}) must be deduced, but they must be not postulated. To
solve the problem means to develop the perturbation theory for the
vacuum amplitudes $R(t)$ and $R(\beta)$, which is correct both
normal phase and condensed phase. Phase transition normal metal -
superconductor has been considered as an example. Since in
Nambu-Gor'kov formalism (the method of anomalous propagators) any
phase transition can be described \citep{matt2}, then our method
can be generalized to the rest transitions (ferromagnetism, waves
of charge and spin density, crystallization and so on). Further we
can apply the developed method in order to calculate free energy
of a superconductor at arbitrary temperatures, spatial
inhomogeneities, magnetic fields and currents, moreover with
nonlocal magnetic response. The obtained expression will be the
generalization of Ginzburg-Landau functional (\ref{1.13}) in above
mentioned sense.

In the section \ref{unstability} we consider the instability of
normal Fermy system at the switching of attraction between
particles. Mathematically this is expressed in the fact that a
two-particle propagator $K$, calculated on free one-particle
propagators in stair approximation, has a pole $\alpha$ which
doesn't belong to a free propagator $K_{0}$ and situated on
complex axis. This means a presence of bound states of particles
in the system (with the binding energy $|\alpha|$) and evolution
of the system in time as $\sim e^{|\alpha|t}$. On the other hand,
a two-particle propagator has a pole at the energy of the bound
state if the bound state of two isolated particles exists. The
residue in this pole is product of Bethe-Solpiter amplitudes
$\eta\eta^{+}$ - the amplitudes of pairing.

In the section \ref{correlations} we generalizes the result of
two-particles problem to a many-particle system. It follows from
the identity principle, that amplitudes of pairing are determined
by dynamics of all particles of the system, and observed values of
the amplitudes of pairing are the result of the averaging over a
system. Thus, the collective (condensate) of pairs exists. In
order to find an one-particle propagator $G_{S}$ and to generalize
the two-particle problem to a many-particle case we proposed the
method of an uncoupling of correlations. The method considers
interaction of a additional fermion with fluctuations of pairing
(formation and decay of the pairs). As a result of such
interaction, the law of quasi-particles' dispersion changes as
$\varepsilon(k)\rightarrow\sqrt{\varepsilon^{2}(k)+\Delta\Delta^{+}}$,
where $\Delta$ and $\Delta^{+}$ are amplitudes of pairing playing
a role of a gap and they are analog of Bethe-Solpiter amplitudes
in two-particle problem. Gor'kov equations and the existence of
anomalous propagators $F$ and $F^{+}$ follow from Dyson equation
for the described above process. This fact means breakdown of a
global gauge symmetry, namely number of particles is not conserved
in the course of a presence of a pairs' condensate. After
calculation of particles' interaction with fluctuations of pairing
all characteristics of a system must be calculated over the new
vacuum with broken symmetry.

For calculation of observed values of the amplitude of pairing
$\Delta$ and $\Delta^{+}$ it is necessary to know a vacuum
amplitude $R(t)$ of a system. The vacuum amplitude must be
calculated over the new vacuum with broken symmetry. This gives
possibility to use the theorem about connection of a vacuum
amplitude with ground state energy of a system. In the section
\ref{ground} the method of uncoupling of correlations is proposed.
The method allows to represent a vacuum amplitude via anomalous
propagators $F$ and $F^{+}$. Thus we obtain a functional of ground
state energy over the fields $\Delta$ and $\Delta^{+}$. An
extremal of the obtained functional is the equation of self
consistency for the parameter $\Delta$ in Nambu-Gor'kov formalism.
This means, that order parameter is averaged Bethe-Solpiter
amplitude over all system. In the section \ref{temperature} we
generalize the results of two previous sections for the case of
nonzero temperature. Using the method of uncoupling of
correlations we calculate a vacuum amplitude $R(\beta)$ and a
functional of free energy $\Omega(\Delta,T)$ over the fields
$\Delta$ and $\Delta^{+}$. In a high-temperature limit
$T\rightarrow T_{C}$ the expansion of free energy in powers of
$|\Delta|$ has a form of Landau expansion. This fact proves, that
the averaged over a system the amplitudes of pairing $\Delta$ and
$\Delta^{+}$ have properties, which are analogous to the
properties of an order parameter in a phenomenological theory. In
the section \ref{nonzeromomentum} we consider the case of a
pairing with nonzero momentum of a pair's center of mass.

In the sections \ref{spaceinhom} and \ref{magnetic} using the
developed method of microscopic description of a phase transition
we obtained the functional of free energy of a spatial
inhomogeneous superconductor in magnetic field. The functional
generalizes Ginzburg-Landau functional in cases of arbitrary
temperatures, arbitrary spatial inhomogeneities and nonlocality of
magnetic response. The free energy has a form of Ginzburg-Landau
functional in the high-temperature limit at the condition of
slowness of gap's change in space. Corresponding equations of
superconductor's state demonstrate the nonlinear connection
between the current and the magnetic field. In the low-temperature
limit in the case of weak field $H\ll H_{C}$ at the condition of
slowness of gap's change the nonlocal connection between the
current and the field appears, which is the long-wave limit of
Pippard law. The last fact proves the nonlocality of the obtained
functional of free energy.

\section{The instability of normal state and two-particle dynamics}\label{unstability}

Let we have a system from $N$ noninteracting fermions in volume
$V$. In ideal Fermy gas propagation of a particle with momentum
$\textbf{k}$, energy $\varepsilon\approx
v_{F}(|\textbf{k}|-k_{F})$ counted off Fermy surface (we are using
system of units, where $\hbar=k_{B}=1$) and spin $\sigma$ is
described by the free propagator:
\begin{eqnarray}\label{2.1}
  G_{0}(\textbf{k},t)=\left\{\begin{array}{cc}
    -i\langle\Phi_{0}|C_{\textbf{k},\sigma}(t)C_{\textbf{k},\sigma}^{+}(0)|\Phi_{0}\rangle, \qquad t>0 \\
    i\langle\Phi_{0}|C_{\textbf{k},\sigma}^{+}(0)C_{\textbf{k},\sigma}(t)|\Phi_{0}\rangle, \qquad t\leq0 \\
  \end{array}\right\}=
-i\theta_{t}A_{0}e^{-i|\varepsilon|t}+i\theta_{-t}B_{0}e^{i|\varepsilon|t},\nonumber\\
G(\textbf{k},t)=\int\frac{d\omega}{2\pi}G(\textbf{k},\omega)e^{-i\omega
t}\Longrightarrow
G_{0}(\textbf{k},\omega)=\frac{1}{\omega-\varepsilon(k)}=\frac{\omega+\varepsilon}{\omega^{2}-\varepsilon^{2}}
=\frac{A_{0}}{\omega-|\varepsilon|}+\frac{B_{0}}{\omega+|\varepsilon|},
\end{eqnarray}
where
\begin{equation}\label{2.2}
A_{0}=\frac{1}{2}\left(1+\frac{\varepsilon}{|\varepsilon|}\right),\qquad
B_{0}=\frac{1}{2}\left(1-\frac{\varepsilon}{|\varepsilon|}\right),\qquad
\theta_{t}=\left\{\begin{array}{c}
  1,\qquad t>0\\
  0,\qquad t<0\\
\end{array}\right\},
\end{equation}
$C_{\textbf{k},\sigma}(t)$ and $C_{\textbf{k},\sigma}^{+}(t)$ are
creation and annihilation operators in Heisenberg representation.
Now let attractive force acts between particles. The force is
described by matrix element of interaction potential:
\begin{equation}\label{2.3}
    \langle
    \textbf{l},-\textbf{l}|\widehat{V}|\textbf{k},-\textbf{k}\rangle=\lambda
    w_{l}w_{k}<0, \qquad w_{k}=\left\{\begin{array}{c}
      1,\qquad \varepsilon(k)<\omega_{D} \\
      0,\qquad \varepsilon(k)>\omega_{D}\\
    \end{array}\right\},
\end{equation}
moreover interacting particles have opposite spins. This model
potential is correct if a range of interaction $r_{0}$ is much
smaller than average distance between particles:
$r_{0}\sqrt[3]{N/V}\sim r_{0}k_{F}\ll 1$, that means "slowness" of
collisions. In turn, account of Fermy statistic, in the limit of
"slow" collisions particles can scatter with opposite spins only.

\begin{figure}[]
\includegraphics[width=17.0cm]{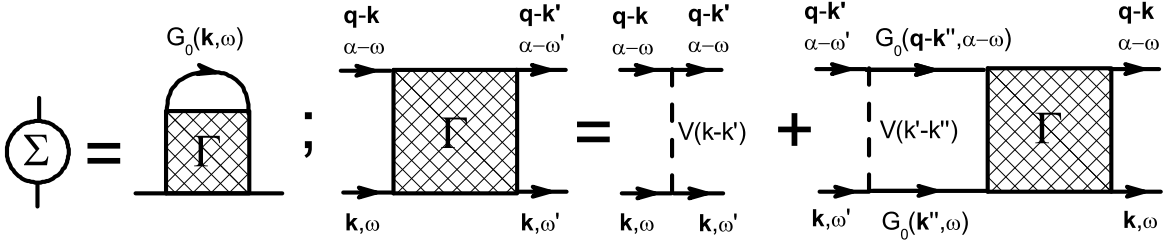}
\caption{The mass operator $\Sigma$ expressed via
$\Gamma(\textbf{q},\alpha)$-matrix. Bethe-Solpiter equation for
$\Gamma(\textbf{q},\alpha)$-matrix.} \label{fig1}
\end{figure}

The mass operator $\Sigma(G_{0})$ in stair approximation is
determined by so called  $\Gamma$-matrix (amplitude of
scattering), which is solution of the equation represented by the
diagram in Fig.\ref{fig1}.
For the case $\textbf{q}=0$ $\Gamma$-matrix has a form:
\begin{equation}\label{2.4}
    \Gamma=\frac{\lambda w_{k'}w_{k}}{1-i\lambda\int\frac{d^{3}k}{(2\pi)^{3}}\frac{d\omega}{2\pi}w_{k}
    G_{0}(\textbf{k},\alpha-\omega),G_{0}(\textbf{k},\omega)}=\frac{\lambda w_{k'}w_{k}}{1-|\lambda|\frac{mk_{F}}{4\pi^{2}}\ln\left|\frac{4\omega_{D}^{2}}{\alpha^{2}}-1\right|}.
\end{equation}
This expression has a pole in the point $\alpha_{0}$. In its
neighborhood the expression has a view:
\begin{equation}\label{2.5}
    \Gamma(\alpha\rightarrow\alpha_{0})=-\frac{2\pi^{2}}{mk_{F}}\frac{\pm i|\alpha_{0}|}{\alpha\pm i|\alpha_{0}|},
    \qquad \alpha^{2}_{0}(\lambda\rightarrow0)=-4\omega_{D}^{2}\exp\left\{-\frac{4\pi^{2}}{mk_{F}|\lambda|}\right\}.
\end{equation}
If with the help of Fourier transformation to pass from
$\omega$-representation to $t$-representation, then we shall have
the expression (\ref{1.10}), which increases infinitely at
$t\rightarrow \infty$. Hence, according to the first diagram in
Fig.\ref{fig1}, we have the same bad behavior of the mass operator
$\Sigma$. The presence of a pole in imaginary axis means
instability of a system. In our opinion, this result can be
interpreted as follows: in a system because of the interaction
(\ref{2.3}) the strong fluctuations exists, but the free
propagator $G_{0}$ doesn't consider these perturbations. In turn,
it leads to the instable solution of Dyson equation. This means,
that besides the interaction between particles $V(k)$ the
interaction of particles with aforesaid fluctuations exists too.
For stability of the solution, dressed propagators $G_{S}$, which
considers scattering on the fluctuations, must be in the equation
for the mass operator $\Sigma$ and scattering amplitude $\Gamma$
in Fig.(\ref{fig1}) instead the free propagators $G_{0}$.

Let's determine a two-particle propagator by the expression:
\begin{eqnarray}\label{2.6}
  &&K(x_{1},x_{2};x_{3},x_{4})=\langle\Psi_{0}|T\widehat{C}(x_{1})\widehat{C}(x_{2})\widehat{C}^{+}(x_{3})\widehat{C}^{+}(x_{4})|\Psi_{0}\rangle\nonumber\\
  &&K_{0}(x_{1},x_{2};x_{3},x_{4})=\langle\Phi_{0}|T\widehat{C}(x_{1})\widehat{C}(x_{2})\widehat{C}^{+}(x_{3})\widehat{C}^{+}(x_{4})|\Phi_{0}\rangle=
G_{0}(x_{1},x_{3})G_{0}(x_{2},x_{4}),
\end{eqnarray}
where $K$ is a propagator in a system with interaction, $K_{0}$ is
a free two-particle propagator and it is equal to a product of
one-particle propagators $G_{0}$, $x\equiv (\xi,t)$. The free and
dressed two-particle propagators are connected by Bethe-Solpiter
equation (in an operator view):
\begin{eqnarray}\label{2.7}
  K=K_{0}+K_{0}i\upsilon K=K_{0}+K_{0}i\Gamma
K_{0}\Rightarrow\Gamma=\upsilon+i\upsilon K_{0}\Gamma,
\end{eqnarray}
where
$\upsilon(x_{1},x_{2};x_{3},x_{4})=V(\textbf{r}_{1}-\textbf{r}_{2})\delta(x_{1}-x_{2})\delta(x_{2}-x_{4})\delta(t_{1}-t_{2})$.
The last equation in (\ref{2.7}) for the scattering amplitude
$\Gamma$ is represented graphically in Fig.\ref{fig1}. We can see
from the equation (\ref{2.7}), that a presence of the pole in
$\Gamma$ means a presence the same pole in the two-particle
propagator $K$. As it is well known
\citep{migdal,naka,mand,macf,kar}, \emph{a two-particle propagator
has a pole structure at the values of energy corresponding to a
bound state}. The pole $|\alpha_{0}|$ (\ref{2.5}) doesn't belong
to the free propagator $K_{0}$, but it appears as a result of the
attraction (\ref{2.3}) $\lambda<0$. Hence, the pole means a
presence of bound states of two particles in a system with the
binding energy $E_{s}\approx|\alpha_{0}|$.

For investigation of the bound state and calculation of of
particles' interaction with above described fluctuations let's
consider the problem of two particles at first. Previously we
considered dynamics of two selected particles being in a field of
the rest particles of a system. Therefore the propagator $K$
determined dynamics of all particles of a system in such
approximation. Now we shall consider the system consisting from
two particles only being in the state $\Phi_{s}$ with the energy
$E_{s}$. Interaction between them is described by the potential
$V(k)\equiv V_{12}$:
\begin{equation}\label{2.8}
    (H_{1}+H_{2}+V_{12})\Phi_{s}=E_{s}\Phi_{s}.
\end{equation}
Let's determine a propagator for two particles as kernel of the
integral operator finding $\Phi(t)$ known $\Phi_{1}(t')$:
\begin{equation}\label{2.9}
    \Phi(\xi_{1},\xi_{2},t,t')=-\int K(\xi_{1},\xi_{2},t;\xi_{1}',\xi_{2}',t')\Phi_{1}(\xi_{1}',\xi_{2}',t')d\xi_{1}'d\xi_{2}'.
\end{equation}
Then the two-particle propagator can be written in Fourier
representation as \citep{migdal}:
\begin{eqnarray}\label{2.14}
  -K(\xi_{1},\xi_{2};\xi_{1}',\xi_{2}';E)=i\sum_{s}\frac{\Phi_{s}(\xi_{1},\xi_{2})\Phi_{s}^{\ast}(\xi_{1}',\xi_{2}')}{E-E_{s}+i\gamma}.
\end{eqnarray}
If the bound state is among states $s$, then the pole of the
function $K(\xi_{1},\xi_{2},E)$ corresponds to the bound state $s$
at a real $E$, where $E$ is equal to energy of the bound state
$E_{s}$.

The residue in a pole $E=E_{s}$ is
$\Phi_{s}(\xi_{1},\xi_{2})\Phi_{s}^{\ast}(\xi_{1}',\xi_{2}')$. As
$K_{0}$ hasn't a pole in $E=E_{s}$, where $E_{s}$ is the energy of
the bound state, then a pole of $K$ means a presence of a pole at
$\Gamma$ according to the equation (\ref{2.7}). For the function
$K(x_{1},x_{2};x_{1}',x_{2})'$ with different times the expression
can be written:
\begin{eqnarray}\label{2.15}
  iK(\xi_{1},\xi_{2},\tau_{1};\xi_{1}',\xi_{2}',\tau_{1}';E)=\sum_{s}\frac{\Pi_{s}(\xi_{1},\xi_{2},\tau_{1};\xi_{1}',\xi_{2}',\tau_{1}')}{E-E_{s}+i\gamma},
\end{eqnarray}
where we denoted that
\begin{eqnarray*}
  t_{1}-t_{2}=\tau_{1}',\qquad t_{1}-t_{2}'=\tau_{1}',\qquad t_{1}+t_{2}=2t,\qquad
  t_{1}'+t_{2}'=2t'.
\end{eqnarray*}
This expression is the formula (\ref{2.14}) at
$\tau_{1}=\tau_{1}'=0$.

\begin{figure}[h]
\includegraphics[width=8.0cm]{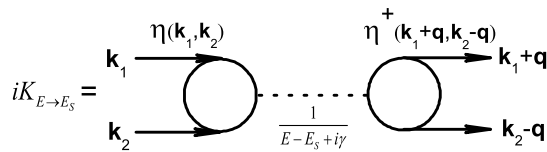}
\caption{The two-particle propagator for isolated pair of
particles in neighborhood of a pole corresponding to the bound
state $E_{s}$.} \label{fig2}
\end{figure}

The residue
$\Pi_{s}(\xi_{1},\xi_{2},\tau_{1};\xi_{1}',\xi_{2}',\tau_{1}')$
for the bound state can be written in multiplicative form as in
the case $\tau_{1}=\tau_{1}'=0$:
\begin{equation}\label{2.16}
\Pi_{s}=\eta_{s}(\xi_{1},\xi_{2},\tau_{1})\eta_{s}^{+}(\xi_{1}',\xi_{2}',\tau_{1}').
\end{equation}
The values $\eta$ and $\eta^{+}$ are Bethe-Solpiter amplitudes
\citep{naka,mand,macf,kar}. They are connected with the wave
functions $\Phi_{s}$ as (in momentum representation $\xi\equiv
\textbf{k}$):
$\Phi_{s}(\textbf{k}_{1},\textbf{k}_{2})=\int\eta_{s}(\textbf{k}_{1},\textbf{k}_{2},\epsilon)\frac{d\epsilon}{2\pi}$.
In the equation for $K$ (\ref{2.7}), $K_{0}$ can be neglected near
a pole corresponding to a bound state. Resulting homogeneous
equation has the solution (\ref{2.16}), moreover the function
$\eta_{s}(\xi_{1},\xi_{2},\tau_{1})$ satisfies the equation
\begin{equation}\label{2.17}
    \eta=iK_{0}\upsilon\eta\Longleftrightarrow\eta_{s}(\textbf{k}_{1},\textbf{k}_{2},\epsilon)=
iG_{0}(\textbf{k}_{1},E_{s}/2-\epsilon)G_{0}(\textbf{k}_{2},E_{s}/2+\epsilon)\int
V(k)\eta_{s}(\textbf{k}_{1}+\textbf{q},\textbf{k}_{2}-\textbf{q},\epsilon')\frac{d\epsilon'}{2\pi}\frac{d\textbf{k}}{(2\pi)^{3}}.
\end{equation}
The term in the sum (\ref{2.15}), corresponding to the bound state
$E_{s}$, can be represented by the diagram in Fig.\ref{fig2},
where the dotted line means the multiplier
$\frac{1}{E-E_{s}+i\gamma}$ - propagation of two particles in
bound state, and the blocks are
$\eta_{s}(\textbf{k}_{1},\textbf{k}_{2},\epsilon)$ and
$\eta_{s}^{+}(\textbf{k}_{1}+\textbf{q},\textbf{k}_{2}-\textbf{q},\epsilon)$
- transition amplitudes in bound state and back. The diagram in
Fig.\ref{fig2} can be interpreted by the following way. Two
particles with momentums $\textbf{k}_{1}$ and $\textbf{k}_{2}$
form a bound state with energy $E_{s}$ with amplitude
$\eta_{s}(\textbf{k}_{1},\textbf{k}_{2})$. Further, the two
particles propagate together. Then the bound state can decay with
amplitude
$\eta_{s}^{+}(\textbf{k}_{1}+\textbf{q},\textbf{k}_{2}-\textbf{q})$.
As a result the two free particles appear with momentums
$\textbf{k}_{1}+\textbf{q}$ and $\textbf{k}_{2}-\textbf{q}$.

\section{The uncoupling of correlations and a multiparticle dynamics.}\label{correlations}

In the previous section we considered dynamics of two isolated
particles. Now we have to generalize the obtained results to the
multi-particle case - propagation of two interacting particles in
a system of identical fermions. This situation differs from the
previous case by the following conditions:
\begin{enumerate}
    \item Each pair of fermions is in field of all the rest particles.
    \item All particles of a system are identical. Moreover, the average size of a pair $l_{0}\sim
1/\sqrt{|\alpha_{0}|2m}\gg\sqrt[3]{V/N}$ is more big than average
distant between particles, that means the wave packages of pairs
overlap strongly.
\end{enumerate}
Mathematically this means, that \emph{the amplitudes $\eta$ and
$\eta^{+}$ are not solution of the equation} (\ref{2.17}),
\emph{which is correct for isolated pair only. Now the amplitudes
are determined by dynamics of all particles of the system, and
their observed value is result of an averaging over a system}.
Pairs in such system are effective, namely two fermions having
formed a bound state with an amplitude
$\eta_{s}(\textbf{k}_{1},\textbf{k}_{2},\epsilon)$
(Fig.\ref{fig2}) are not fixed pair: one from partners in a pair
can leave the bound state with a fermion from another pair with
amplitude
$\eta_{s}^{+}(\textbf{k}_{1}+\textbf{q},\textbf{k}_{2}-\textbf{q},\epsilon)$.
Thus, the collective of pairs (condensate) exists.

\begin{figure}[h]
\includegraphics[width=12cm]{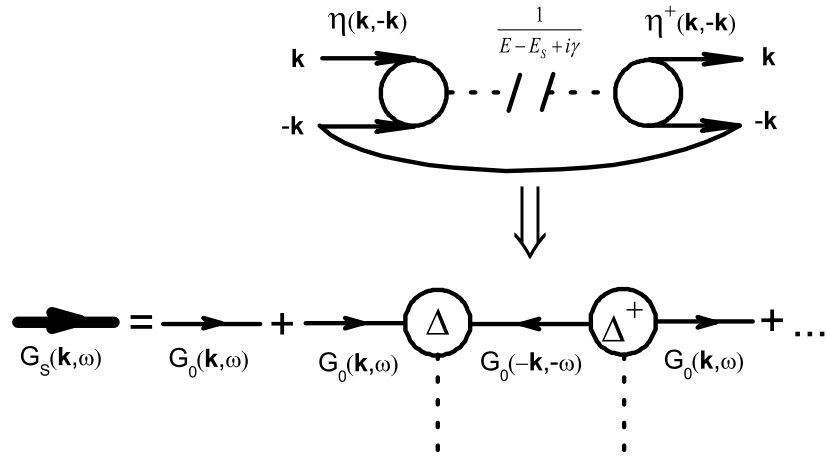}
\caption{The procedure of uncoupling of correlations for a
two-particle propagator of a pair being in a field of all rest
fermion of a system. The result of the uncoupling is the dressed
one-particle propagator $G_{S}(\textbf{k},\omega)$, as a
consequence of interaction of a free fermion with fluctuations of
pairing.} \label{fig3}
\end{figure}

Let's consider the two-particle propagator $K_{E\rightarrow
E_{s}}$ represented in Fig.\ref{fig2}. As it has been noted
earlier, fermions with opposite momentums and and opposite spins
form a pair. Let's suppose, that corresponding amplitudes of
pairing don't depend on time. In order to obtain an one-particle
propagator $G_{S}$ we shall use the method of uncoupling of
correlations considered in the Appendix \ref{appendixA}, and we
shall be acting analogously to Fig.\ref{fig11}. The procedure of
an uncoupling is represented in Fig.\ref{fig3}. We connect the
entering line and the outgoing line corresponding to particles
with momentum $-\textbf{k}$ and energy parameter $-\omega$ each.
As a result we have the intermediate propagator $G_{0}$. Since
partners in each pair are not fixed in consequence of the identity
principle and strong intersection of wave packages of pairs, that
is a condensate of pairs exists, then it is necessary to cut the
dotted line - the propagator of a pair
$\frac{1}{E-E_{s}+i\gamma}$. Then the points of a joining of the
dotted lines correspond to interaction with the effective field
(in accordance with the rules of diagram technics in the Appendix
\ref{appendixA}).  The fluctuation of pairing play a role of the
above mentioned effective field. The amplitudes of such
interaction we denote as $-i\Delta(\textbf{k},-\textbf{k})$ and
$i\Delta^{+}(\textbf{k},-\textbf{k})$. These values correspond to
the amplitudes of the two-particle problem
$\eta_{s}(\textbf{k},-\textbf{k})$ and
$\eta_{s}^{+}(\textbf{k},-\textbf{k})$ to the extent that their
observed values is result of averaging over a system
$\langle\eta_{s}(\textbf{k},-\textbf{k})\rangle\sim\Delta$ è
$\langle\eta_{s}^{+}(\textbf{k},-\textbf{k})\rangle\sim\Delta^{+}$
in consequence of statistical correlations between pairs and they
are determined by dynamics of all system's particles.

The result of the procedure of uncoupling of correlations means
the follows. Let an additional particle with momentum
$\textbf{k},\omega$ propagates through a system of identical
fermions. In the process of propagation a particle can form bound
states with other fermions according to the following mechanism.
Some pair of fermions decays in components with momentums
$-\textbf{k},-\omega$ and $\textbf{k},\omega$ with amplitude
$i\Delta^{+}$. Second particle of the decayed pair is in state of
the additional particle ($\textbf{k},\omega$) and it is identical
to the additional particle. The second particle propagates through
a system further. First particle of decayed pair forms bound state
with the initial additional particle with amplitude $-i\Delta$.
Anew formed pair replenishes the condensate of pairs in a system.
Thus, the dressed propagator $G_{S}$ takes into account
interaction of a particle, initially described by free propagator
$G_{0}$, with fluctuations of pairing. Intensity of the
interaction is the amplitudes $-i\Delta$ and $i\Delta^{+}$.

\begin{figure}[h]
\includegraphics[width=8.5cm]{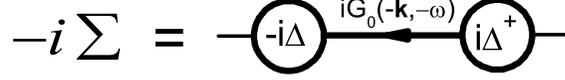}
\caption{The diagram for the mass operator  $\Sigma$ describing
interaction of a fermion with fluctuations of pairing.}
\label{fig4}
\end{figure}

Starting from the aforesaid, we can write the mass operator for
such process (Fig.\ref{fig4}) as
\begin{eqnarray}\label{3.1}
    -i\Sigma=-i\Delta
iG_{0}(-k,-\omega)i\Delta^{+}\Rightarrow\Sigma=\frac{\Delta\Delta^{+}}{\omega+\varepsilon(k)}.
\end{eqnarray}
This mass operator has been proposed in \citep{migdal,pines},
however an existence of the amplitudes $\Delta,\Delta^{+}$ and the
equation of self-consistency were postulated (as the anomalous
averages). From Dyson equation (\ref{A6}) we can obtain the
dressed one-particle propagator:
\begin{eqnarray}\label{3.2}
&&G_{S}(\textbf{k},\omega)=\frac{1}{\omega-\varepsilon-\Sigma}=\frac{\omega+\varepsilon}{\omega^{2}-E^{2}}=\frac{A_{S}}{\omega-E}+\frac{B_{S}}{\omega+E}\nonumber\\
&&G_{S}(\textbf{k},t)=-i\theta_{t}A_{S}e^{-iEt}+i\theta_{-t}B_{S}e^{iEt}\\
&&A_{S}=\frac{1}{2}\left(1+\frac{\varepsilon}{E}\right),\qquad
B_{S}=\frac{1}{2}\left(1-\frac{\varepsilon}{E}\right),\nonumber
\end{eqnarray}
where
\begin{equation}\label{3.3}
    E(k)=\sqrt{\varepsilon^{2}(k)+\Delta^{2}}
\end{equation}
is dispersion law of dressed particles (quasi-particles). The
amplitude $\Delta$ is named by gap, because minimal work for
creation of one-particle excitations is $2\Delta$. In accordance
with the definition of one-particle propagator we can write:
\begin{eqnarray}\label{3.4}
  &&G_{S}(\textbf{k},t)=\left\{\begin{array}{cc}
    -i\langle\Psi_{0}|C_{\textbf{k},\sigma}(t)C_{\textbf{k},\sigma}^{+}(0)|\Psi_{0}\rangle, \qquad t>0 \\
    i\langle\Psi_{0}|C_{\textbf{k},\sigma}^{+}(0)C_{\textbf{k},\sigma}(t)|\Psi_{0}\rangle, \qquad t\leq0 \\
  \end{array}\right\},
\end{eqnarray}
where the system is placed in another ground state $\Psi_{0}$. In
the state $\Psi_{0}$ interaction of particles with fluctuation of
pairing (existence of condensate of pairs) is taken into account,
moreover $\Psi_{0}(\Delta=0)=\Phi_{0}$, $G_{S}(\Delta=0)=G_{0}$. A
propagator defines occupations number of quasi-particles
$n_{\textbf{k}}$ by the following way:
\begin{eqnarray}\label{3.5}
n_{\textbf{k}}=\langle\Psi_{0}|C_{\textbf{k}}^{+}C_{\textbf{k}}|\Psi_{0}\rangle=-i\lim_{t\rightarrow0^{-}}G_{S}(\textbf{k},t)=B_{S},
\qquad
1-n_{\textbf{k}}=\langle\Psi_{0}|C_{\textbf{k}}C_{\textbf{k}}^{+}|\Psi_{0}\rangle=i\lim_{t\rightarrow0^{+}}G_{S}(\textbf{k},t)=A_{S}.
\end{eqnarray}
Hence, we can suppose, that
\begin{equation}\label{3.6}
C_{\textbf{k}}^{+}|\Psi_{0}\rangle=\sqrt{A_{S}}|\Psi_{0},1_{\textbf{k}}^{p}\rangle,
\qquad
C_{\textbf{k}}|\Psi_{0}\rangle=\sqrt{B_{S}}|\Psi_{0},1_{\textbf{k}}^{h}\rangle,
\end{equation}
where $|\Psi_{0},1_{\textbf{k}}^{p}\rangle$ and
$|\Psi_{0},1_{\textbf{k}}^{h}\rangle$ are states with one added
particle and one removed particle with momentum $\textbf{k}$
accordingly. Hamiltonian of a system of free quasi-particle has a
view:
\begin{equation}\label{3.7}
\widehat{H}_{0}=\sum_{\textbf{k}}E(k)C_{\textbf{k}}^{+}C_{\textbf{k}},
\end{equation}
Chemical potential of a quasi-particles' system equals to zero.
Therefore the grand potential $\Omega$ coincides with Helmholtz
free energy in a superconductive state. The states
$|\Psi_{0}\rangle$, $|\Psi_{0},1_{\textbf{k}}^{p}\rangle$ and
$|\Psi_{0},1_{\textbf{k}}^{h}\rangle$ are eigenvectors of the
Hamiltonian (\ref{3.7}):
\begin{equation}\label{3.7a}
   \widehat{H}_{0}|\Psi_{0}\rangle=\Omega_{0}|\Psi_{0}\rangle,
\qquad
\widehat{H}_{0}|\Psi_{0},1_{\textbf{k}}^{p}\rangle=(\Omega_{0}+E(k))|\Psi_{0},1_{\textbf{k}}^{p}\rangle,
\qquad
\widehat{H}_{0}|\Psi_{0},1_{\textbf{k}}^{h}\rangle=(\Omega_{0}+E(k))|\Psi_{0},1_{\textbf{k}}^{h}\rangle.
\end{equation}
Then, using the definition (\ref{3.4}), we can find:
\begin{eqnarray}\label{3.8}
G_{S}(\textbf{k},t>0)&=&-i\langle\Psi_{0}|C_{\textbf{k}}(t)C_{\textbf{k}}^{+}(0)|\Psi_{0}\rangle=
-i\langle\Psi_{0}|e^{i\widehat{H}_{0}t}C_{\textbf{k}}e^{-i\widehat{H}_{0}t}C_{\textbf{k}}^{+}|\Psi_{0}\rangle\nonumber\\
&=&-i\langle\Psi_{0},1_{\textbf{k}}^{p}|e^{i\Omega_{0}t}\sqrt{A_{S}}e^{-i(\Omega_{0}+E(k))t}\sqrt{A_{S}}|\Psi_{0},1_{\textbf{k}}^{p}\rangle\nonumber\\
&=&-iA_{S}e^{-iEt}\langle\Psi_{0},1_{\textbf{k}}^{p}|\Psi_{0},1_{\textbf{k}}^{p}\rangle=-iA_{S}e^{-iEt},
\end{eqnarray}
that coincides with (\ref{3.2}). For $t\leq0$ the proof is
analogous.

Dyson equation can be represented in other form. Let's use the
definition $(\omega-\varepsilon)G_{0}=1$. On the other hand we
have $G_{0}=G_{S}/(1+G_{S}\Sigma)$. Moreover, let's introduce the
notations
\begin{eqnarray}\label{3.9}
-G_{S}\Sigma\equiv\Delta F^{+}, \qquad
-G_{S}\Sigma\equiv\Delta^{+}F.
\end{eqnarray}
Then, we can obtain the set of equations:
\begin{eqnarray}
  &&(\omega-\varepsilon(k))G_{S}(\textbf{k},\omega)+\Delta F^{+}(\textbf{k},\omega)=1\label{3.10}\\
  &&(\omega+\varepsilon(k))F^{+}(\textbf{k},\omega)+\Delta^{+}G_{S}(\textbf{k},\omega)=0\label{3.11}.
\end{eqnarray}
These equations are Gor'kov equations in momentum representation.
However, unlike phenomenological approach (where existence of the
anomalous propagator $F$ and the equation for order parameter are
postulated) these equations are obtained by microscopic way with
help of the procedure of uncoupling of correlations. From the
equations (\ref{3.10},\ref{3.11}) we can find, that
\begin{equation}\label{3.12}
F^{+}=\frac{-\Delta^{+}}{\omega^{2}-E^{2}}, \qquad
F=\frac{-\Delta}{\omega^{2}-E^{2}}.
\end{equation}
The anomalous propagators describe creation of two fermions from
the condensate of pairs - $F^{+}$, formation a pair by two
particles with leaving to the condensate - $F$. Moreover, $F$ and
$F^{+}$ are the infinity sum of the serial processes of creation
and annihilation of pairs described by amplitudes $\Delta$ and
$\Delta^{+}$. Mathematically this is expressed in the fact that
\begin{eqnarray}\label{3.13}
  F^{+}_{\alpha\beta}(\textbf{k},t)&=&\frac{\Delta^{+}}{\sqrt{\Delta^{+}\Delta}}\left\{\begin{array}{cc}
    i\langle\Psi_{0}|C_{-\textbf{k},\beta}^{+}(t)C_{\textbf{k},\alpha}^{+}(0)|\Psi_{0}\rangle, \qquad t>0 \\
    i\langle\Psi_{0}|C_{\textbf{k},\alpha}^{+}(0)C_{-\textbf{k},\beta}^{+}(t)|\Psi_{0}\rangle, \qquad t\leq0 \\
  \end{array}\right\}\nonumber\\
&=&g_{\alpha\beta}\left(i\theta_{t}\frac{\Delta^{+}}{\sqrt{\Delta^{+}\Delta}}\sqrt{A_{S}B_{S}}e^{-iEt}
+i\theta_{-t}\frac{\Delta^{+}}{\sqrt{\Delta^{+}\Delta}}\sqrt{A_{S}B_{S}}e^{iEt}\right),\\
F_{\alpha\beta}(\textbf{k},t)&=&\frac{\Delta}{\sqrt{\Delta^{+}\Delta}}\left\{\begin{array}{cc}
    i\langle\Psi_{0}|C_{\textbf{k},\alpha}(t)C_{-\textbf{k},\beta}(0)|\Psi_{0}\rangle, \qquad t>0 \\
    i\langle\Psi_{0}|C_{-\textbf{k},\beta}(0)C_{\textbf{k},\alpha}(t)|\Psi_{0}\rangle, \qquad t\leq0 \\
  \end{array}\right\}\nonumber\\
&=&g_{\alpha\beta}\left(i\theta_{t}\frac{\Delta}{\sqrt{\Delta^{+}\Delta}}\sqrt{A_{S}B_{S}}e^{-iEt}
+i\theta_{-t}\frac{\Delta}{\sqrt{\Delta^{+}\Delta}}\sqrt{A_{S}B_{S}}e^{iEt}\right),
\qquad g_{\alpha\beta}=\left(%
\begin{array}{cc}
  0 & 1 \\
  1 & 0 \\
\end{array}%
\right)\label{3.13a}.
\end{eqnarray}
Let's prove the formulas (\ref{3.13},\ref{3.13a}). For this let's
consider the expression:
\begin{eqnarray}\label{3.14}
&&\langle\Psi_{0}|C_{-\textbf{k},\beta}^{+}(t)C_{\textbf{k},\alpha}^{+}(0)|\Psi_{0}\rangle=
\langle\Psi_{0}|e^{i\widehat{H}_{0}t}C_{-\textbf{k},\beta}^{+}e^{-i\widehat{H}_{0}t}C_{\textbf{k},\alpha}^{+}|\Psi_{0}\rangle\nonumber\\
&&=\langle\Psi_{0},1_{-\textbf{k},\alpha}^{h}|e^{i\Omega_{0}t}\sqrt{B_{S}}e^{-i(\Omega_{0}+E(k))t}\sqrt{A_{S}}|\Psi_{0},1_{\textbf{k},\beta}^{p}\rangle\nonumber\\
&&=\sqrt{A_{S}B_{S}}e^{-iE(k)t}\langle\Psi_{0},1_{-\textbf{k},\alpha}^{h}|\Psi_{0},1_{\textbf{k},\beta}^{p}\rangle=\sqrt{A_{S}B_{S}}e^{-iE(k)t},
\end{eqnarray}
that corresponds to (\ref{3.13}). In the last equality the fact
has been used, that the states, created by addition of a particle
to a state $\textbf{k},\alpha$ or removing of a particle from a
state $-\textbf{k},\beta$ at $\alpha\neq\beta$, are identical in
the course of existence of the pair condensate. The rest cases is
proved analogously. It is not difficult to see, that
$F(\Delta=0)=F^{+}(\Delta=0)=0$, because $A_{0}(k)B_{0}(k)=0$.
From (\ref{3.13}) we can see, \emph{that the existence of nonzero
anomalous propagators $F$ and $F^{+}$ means breakdown of global
gauge symmetry in a system, that is number of particles is not
conserved in the course of existence of a pair condensate}. Hence
the states $\Phi_{0}$ and $\Psi_{0}$ have different symmetries:
\begin{equation}\label{3.15}
    \langle\Phi_{0}|\Psi_{0}\rangle=0.
\end{equation}
However distribution function over $N$ has a maximum at the
average number of particles $\langle N\rangle$ determined by the
expression:
\begin{eqnarray}\label{3.16}
\langle
N\rangle=2\sum_{\textbf{k}}n_{\textbf{k}}=-2i\int\frac{d^{3}k}{(2\pi)^{2}}\frac{d\omega}{2\pi}\lim_{t\rightarrow0^{-}}G_{S}(\textbf{k},t)
=2\int\frac{d^{3}k}{(2\pi)^{2}}B_{S}(k)\approx
N=2\int\frac{d^{3}k}{(2\pi)^{2}}B_{0}(k).
\end{eqnarray}

Now let's return to the left part of the expression (\ref{2.4})
for $\Gamma$-matrix. After considering of particles' interaction
with fluctuations of pairing we have to substitute dressed
propagators $G_{S}$ instead free propagators $G_{0}$ in the
formula (\ref{2.4}). It is not difficult to verify, that
\textit{$\Gamma$ hasn't poles at any $\alpha$ and $\Delta$. This
means, that the problem of instability of a system is removed, and
we can use dressed propagator for further calculations
confidently. Moreover, the absence of poles means the absence of
bound states, because we have taken into account them in the
specter of quasi-particles $E(k)$} (\ref{3.3}).

It is necessary to note, that the mass operator (\ref{3.1}) and
Gor'kov equation (\ref{3.10},\ref{3.11}) haven't parameter of
interaction between particles. This means, that the amplitudes of
pairing $\Delta$ and $\Delta^{+}$ exists regardless of interaction
between particles and its type. However, as we shall see below,
the interaction determines the average value of the amplitudes,
that is observed in experiment. This average value is not zero in
the case of attraction between particle only.

\section{Ground state energy.}\label{ground}

\subsection{Summary kinetic energy of particles.}

In order to calculate ground state energy it is necessary to know
kinetic energy of particles and energy of their interaction. The
operator of kinetic energy of all particles of a system has a
form:
\begin{equation}\label{4.1}
    \widehat{W}=\sum_{\textbf{k},\alpha}v_{F}(k-k_{F})C_{\textbf{k},\alpha}^{+}C_{\textbf{k},\alpha}
=2\sum_{\textbf{k}}\varepsilon(k)C_{\textbf{k},\alpha}^{+}C_{\textbf{k},\alpha}.
\end{equation}
Then the corresponding average value is
\begin{eqnarray}\label{4.2}
    \langle
W\rangle&=&-2i\sum_{\textbf{k}}G(\textbf{k},t\rightarrow0^{-})\varepsilon(k)
=-2i\lim_{t\rightarrow0^{-}}\sum_{\textbf{k}}\int\frac{d\omega}{2\pi}G(\textbf{k},\omega)e^{-i\omega t}\varepsilon(k)\nonumber\\
&=&2\sum_{\textbf{k}}B(k)\varepsilon(k)=V\nu_{F}\int_{-v_{F}k_{F}}^{\infty}B(\varepsilon)\varepsilon
d\varepsilon.
\end{eqnarray}
Here $\nu_{F}=\frac{k_{F}^{2}}{\pi^{2}v_{F}}$ is density of states
on Fermy surface. Since the interaction
$V_{\textbf{l}-\textbf{lk}-\textbf{k}}$ (\ref{2.3}) exists in the
layer $-\omega_{D}<\varepsilon(k)<\omega_{D}$ only, we can suppose
that
\begin{eqnarray}\label{4.3}
G=\left[
\begin{array}{cc}
  G_{0}; & |\varepsilon(k)|>\omega_{D} \\
  G_{S}; & |\varepsilon(k)|<\omega_{D} \\
\end{array}%
\right], \qquad A(k)=\left[
\begin{array}{cc}
  A_{0}(k); & |\varepsilon(k)|>\omega_{D} \\
  A_{S}(k); & |\varepsilon(k)|<\omega_{D} \\
\end{array}%
\right], \qquad B(k)=\left[
\begin{array}{cc}
  B_{0}(k); & |\varepsilon(k)|>\omega_{D} \\
  B_{S}(k); & |\varepsilon(k)|<\omega_{D} \\
\end{array}%
\right].
\end{eqnarray}
Then we can separate a normal part and a superconductive part of
the kinetic energy:
\begin{eqnarray}\label{4.4}
    &&\langle W\rangle=V\nu_{F}\int_{-v_{F}k_{F}}^{-\omega_{D}}B_{0}\varepsilon
d\varepsilon+V\nu_{F}\int_{-\omega_{D}}^{\omega_{D}}B_{S}\varepsilon
d\varepsilon+V\nu_{F}\int_{\omega_{D}}^{\infty}B_{0}\varepsilon
d\varepsilon\nonumber\\
&&=W_{n}+V\nu_{F}\int_{-\omega_{D}}^{\omega_{D}}B_{S}\varepsilon
d\varepsilon-V\nu_{F}\int_{-\omega_{D}}^{\omega_{D}}B_{0}\varepsilon
d\varepsilon
=W_{n}-V\frac{\nu_{F}}{2}\int_{-\omega_{D}}^{\omega_{D}}\left(\frac{\varepsilon^{2}}{E}-\frac{\varepsilon^{2}}{|\varepsilon|}\right)\nonumber\\
&&=W_{n}+V\frac{\nu_{F}}{2}\left(\omega_{D}^{2}-\omega_{D}\sqrt{\omega_{D}^{2}+\Delta^{2}}+\Delta^{2}
\texttt{arcsinh}\frac{\omega_{D}}{\Delta}\right).
\end{eqnarray}
We can see, that the pairing leads to a loss in the kinetic
energy.

\subsection{Vacuum amplitude.}

Since we took account of interaction of particles with
fluctuations of pairing and we discovered that the ground state of
a system $|\Psi_{0}\rangle$ has other symmetry as compared with
the initial state $|\Phi_{0}\rangle$, hence the vacuum amplitude
of a system can be written as:
\begin{eqnarray}\label{4.5}
R(t)&=&\langle\Psi_{0}|
\widetilde{U}(t-t_{0})|\Psi_{0}\rangle_{t_{0}=0}=\langle\Psi_{0}|
U(t)|\Psi_{0}\rangle e^{iW_{0}t}\nonumber\\
&=&\sum_{n=0}^{\infty}\frac{(-i)^{n}}{n!}\int_{0}^{t}dt_{1}\ldots\int_{0}^{t}dt_{n}
\left\langle\Psi_{0}|
T\left\{\widehat{H}_{I}(t_{1})\ldots\widehat{H}_{I}(t_{n})\right\}|\Psi_{0}\right\rangle,
\end{eqnarray}
where
$\widehat{H}_{I}(t)=e^{+it\hat{H_{0}}}\widehat{V}e^{-it\hat{H_{0}}}$
is operator of particles' interaction in an interaction
representation, $W_{0}$ - ground state energy of a system without
interaction. The averaging $\langle\rangle_{0}\equiv
\langle\Psi_{0}|\ldots|\Psi_{0}\rangle$ is realized by ensemble of
\emph{noninteracting quasi particles}. The Hamiltonian of such
system has the following form:
\begin{equation}\label{4.6}
\widehat{H}_{0}+\widehat{V}=\sum_{\alpha}\sum_{\textbf{k}}E(k)C_{\textbf{k},\alpha}^{+}C_{\textbf{k},\alpha}
+\frac{1}{2V}\sum_{\alpha,\beta,\gamma,\delta}\sum_{\textbf{k},\textbf{l},\textbf{m},\textbf{n}}V_{klmn}C_{\textbf{l},\beta}^{+}C_{\textbf{k},\alpha}^{+}C_{\textbf{m},\gamma}C_{\textbf{n},\delta},
\end{equation}
where momentum is conserved
$\textbf{k}+\textbf{l}=\textbf{m}+\textbf{n}$ and spin is
conserved $\alpha+\beta=\gamma+\delta$, $V$ is volume of a system.
The sequence order of indexes of matrix elements and of creation
and annihilation operators is important.

The expressions for several first orders ($n=0,1,2...$) in the
expansion of vacuum amplitude are (let us suppose $t_{2}>t_{1}$
for definiteness):
\begin{eqnarray}\label{4.7}
&&R_{0}(t)=\langle\Psi_{0}|\Psi_{0}\rangle=1\nonumber\\
&&R_{1}(t)=\frac{1}{1!}\frac{1}{V}\int_{0}^{t}dt_{1}\sum_{\alpha,\beta,\gamma,\delta}\sum_{\textbf{k},\textbf{l},\textbf{m},\textbf{n}}\left(-\frac{i}{2}V_{klmn}\right)
\langle\Psi_{0}|C_{\textbf{l},\beta}^{+}(t_{1})C_{\textbf{k},\alpha}^{+}(t_{1})C_{\textbf{m},\gamma}(t_{1})C_{\textbf{n},\delta}(t_{1})|\Psi_{0}\rangle
\nonumber\\
&&R_{2}(t)=\frac{1}{2!}\frac{1}{V^{2}}\int_{0}^{t}dt_{1}\int_{0}^{t}dt_{2}\sum_{\alpha,\beta,\gamma,\delta}\sum_{\textbf{k},\textbf{l},\textbf{m},\textbf{n}}\left(-\frac{i}{2}V_{klmn}\right)
\sum_{\alpha',\beta',\gamma',\delta'}\sum_{\textbf{k}',\textbf{l}',\textbf{m}',\textbf{n}'}\left(-\frac{i}{2}V_{k'l'm'n'}\right)\nonumber\\
&&\times\langle\Psi_{0}|C_{\textbf{l}',\beta'}^{+}(t_{2})C_{\textbf{k}',\alpha'}^{+}(t_{2})C_{\textbf{m}',\gamma'}(t_{2})C_{\textbf{n}',\delta'}(t_{2})C_{\textbf{l},\beta}^{+}(t_{1})C_{\textbf{k},\alpha}^{+}(t_{1})C_{\textbf{m},\gamma}(t_{1})C_{\textbf{n},\delta}(t_{1})|\Psi_{0}\rangle,
\end{eqnarray}
where $C^{+}(t)=e^{+it\hat{H_{0}}}C^{+}e^{-it\hat{H_{0}}}$,
$C(t)=e^{+it\hat{H_{0}}}Ce^{-it\hat{H_{0}}}$ are operators of
creation and annihilation in Heisenberg representation, that
coincides with interaction representation for an ensemble of
noninteracting particles.

\begin{figure}[h]
\includegraphics[width=8cm]{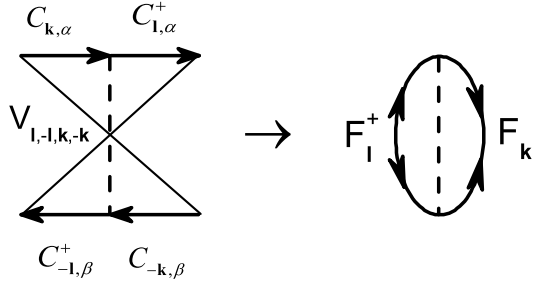}
\caption{The uncoupling of correlations in a vacuum amplitude for
the process of scattering of two fermions by the potential of
interaction $V_{\textbf{l},-\textbf{l},\textbf{k},-\textbf{k}}$
from the initial state $\textbf{k},\alpha$ and $-\textbf{k},\beta$
to the final states $\textbf{l},\alpha$ and $-\textbf{l},\beta$.
As a result we have the anomalous transition amplitude
"vacuum-vacuum".} \label{fig5}
\end{figure}

In Appendix \ref{appendixB} we propose the method of uncoupling of
correlations for approximate calculation of a vacuum amplitude
$R(t)$. As an example Hartree-Fock normal processes have been
considered there. We shall generalize this method for anomalous
processes here. In our case the particles interact by the
potential (\ref{2.3})
$V_{\textbf{l},-\textbf{l},\textbf{k},-\textbf{k}}$. Hence, the
vacuum amplitude has a form:
\begin{eqnarray}\label{4.8}
&&R(t)=1+\frac{1}{1!}\frac{1}{V}\int_{0}^{t}dt_{1}\sum_{\alpha,\beta}\sum_{\textbf{k},\textbf{l}}\left(-\frac{i}{2}V_{\textbf{l},-\textbf{l},\textbf{k},-\textbf{k}}\right)
\langle\Psi_{0}|C_{-\textbf{l},\beta}^{+}(t_{1})C_{\textbf{l},\alpha}^{+}(t_{1})C_{\textbf{k},\alpha}(t_{1})C_{-\textbf{k},\beta}(t_{1})|\Psi_{0}\rangle
\nonumber\\
&&+\frac{1}{2!}\frac{1}{V^{2}}\int_{0}^{t}dt_{2}\int_{0}^{t}dt_{1}\sum_{\alpha,\beta}\sum_{\textbf{k},\textbf{l}}\left(-\frac{i}{2}V_{\textbf{l},-\textbf{l},\textbf{k},-\textbf{k}}\right)
\sum_{\alpha',\beta'}\sum_{\textbf{k}',\textbf{l}'}\left(-\frac{i}{2}V_{\textbf{l}',-\textbf{l}',\textbf{k}',-\textbf{k}'}\right)\nonumber\\
&&\times\langle\Psi_{0}|C_{-\textbf{l}',\beta'}^{+}(t_{2})C_{\textbf{l}',\alpha'}^{+}(t_{2})C_{\textbf{k}',\alpha'}(t_{2})C_{-\textbf{k}',\beta'}(t_{2})C_{-\textbf{l},\beta}^{+}(t_{1})C_{\textbf{l},\alpha}^{+}(t_{1})C_{\textbf{k},\alpha}(t_{1})C_{-\textbf{k},\beta}(t_{1})|\Psi_{0}\rangle
+...
\end{eqnarray}
We can uncouple correlations by the following way taking into
account anticommutation of the operators $C$ and $C^{+}$:
\begin{eqnarray}\label{4.9}
&&R(t)\approx
1+\frac{1}{1!}(-1)^{2}\frac{1}{V}\int_{0}^{t}dt_{1}\sum_{\alpha,\beta}\sum_{\textbf{k},\textbf{l}}\left(-\frac{i}{2}V_{\textbf{l},-\textbf{l},\textbf{k},-\textbf{k}}\right)
\langle\Psi_{0}|C_{\textbf{l},\alpha}^{+}(t_{1})C_{-\textbf{l},\beta}^{+}(t_{1})|\Psi_{0}\rangle
\langle\Psi_{0}|C_{-\textbf{k},\beta}(t_{1})C_{\textbf{k},\alpha}(t_{1})|\Psi_{0}\rangle
\nonumber\\
&&+\frac{1}{2!}(-1)^{4}\frac{1}{V^{2}}\int_{0}^{t}dt_{2}\int_{0}^{t}dt_{1}\sum_{\alpha,\beta}\sum_{\textbf{k},\textbf{l}}\left(-\frac{i}{2}V_{\textbf{l},-\textbf{l},\textbf{k},-\textbf{k}}\right)
\sum_{\alpha',\beta'}\sum_{\textbf{k}',\textbf{l}'}\left(-\frac{i}{2}V_{\textbf{l}',-\textbf{l}',\textbf{k}',-\textbf{k}'}\right)\nonumber\\
&&\times\langle\Psi_{0}|C_{\textbf{l}',\alpha'}^{+}(t_{2})C_{-\textbf{l}',\beta'}^{+}(t_{2})|\Psi_{0}\rangle\langle\Psi_{0}|
C_{-\textbf{k}',\beta'}(t_{2})C_{\textbf{k}',\alpha'}(t_{2})|\Psi_{0}\rangle\langle\Psi_{0}|
C_{\textbf{l},\alpha}^{+}(t_{1})C_{-\textbf{l},\beta}^{+}(t_{1})|\Psi_{0}\rangle\langle\Psi_{0}|
C_{-\textbf{k},\beta}(t_{1})C_{\textbf{k},\alpha}(t_{1})|\Psi_{0}\rangle
\nonumber\\
&&+\ldots=1+R_{1}+\frac{1}{2!}R_{1}^{2}+\ldots=\exp(R_{1})
\end{eqnarray}
As a result of the uncoupling of correlations we can see, that
anomalous processes $CC$ and $C^{+}C^{+}$ give contribution to the
vacuum amplitude of a system with the interaction (\ref{2.3})
only. Another combinations of the uncoupling with an obtaining of
normal propagator $C^{+}C$ don't conserve a momentum. Such
representation of the vacuum amplitude by uncoupled correlations
is analogous to Fock approximation for normal processes, and it
means a neglect of dynamic correlation between pairs. Then $R(t)$
can be written as
\begin{eqnarray}\label{4.10}
    \ln R(t)=R_{1}(t)&=&\frac{1}{V}\int_{0}^{t}dt_{1}\sum_{\alpha,\beta}\sum_{\textbf{k},\textbf{l}}\left(-\frac{i}{2}V_{\textbf{l},-\textbf{l},\textbf{k},-\textbf{k}}\right)
\frac{\sqrt{\Delta^{+}\Delta}}{\Delta^{+}}(-i)F^{+}_{\alpha\beta}(\textbf{l},t_{1}-t_{1})
\frac{\sqrt{\Delta^{+}\Delta}}{\Delta}(-i)F_{\alpha\beta}(\textbf{k},t_{1}-t_{1})\nonumber\\
&=&\frac{2}{V}\sum_{\textbf{k},\textbf{l}}\left(-\frac{i}{2}V_{\textbf{l},-\textbf{l},\textbf{k},-\textbf{k}}\right)
(-i)F^{+}(\textbf{l},t_{1}-t_{1})(-i)F(\textbf{k},t_{1}-t_{1})t\nonumber\\
&=&\frac{i\lambda}{V}\sum_{\textbf{l}}\int\frac{d\omega}{2\pi}w_{l}F^{+}(\textbf{l},\omega)\sum_{\textbf{k}}\int\frac{d\omega}{2\pi}w_{k}F(\textbf{k},\omega)t,
\end{eqnarray}
where a summation over spin variables gave the multiplier 2. By
analogy with Fig.\ref{fig12} in Appendix \ref{appendixB}, the
process of uncoupling of correlations we can represent by
graphically in Fig.\ref{fig5}, where in the diagram of a
scattering of two fermions from the states $\textbf{k}$ and
$-\textbf{k}$ into the state $\textbf{l}$ and $-\textbf{l}$ as a
result of the interaction
$V_{\textbf{l},-\textbf{l},\textbf{k},-\textbf{k}}$, we connect
the lines corresponding to oppositely directed momentums. As a
result we have two anomalous propagators: $F(\textbf{k},\omega)$
and $F^{+}(\textbf{l},\omega)$. The obtained diagram means the
following. Two fermions with opposite momentums and opposite spins
appear from a condensate of pairs, interact with each other by the
potential $V_{\textbf{l},-\textbf{l},\textbf{k},-\textbf{k}}$ and
go back into the condensate of pairs.

Using the formulas (\ref{3.13}) and (\ref{3.13a}) we can rewrite
$R(t)$ in the following form:
\begin{eqnarray}\label{4.11}
\ln
R(t)&=&\frac{-i\lambda}{V}\left(V\frac{\nu_{F}}{2}\right)^{2}t\int_{-\omega_{D}}^{\omega_{D}}\sqrt{A(\varepsilon)B(\varepsilon)}d\varepsilon
\int_{-\omega_{D}}^{\omega_{D}}\sqrt{A(\varepsilon)B(\varepsilon)}d\varepsilon
=-i\lambda
V\left(\frac{\nu_{F}}{2}\right)^{2}\Delta^{2}\texttt{arcsinh}\left(\frac{\omega_{D}}{\Delta}\right)t.
\end{eqnarray}
This formula was obtained from such arguments. Since the area of
action of the potential is limited by the layer $2\omega_{D}$ in a
neighborhood of Fermy surface, then the amplitudes $\Delta$ and
$\Delta^{+}$ is not equal to zero in this area only. As it was
pointed before, presented manner of uncoupling of correlations is
analogous to Fock exchange interaction for normal processes. In
Hartree-Fock approximation a decay of quasi-particles is absent
\citep{matt}. This means, that the amplitude of pairing is real:
$\Delta=\Delta^{+}$ in a momentum space.

For calculation of the contribution of interaction to internal
energy it is necessary to use the theorem (\ref{1.7}), which
connect a vacuum amplitude with a ground state energy:
\begin{eqnarray}\label{4.12}
&&\Omega_{\lambda}=i\frac{d}{dt}\ln R(t)
=-\frac{\lambda}{V}\sum_{\textbf{l}}\int\frac{d\omega}{2\pi}w_{l}F^{+}(\textbf{l},\omega)\sum_{\textbf{k}}\int\frac{d\omega}{2\pi}w_{k}F(\textbf{k},\omega)
\nonumber\\
&&=\lambda
V\left(\frac{\nu_{F}}{2}\right)^{2}\int_{-\omega_{D}}^{\omega_{D}}\frac{\Delta}{2E}d\varepsilon
\int_{-\omega_{D}}^{\omega_{D}}\frac{\Delta}{2E}d\varepsilon=\lambda
V\left(\frac{\nu_{F}}{2}\right)^{2}\Delta^{2}\texttt{arcsinh}\left(\frac{\omega_{D}}{\Delta}\right).
\end{eqnarray}
Since $\lambda<0$, then the interaction tries to reconstruct a
system so, that the gap $\Delta$ is as much as possible. We can
use the theorem because interaction of particle with pairing
fluctuation, changing a symmetry of a system:
$\Phi_{0}\rightarrow\Psi_{0}$, $G_{0}\rightarrow G_{S}$, was
considered before switching of the interaction
$V_{\textbf{l},-\textbf{l},\textbf{k},-\textbf{k}}$. Switching of
the interaction
$V_{\textbf{l},-\textbf{l},\textbf{k},-\textbf{k}}$, transferring
the state $\Psi_{0}$ in some other $\Psi_{0}'$, doesn't change
symmetry of a system: $\langle\Psi_{0}|\Psi_{0}'\rangle\neq0$.
This means that the adiabatic hypothesis is correct in the case of
phase transition even.

\subsection{Internal energy.}

On the basis of above obtained results we can write expression for
internal energy of a system (at the temperature $T=0$ internal
energy coincides with free energy):
\begin{eqnarray}\label{4.13}
    \Omega&=&\langle W\rangle+\Omega_{\lambda}=-2i\lim_{t\rightarrow0^{-}}\sum_{\textbf{k}}\frac{d\omega}{2\pi}G(\textbf{k},\omega)e^{-i\omega
t}\varepsilon(k)-\frac{\lambda}{V}\sum_{\textbf{k}}\int\frac{d\omega}{2\pi}w_{k}F^{+}(\textbf{k},\omega)\sum_{\textbf{k}}\int\frac{d\omega}{2\pi}w_{k}F(\textbf{k},\omega).
\end{eqnarray}
As it was pointed before, the gap is real $\Delta=\Delta^{+}$,
then $F(k,\omega)=F^{+}(\textbf{k},\omega)$. We can see, that the
energy depends on the unknown amplitude of pairing $\Delta$, which
corresponds to Bethe-Salpeter amplitude $\eta$ in the two-particle
problem. The amplitude $\Delta$ is determined by dynamics of all
particles of a system, and its observed value is a result of
averaging over a system. The procedure of averaging means
mathematically, that the observer value of $\Delta$ minimizes the
internal energy:
\begin{eqnarray}\label{4.15}
    \frac{d\Omega}{d\Delta}=0\Longrightarrow(-i)\Delta=\frac{\lambda}{V}\sum_{\textbf{k}}\int\frac{d\omega}{2\pi}w_{k}F(\textbf{k},\omega),
\end{eqnarray}
that coincides with (\ref{1.17}). It means that \textit{the
self-consistency equation for order parameter in Nambu-Gor'kov
formalism is an extremal of the obtained free energy functional}
(\ref{4.13}), \textit{and the order parameter is the averaged
Bethe-Salpeter amplitude over a system}.

The functional (\ref{4.13}) can be written in an explicit form in
quadratures:
\begin{eqnarray}\label{4.16}
\Omega=\Omega_{n}-V\frac{\nu_{F}}{2}\int_{-\omega_{D}}^{\omega_{D}}\left(\frac{\varepsilon^{2}}{E}-\frac{\varepsilon^{2}}{|\varepsilon|}\right)
+V\frac{\nu_{F}}{2}g\int_{-\omega_{D}}^{\omega_{D}}\frac{\Delta}{2E}d\varepsilon\int_{-\omega_{D}}^{\omega_{D}}\frac{\Delta}{2E}d\varepsilon,
\end{eqnarray}
where $\Omega_{n}$ is the energy of a normal phase,
$g=\lambda\frac{\nu_{F}}{2}$ is the effective interaction
constant. Value of the energy on the extremal (\ref{4.15}) is
\begin{eqnarray}\label{4.17}
\Omega_{min}=\Omega_{n}+V\frac{\nu_{F}}{2}(\omega_{D}^{2}-\omega_{D}\sqrt{\omega_{D}^{2}+\Delta^{2}}).
\end{eqnarray}
Thus, we solved the basic problem of statistical mechanics (at
zero temperature): the calculation of a partition function (free
energy) and, in particular, of a vacuum amplitude in a system of
interacting particles for case, when the interaction causes a
phase transition, that is symmetry of a system changes at a
switching of the interaction. Unlike other methods, \emph{this
result was obtained from first principles without introducing any
artificial parameters of type of order parameter, but starting
from parameters of the Hamiltonian only}.

\section{Nonzero temperatures.}\label{temperature}
\subsection{Normal and anomalous propagators.}

In the sections \ref{correlations} and \ref{ground} we have
described microscopic mechanism of formation of long-range order
in a system at zero temperature. In this section we shall
formulate the approach for case of nonzero temperatures. Let we
have a system from $N$ noninteracting fermions being in volume $V$
at temperature $T$. Then we must use Matsubara propagators, where
time $t$ is complex: $t\rightarrow-i\tau$, $\tau\in[0,\beta]$. In
ideal Fermy gas propagation of a particle with momentum
$\textbf{k}$, energy $\varepsilon\approx
v_{F}(|\textbf{k}|-k_{F})$ and spin $\sigma$ is described by the
free propagator:
\begin{eqnarray}\label{5.1}
  &&G_{0}(\textbf{k},\tau=\tau_{2}-\tau_{1})=\left\{\begin{array}{cc}
    -i\texttt{Sp}\left\{\widehat{\rho}_{0}C_{\textbf{k},\sigma}(\tau_{2})C_{\textbf{k},\sigma}^{+}(\tau_{1})\right\}, \qquad \tau>0 \\
    i\texttt{Sp}\left\{\widehat{\rho}_{0}C_{\textbf{k},\sigma}^{+}(\tau_{1})C_{\textbf{k},\sigma}(\tau_{2})\right\}, \qquad \tau\leq0 \\
  \end{array}\right\}=\nonumber\\
&&-i\theta_{\tau}(g_{0}^{+}A_{0}e^{-|\varepsilon|\tau}+g_{0}^{-}B_{0}e^{|\varepsilon|\tau})+
i\theta_{-\tau}(g_{0}^{-}A_{0}e^{-|\varepsilon|\tau}+g_{0}^{+}B_{0}e^{|\varepsilon|\tau}),\nonumber\\
&&G(\textbf{k},\tau)=\frac{1}{\beta}\sum_{n=-\infty}^{n=+\infty}G(\textbf{k},\omega_{n})e^{-i\omega_{n}\tau},
\qquad G(\textbf{k},\omega_{n})=\frac{1}{2}\int_{-\beta}^{\beta}G(\textbf{k},\tau)e^{i\omega_{n}\tau}d\tau\nonumber\\
&&G_{0}(\textbf{k},\omega_{n})=\frac{i}{i\omega_{n}-\varepsilon(k)}=i\frac{i\omega_{n}+\varepsilon}{(i\omega_{n})^{2}-\varepsilon^{2}}
=i\frac{A_{0}}{i\omega_{n}-|\varepsilon|}+i\frac{B_{0}}{i\omega_{n}+|\varepsilon|},
\end{eqnarray}
where
\begin{equation}\label{5.2}
g_{0}^{+}=\frac{1}{e^{-|\varepsilon|\beta}+1},\qquad
g_{0}^{-}=\frac{1}{e^{|\varepsilon|\beta}+1},\qquad
\omega_{n}=\frac{(2n+1)\pi}{\beta},
\end{equation}
$C_{\textbf{k},\sigma}(\tau)$ and
$C_{\textbf{k},\sigma}^{+}(\tau)$ are operators of creation and
annihilation in Heisenberg representation. $\widehat{\rho}_{0}$ is
density matrix of noninteracting particles:
\begin{equation}\label{5.3}
    \widehat{\rho}_{0}=\exp\left\{\frac{\Omega-\widehat{H}_{0}+\mu
\widehat{N}}{T}\right\}
=\exp\left\{\frac{\Omega-\sum_{\textbf{k},\sigma}\frac{k^{2}}{2m}C_{\textbf{k},\sigma}C_{\textbf{k},\sigma}^{+}+\mu
\widehat{N}}{T}\right\}
=\exp\left\{\frac{\Omega-\sum_{\textbf{k},\sigma}\varepsilon(k)C_{\textbf{k},\sigma}C_{\textbf{k},\sigma}^{+}}{T}\right\},
\end{equation}
where $\varepsilon(k)=\frac{k^{2}}{2m}-\mu\approx v_{F}(k-k_{F})$
is kinetic energy of particles counted off from Fermy surface.

Now let an attracting force acts between particles. The force is
described by the matrix element of interaction (\ref{2.3}). In
this case the instability of a system appears again with regard to
a pairing (as in the section \ref{unstability}) and
$\Gamma$-matrix has a pole structure:
\begin{equation}\label{5.4}
    \Gamma(0,0)=\frac{\lambda}{1+\lambda\frac{mk_{F}}{2\pi^{2}}\ln\frac{2\gamma\omega_{D}}{\pi
T}}\approx-\frac{2\pi^{2}}{mk_{F}}\frac{T_{C}}{T-T_{C}},\qquad
T_{C}=\frac{2\gamma}{\pi}\omega_{D}\left(-\frac{1}{2|\lambda|\nu_{F}}\right),
\end{equation}
where $\nu_{F}=\frac{mk_{F}}{\pi^{2}}$ is density of states on
Fermy surface. We can see that bound states exist in a system
while temperature is not higher than critical temperature $T_{C}$
- at higher temperatures particles have large kinetic energy, so
that an attraction between them leads to a scattering only.

Generalization of the two particle problem on the multiparticle
case is done analogously to the section \ref{correlations}. In the
case of nonzero temperature the mass operator has a form:
\begin{equation}\label{5.5}
    -\Sigma(\textbf{k},\omega_{n})=(-\Delta)iG_{0}^{+}(-\textbf{k},\omega_{n})(-\Delta^{+})=\frac{-\Delta\Delta^{+}}{i\omega_{n}+\varepsilon(k)}.
\end{equation}
Then it follows from Dyson equation, that a dressed propagator has
a view:
\begin{eqnarray}\label{5.6}
&&\frac{1}{G_{0}}=\frac{1}{G_{S}}-i\Sigma\Rightarrow
G_{S}(\textbf{k},\omega_{n})=\frac{i}{i\omega_{n}-\varepsilon(k)-\Sigma(\textbf{k},\omega_{n})}\nonumber\\
&&=i\frac{i\omega_{n}+\varepsilon}{(i\omega_{n})^{2}-E^{2}(k)}=i\frac{A_{S}}{i\omega_{n}-|\varepsilon|}+i\frac{B_{S}}{i\omega_{n}+|\varepsilon|}.
\end{eqnarray}
It can be written with help of a total definition of Green
function in $(\textbf{k},t)$-space:
\begin{eqnarray}\label{5.7}
  &&G_{S}(\textbf{k},\tau)=\tau_{2}-\tau_{1})=\left\{\begin{array}{cc}
    -i\texttt{Sp}\left\{\widehat{\varrho}_{0}C_{\textbf{k},\sigma}(\tau_{2})C_{\textbf{k},\sigma}^{+}(\tau_{1})\right\}, \qquad \tau>0 \\
    i\texttt{Sp}\left\{\widehat{\varrho}_{0}C_{\textbf{k},\sigma}^{+}(\tau_{1})C_{\textbf{k},\sigma}(\tau_{2})\right\}, \qquad \tau\leq0 \\
  \end{array}\right\}\nonumber\\
&&=-i\theta_{\tau}(g_{S}^{+}A_{S}e^{-E\tau}+g_{S}^{-}B_{S}e^{E\tau})+
i\theta_{-\tau}(g_{S}^{-}A_{S}e^{-E\tau}+g_{S}^{+}B_{S}e^{E\tau}),
\end{eqnarray}
where
\begin{equation}\label{5.8}
g_{S}^{+}=\frac{1}{e^{-E\beta}+1},\qquad
g_{S}^{-}=\frac{1}{e^{E\beta}+1}.
\end{equation}
$\widehat{\varrho}_{0}$ is the density matrix of noninteracting
\emph{quasi-particles}:
\begin{equation}\label{5.9}
    \widehat{\varrho}_{0}=\exp\left\{\frac{\Omega-\sum_{\textbf{k},\sigma}E(k)C_{\textbf{k},\sigma}C_{\textbf{k},\sigma}^{+}}{T}\right\}.
\end{equation}
It should be noted, that the state described by the density matrix
$\varrho_{0}$ has another symmetry in comparison with the initial
state $\rho_{0}$. Occupation numbers $n(k)$ is determined by the
following manner:
\begin{equation}\label{5.10}
    n(k)=-i\lim_{\tau\rightarrow 0^{-}}G(\textbf{k},\tau)=g^{-}A+g^{+}B,
\qquad\lim_{T\rightarrow 0}n(k)=B(k).
\end{equation}

Let's introduce the designations:
\begin{equation}\label{5.11}
   -G\Sigma\equiv\Delta F^{+}, \qquad -G^{+}\Sigma^{+}\equiv\Delta^{+}F..
\end{equation}
Then Dyson equation can be rewritten in a form of Gor'kov
equations:
\begin{eqnarray}
  &&(i\omega_{n}-\varepsilon)G+\Delta F^{+}=i \label{5.12}\\
  &&(i\omega_{n}+\varepsilon)F^{+}+G\Delta=0\label{5.12a}.
\end{eqnarray}
The expressions for anomalous propagators follow from Gor'kov
equations:
\begin{equation}\label{5.13}
    F^{+}(\textbf{k},\omega_{n})=\frac{-i\Delta^{+}}{(i\omega_{n})^{2}-E^{2}(k)}, \qquad
F(\textbf{k},\omega_{n})=(F^{+}(\textbf{k},\omega_{n}))^{+}=\frac{i\Delta}{(i\omega_{n})^{2}-E^{2}(k)},
\end{equation}
These expressions are analogous to the expressions (\ref{3.12}).
We can write the anomalous propagators in (\textbf{k},t)-space and
in a form of a vacuum average of creation and annihilation
operators:
\begin{eqnarray}\label{5.14}
  F^{+}_{\alpha\beta}(\textbf{k},\tau)&=&\frac{\Delta^{+}}{\sqrt{\Delta^{+}\Delta}}\left\{\begin{array}{cc}
    i\texttt{Sp}\{\widehat{\varrho}_{0}C_{-\textbf{k},\beta}^{+}(\tau_{2})C_{\textbf{k},\alpha}^{+}(\tau_{1})\}, \qquad \tau>0 \\
    i\texttt{Sp}\{\widehat{\varrho}_{0}C_{\textbf{k},\alpha}^{+}(\tau_{1})C_{-\textbf{k},\beta}^{+}(\tau_{2})\}, \qquad \tau\leq0 \\
  \end{array}\right\}\nonumber\\
&=&ig_{\alpha\beta}\frac{\Delta^{+}}{\sqrt{\Delta^{+}\Delta}}\sqrt{A_{S}B_{S}}
\left[\left(g_{S}^{+}e^{-E\tau}-g_{S}^{-}e^{E\tau}\right)\theta_{\tau}-\left(g_{S}^{+}e^{E\tau}-g_{S}^{-}e^{-E\tau}\right)\theta_{-\tau}\right],\\
F_{\alpha\beta}(\textbf{k},\tau)&=&\frac{\Delta}{\sqrt{\Delta^{+}\Delta}}\left\{\begin{array}{cc}
    -i\texttt{Sp}\{\widehat{\varrho}_{0}C_{\textbf{k},\alpha}(\tau_{2})C_{-\textbf{k},\beta}(\tau_{1})\}, \qquad \tau>0 \\
    -i\texttt{Sp}\{\widehat{\varrho}_{0}C_{-\textbf{k},\beta}(\tau_{1})C_{\textbf{k},\alpha}(\tau_{2})\}, \qquad \tau\leq0 \\
  \end{array}\right\}\nonumber\\
&=&ig_{\alpha\beta}\frac{\Delta}{\sqrt{\Delta^{+}\Delta}}\sqrt{A_{S}B_{S}}
\left[-\left(g_{S}^{+}e^{-E\tau}-g_{S}^{-}e^{E\tau}\right)\theta_{\tau}+\left(g_{S}^{+}e^{E\tau}-g_{S}^{-}e^{-E\tau}\right)\theta_{-\tau}\right]
\label{5.14a},
\end{eqnarray}
that is analogous to the expressions (\ref{3.13}) and
(\ref{3.13a}).

\subsection{Kinetic energy and entropy.}

In order to calculate a free energy it is necessary to know
kinetic energy of particles of a system, energy of their
interaction and entropy of collective excitations. Average kinetic
energy of all particles of a system is
\begin{eqnarray}\label{5.15}
    \langle
W\rangle&=&-2i\sum_{\textbf{k}}G(\textbf{k},\tau\rightarrow0^{-})\varepsilon(k)
=-\frac{2i}{\beta}\lim_{\tau\rightarrow0^{-}}\sum_{\textbf{k}}\sum_{n=-\infty}^{n=+\infty}G(\textbf{k},\omega_{n})e^{-i\omega_{n} t}\varepsilon(k)\nonumber\\
&=&2\sum_{\textbf{k}}(g^{-}A+g^{+}B)\varepsilon(k)=V\nu_{F}\int_{-v_{F}k_{F}}^{\infty}(g^{-}A+g^{+}B)\varepsilon
d\varepsilon.
\end{eqnarray}
Since the interaction $V_{\textbf{l}-\textbf{lk}-\textbf{k}}$
(\ref{2.3}) exists only in the layer
$-\omega_{D}<\varepsilon(k)<\omega_{D}$, then we can suppose that
\begin{eqnarray}\label{5.16}
g^{-}=\left[
\begin{array}{cc}
  g_{0}^{-}; & |\varepsilon(k)|>\omega_{D} \\
  g_{S}^{-}; & |\varepsilon(k)|<\omega_{D} \\
\end{array}%
\right], \qquad g^{+}=\left[
\begin{array}{cc}
  g_{0}^{+}; & |\varepsilon(k)|>\omega_{D} \\
  g_{S}^{+}; & |\varepsilon(k)|<\omega_{D} \\
\end{array}%
\right].
\end{eqnarray}
Hence, one may write expression for kinetic energy separating
normal and superconductive parts by analogy (\ref{4.4}):
\begin{eqnarray}\label{5.17}
&&\langle
W\rangle=W_{n}-V\frac{\nu_{F}}{2}\int_{-\omega_{D}}^{\omega_{D}}\left(g_{0}^{-}-g_{0}^{+}\right)\frac{\varepsilon^{2}}{|\varepsilon|}d\varepsilon+
V\frac{\nu_{F}}{2}\int_{-\omega_{D}}^{\omega_{D}}\left(g_{S}^{-}-g_{S}^{+}\right)\frac{\varepsilon^{2}}{E}d\varepsilon\nonumber\\
&&=W_{n}+V\frac{\nu_{F}}{2}\int_{-\omega_{D}}^{\omega_{D}}\tanh\left(\frac{\beta|\varepsilon|}{2}\right)\frac{\varepsilon^{2}}{|\varepsilon|}d\varepsilon-
V\frac{\nu_{F}}{2}\int_{-\omega_{D}}^{\omega_{D}}\tanh\left(\frac{\beta
E}{2}\right)\frac{\varepsilon^{2}}{E}d\varepsilon.
\end{eqnarray}
In the limit of low temperatures $\beta\rightarrow\infty$ this
expression reduces to the expression (\ref{4.4}). If we suppose
that $\Delta=0$, then we shall have $W=W_{n}$.

At temperature $T\neq 0$ a gas of collective excitation exists -
boholons with the spectrum
$E=\sqrt{\varepsilon^{2}(k)+\Delta^{2}}$. Since boholons are
product of decay of Cooper pairs on fermions, hence occupation
numbers of states by boholons are
\begin{equation}\label{5.18}
    f_{S}(k)=\frac{1}{e^{\beta E}+1}=\frac{1}{2}\left(1-\tanh\left(\frac{\beta
E}{2}\right)\right).
\end{equation}
Then entropy of a system is
\begin{eqnarray}\label{5.19}
S&=&-2\sum_{\textbf{k}}\left[f(k)\ln f(k)+(1-f(k))\ln (1-f(k))\right]\nonumber\\
&=&S_{0}-2V\frac{\nu_{F}}{2}\int_{-\omega_{D}}^{\omega_{D}}\left[f_{S}\ln
f_{S}+(1-f_{S})\ln
(1-f_{S})\right]d\varepsilon+2V\frac{\nu_{F}}{2}\int_{-\omega_{D}}^{\omega_{D}}\left[f_{0}\ln
f_{0}+(1-f_{0})\ln (1-f_{0})\right]d\varepsilon.
\end{eqnarray}
Here we separated the normal part again, where
$f_{0}=(e^{\beta|\varepsilon|}+1)^{-1}$, so that $S=S_{n}$ at
$\Delta=0$. The multiplier "2" appeared as result of summation
over spin states.

\subsection{Vacuum amplitude.}

In the previous subsections we considered the interaction of
particles with fluctuations of pairing, and we found, that the
state of a system described by the density matrix
$\widehat{\varrho}_{0}$ has another symmetry in comparison with
the initial state $\widehat{\rho}_{0}$. Hence vacuum amplitude of
a system can be written in a form:
\begin{eqnarray}\label{5.20}
R(\beta)=\langle
\widehat{U}(\beta)\rangle_{0}=\texttt{Sp}\left(\widehat{\varrho}_{0}\widetilde{U}(\beta)\right)=\sum_{n=0}^{\infty}\frac{(-1)^{n}}{n!}\int_{0}^{\beta}d\tau_{1}\ldots\int_{0}^{\beta}d\tau_{n}
Sp\left(\widehat{\varrho}_{0}
T\left\{\widehat{H}_{1}(\tau_{I})\ldots\widehat{H}_{I}(\tau_{n})\right\}\right),
\end{eqnarray}
where
$\widehat{H}_{I}(\tau)=e^{+\tau\hat{H_{0}}}\widehat{V}e^{-\tau\hat{H_{0}}}$
is the interaction operator of particles in interaction
representation. The averaging
$\texttt{Sp}\left(\widehat{\varrho}_{0}\widetilde{U}(\beta)\right)$
is made over ensemble of \emph{noninteracting quasi-particles}.
The potential $V_{\textbf{l},-\textbf{l},\textbf{k},-\textbf{k}}$
acts between particles (\ref{2.3}). Hence we can write (by analogy
with (\ref{4.8})) expended expression for vacuum amplitude:
\begin{eqnarray}\label{5.21}
&&R(\beta)=1+\frac{1}{1!}\frac{1}{V}\int_{0}^{\beta}d\tau_{1}\sum_{\alpha,\gamma}\sum_{\textbf{k},\textbf{l}}\left(-\frac{1}{2}V_{\textbf{l},-\textbf{l},\textbf{k},-\textbf{k}}\right)
\texttt{Sp}\left\{\widehat{\varrho}_{0}C_{-\textbf{l},\gamma}^{+}(\tau_{1})C_{\textbf{l},\alpha}^{+}(\tau_{1})C_{\textbf{k},\alpha}(\tau_{1})C_{-\textbf{k},\gamma}(\tau_{1})\right\}
\nonumber\\
&&+\frac{1}{2!}\frac{1}{V^{2}}\int_{0}^{\beta}d\tau_{2}\int_{0}^{\beta}d\tau_{1}\sum_{\alpha,\gamma}\sum_{\textbf{k},\textbf{l}}\left(-\frac{1}{2}V_{\textbf{l},-\textbf{l},\textbf{k},-\textbf{k}}\right)
\sum_{\alpha',\gamma'}\sum_{\textbf{k}',\textbf{l}'}\left(-\frac{1}{2}V_{\textbf{l}',-\textbf{l}',\textbf{k}',-\textbf{k}'}\right)\nonumber\\
&&\times\texttt{Sp}\left\{\widehat{\varrho}_{0}C_{-\textbf{l}',\gamma'}^{+}(\tau_{2})C_{\textbf{l}',\alpha'}^{+}(\tau_{2})C_{\textbf{k}',\alpha'}(\tau_{2})C_{-\textbf{k}',\gamma'}(\tau_{2})C_{-\textbf{l},\gamma}^{+}(\tau_{1})C_{\textbf{l},\alpha}^{+}(\tau_{1})C_{\textbf{k},\alpha}(\tau_{1})C_{-\textbf{k},\gamma}(\tau_{1})\right\}
\nonumber\\
&&+...,
\end{eqnarray}
where we took into account that
$\texttt{Sp}\left\{\widehat{\varrho}_{0}\right\}=1$. Then, by
analogy with the equation (\ref{4.9}), we can uncouple
correlations by the following way taking into account
anticommutation of operators $C$ and $C^{+}$:
\begin{eqnarray}\label{5.22}
&&R(\beta)\approx
1+\frac{1}{1!}(-1)^{2}\frac{1}{V}\int_{0}^{\beta}d\tau_{1}\sum_{\alpha,\gamma}\sum_{\textbf{k},\textbf{l}}\left(-\frac{1}{2}V_{\textbf{l},-\textbf{l},\textbf{k},-\textbf{k}}\right)
\texttt{Sp}\left\{\widehat{\varrho}_{0}C_{\textbf{l},\alpha}^{+}(\tau_{1})C_{-\textbf{l},\gamma}^{+}(\tau_{1})\right\}
\texttt{Sp}\left\{\widehat{\varrho}_{0}C_{-\textbf{k},\gamma}(\tau_{1})C_{\textbf{k},\alpha}(\tau_{1})\right\}
\nonumber\\
&&+\frac{1}{2!}(-1)^{4}\frac{1}{V^{2}}\int_{0}^{\beta}d\tau_{2}\int_{0}^{\beta}d\tau_{1}\sum_{\alpha,\gamma}\sum_{\textbf{k},\textbf{l}}\left(-\frac{1}{2}V_{\textbf{l},-\textbf{l},\textbf{k},-\textbf{k}}\right)
\sum_{\alpha',\gamma'}\sum_{\textbf{k}',\textbf{l}'}\left(-\frac{1}{2}V_{\textbf{l}',-\textbf{l}',\textbf{k}',-\textbf{k}'}\right)\nonumber\\
&&\times\texttt{Sp}\left\{\widehat{\varrho}_{0}C_{\textbf{l}',\alpha'}^{+}(\tau_{2})C_{-\textbf{l}',\gamma'}^{+}(\tau_{2})\right\}\texttt{Sp}\left\{\widehat{\varrho}_{0}
C_{-\textbf{k}',\gamma'}(\tau_{2})C_{\textbf{k}',\alpha'}(\tau_{2})\right\}\texttt{Sp}\left\{\widehat{\varrho}_{0}
C_{\textbf{l},\alpha}^{+}(\tau_{1})C_{-\textbf{l},\gamma}^{+}(\tau_{1})\right\}\texttt{Sp}\left\{\widehat{\varrho}_{0}
C_{-\textbf{k},\gamma}(\tau_{1})C_{\textbf{k},\alpha}(\tau_{1})\right\}
\nonumber\\
&&+\ldots=1+R_{1}+\frac{1}{2!}R_{1}^{2}+\ldots=\exp(R_{1})
\end{eqnarray}
Let's take into account that our approximation is analogous to
Fock approximation for normal processes. Then we can suppose
$\Delta=\Delta^{+}$, hence $F=-F^{+}$. Then $R(t)$ can be written
as
\begin{eqnarray}\label{5.23}
    \ln R(\beta)=R_{1}(\beta)&=&\frac{1}{V}\int_{0}^{\beta}d\tau_{1}\sum_{\alpha,\gamma}\sum_{\textbf{k},\textbf{l}}\left(-\frac{1}{2}V_{\textbf{l},-\textbf{l},\textbf{k},-\textbf{k}}\right)
iF_{\alpha\gamma}(\textbf{l},\tau_{1}-\tau_{1})
iF_{\alpha\beta}(\textbf{k},\tau_{1}-\tau_{1})\nonumber\\
&=&\frac{2}{V}\sum_{\textbf{k},\textbf{l}}\left(\frac{1}{2}V_{\textbf{l},-\textbf{l},\textbf{k},-\textbf{k}}\right)
F(\textbf{l},\tau\rightarrow 0^{-})F(\textbf{k},\tau\rightarrow 0^{-})\beta\nonumber\\
&=&\frac{\beta\lambda}{V}\sum_{\textbf{k}}w_{k}\frac{1}{\beta}\sum_{n=-\infty}^{n=+\infty}F(\textbf{k},\omega_{n})
\sum_{\textbf{k}}w_{k}\frac{1}{\beta}\sum_{n=-\infty}^{n=+\infty}F(\textbf{k},\omega_{n})\nonumber\\
&=&-\beta\lambda
V\left(\frac{\nu_{F}}{2}\right)^{2}\int_{-\omega_{D}}^{\omega_{D}}\tanh\left(\frac{\beta
E}{2}\right)\frac{\Delta}{2E}d\varepsilon\int_{-\omega_{D}}^{\omega_{D}}\tanh\left(\frac{\beta
E}{2}\right)\frac{\Delta}{2E}d\varepsilon.
\end{eqnarray}
This equation is represented graphically as well as in
Fig.\ref{fig6}. In order to calculate a contribution of
interaction in free energy we can use the formula (\ref{1.12}):
\begin{eqnarray}\label{5.24}
\Omega_{\lambda}&=&-\frac{1}{\beta}\ln R(\beta)
=-\frac{\lambda}{V}\sum_{\textbf{k}}w_{k}\frac{1}{\beta}\sum_{n=-\infty}^{n=+\infty}F(\textbf{k},\omega_{n})
\sum_{\textbf{k}}w_{k}\frac{1}{\beta}\sum_{n=-\infty}^{n=+\infty}F(\textbf{k},\omega_{n})
\nonumber\\
&=&\lambda
V\left(\frac{\nu_{F}}{2}\right)^{2}\int_{-\omega_{D}}^{\omega_{D}}\tanh\left(\frac{\beta
E}{2}\right)\frac{\Delta}{2E}d\varepsilon\int_{-\omega_{D}}^{\omega_{D}}\tanh\left(\frac{\beta
E}{2}\right)\frac{\Delta}{2E}d\varepsilon.
\end{eqnarray}
In the limit of low temperatures $\beta\rightarrow\infty$ this
equation transforms to the equation (\ref{4.12}). If we suppose
that $\Delta=0$, then we shall have $\Omega_{\lambda}=0$.

\subsection{Free energy.}

Starting from the above found results we can write the expression
a for free energy of a system:
\begin{eqnarray}\label{5.25}
&&\Omega=\langle
W\rangle-\frac{1}{\beta}S+\Omega_{\lambda}=-\frac{2i}{\beta}\lim_{\tau\rightarrow0^{-}}\sum_{\textbf{k}}\sum_{n=-\infty}^{n=+\infty}G(\textbf{k},\omega_{n})e^{-i\omega_{n}
t}\varepsilon(k)\nonumber\\
&&+\frac{2}{\beta}\sum_{\textbf{k}}\left[f(k)\ln f(k)+(1-f(k))\ln
(1-f(k))\right]
-\frac{\lambda}{V}\sum_{\textbf{k}}w_{k}\frac{1}{\beta}\sum_{n=-\infty}^{n=+\infty}F(\textbf{k},\omega_{n})
\sum_{\textbf{k}}w_{k}\frac{1}{\beta}\sum_{n=-\infty}^{n=+\infty}F(\textbf{k},\omega_{n})
\end{eqnarray}
We can see, that the energy depends on the unknown amplitude of
pairing $\Delta$, which corresponds to Bethe-Salpeter amplitude
$\eta$ in two-particle problem. The observed value of $\Delta$
must minimize the free energy:
\begin{eqnarray}\label{5.27}
\frac{d\Omega}{d\Delta}=0\Longrightarrow(-i)\Delta=\frac{\lambda}{V\beta}\sum_{\textbf{k}}\sum_{n=-\infty}^{n=+\infty}w_{k}F(\textbf{k},\omega_{n}),
\end{eqnarray}
that coincides with (\ref{1.17}). As in the previous section the
equation of a self-consistence for order parameter in
Nambu-Gor'kov formalism is the extremal of the functional of free
energy (\ref{5.25}), and the order parameter is averaged
Bethe-Salpeter amplitude over a system.

The functional (\ref{5.25}) can be written in an explicit form in
quadratures:
\begin{eqnarray}\label{5.27a}
\Omega=&&\Omega_{n}+V\frac{\nu_{F}}{2}\int_{-\omega_{D}}^{\omega_{D}}\left[\tanh\left(\frac{\beta|\varepsilon|}{2}\right)\frac{\varepsilon^{2}}{|\varepsilon|}-
\tanh\left(\frac{\beta
E}{2}\right)\frac{\varepsilon^{2}}{E}\right]d\varepsilon\nonumber\\
&&+\frac{2V}{\beta}\frac{\nu_{F}}{2}\int_{-\omega_{D}}^{\omega_{D}}\left[f_{S}\ln
f_{S}+(1-f_{S})\ln (1-f_{S})-f_{0}\ln f_{0}-(1-f_{0})\ln
(1-f_{0})\right]d\varepsilon\nonumber\\
&&+V\frac{\nu_{F}}{2}g\int_{-\omega_{D}}^{\omega_{D}}\tanh\left(\frac{\beta
E}{2}\right)\frac{\Delta}{2E}d\varepsilon\int_{-\omega_{D}}^{\omega_{D}}\tanh\left(\frac{\beta
E}{2}\right)\frac{\Delta}{2E}d\varepsilon,
\end{eqnarray}
where $\Omega_{n}$ is the energy of a normal phase,
$g=\lambda\frac{\nu_{F}}{2}$ is the effective interaction
constant, $V$ is volume of a system. $g$ can be expressed via
critical temperature $\beta_{C}$ with help of the equation
$\Delta(\beta_{C})=0$ as following:
\begin{equation}\label{5.28}
    1=-g\int_{-\omega_{D}}^{\omega_{D}}\tanh\left(\frac{\beta_{c}|\varepsilon|}{2}\right)\frac{1}{2|\varepsilon|}d\varepsilon.
\end{equation}
If to suppose $\Delta=0$, then we shall have $\Omega=\Omega_{n}$.
The equilibrium value of $\Delta$ is determined by balance of
kinetic energy, entropy and energy of interaction, that
corresponds to a minimum of the free energy.

Let's consider a low-temperature limit of the free energy
(\ref{5.27}): $\Delta\beta\gg 1$ at $T\rightarrow 0$. This means,
that the value $\Delta-\Delta_{0}$ can be parameter of expansion,
where $\Delta_{0}=\Delta(T=0)$ is equilibrium value of the gap
(amplitude of pairing) at zero temperature. Due a rapid
convergence of integration elements in (\ref{5.27}), the limits of
integration can be $-\infty,+\infty$. Then a low-temperature
expansion has a form:
\begin{equation}\label{5.29}
    \Omega=\Omega_{n}+V\left(\alpha_{0}(T)+b_{0}(T)\Delta+d_{0}\Delta^{2}\right),
\end{equation}
where coefficients of the expansion are
\begin{eqnarray}\label{5.30}
 \alpha_{0}(T) &=& -\frac{\nu_{F}}{4}\Delta_{0}^{2}+2\nu_{F}T^{2}-\frac{\nu_{F}}{2}\sqrt{8\pi\Delta_{0}T^{3}}e^{-\frac{\Delta_{0}}{T}}+
\frac{\nu_{F}}{2}(g+1)\Delta_{0}^{2}-\frac{\nu_{F}}{2}(g+1)\sqrt{8\pi\Delta_{0}^{3}T}e^{-\frac{\Delta_{0}}{T}} \nonumber\\
  b_{0}(T) &=&
  -\frac{\nu_{F}}{2}2(g+1)\Delta_{0}+\frac{\nu_{F}}{2}(g+1)\sqrt{8\pi\Delta_{0}T}e^{-\frac{\Delta_{0}}{T}},
  \qquad  d_{0} = \frac{\nu_{F}}{2}2(g+1)
\end{eqnarray}
Since in the expression $\frac{b_{0}(T)}{2d_{0}}$ the multipliers
$g+1$ is cancelled and, as a rule, $|g|\ll 1$, we can suppose that
$g+1\approx 1$. Then expressions (\ref{5.30}) are simplified:
\begin{eqnarray}\label{5.30a}
 \alpha_{0}(T) &=& \frac{\nu_{F}}{4}\Delta_{0}^{2}+2\nu_{F}T^{2}-\frac{\nu_{F}}{2}\sqrt{8\pi\Delta_{0}T^{3}}e^{-\frac{\Delta_{0}}{T}}+
-\frac{\nu_{F}}{2}\sqrt{8\pi\Delta_{0}^{3}T}e^{-\frac{\Delta_{0}}{T}} \nonumber\\
  b_{0}(T) &=& -\frac{\nu_{F}}{2}2\Delta_{0}+\frac{\nu_{F}}{2}\sqrt{8\pi\Delta_{0}T}e^{-\frac{\Delta_{0}}{T}}, \qquad d_{0}=\nu_{F}
\end{eqnarray}

Let's consider a high-temperature limit of the free energy:
$\Delta\beta_{C}\ll 1$ at $T\rightarrow T_{C}$. Expansion in
powers of $\Delta$ gives:
\begin{equation}\label{5.32}
    \Omega=\Omega_{n}+V\left(\alpha(T)\Delta^{2}+\frac{1}{2}b\Delta^{4}+\frac{1}{3}d\Delta^{6}\right),
\end{equation}
where the coefficients of the expansion are
\begin{eqnarray}\label{5.33}
  \alpha(T) &=& \frac{\nu_{F}}{2}\frac{T-T_{c}}{T_{c}}\nonumber\\
  b &=& \frac{\nu_{F}}{2}\frac{7\zeta(3)}{8\pi^{2}T_{c}^{2}}\\
  d &=&
\frac{\nu_{F}}{2}\left(\frac{52.31\zeta(5)}{\pi^{4}}+4.83\right)\frac{1}{4!T_{c}^{4}}\nonumber.
\end{eqnarray}
This expansion has a form of Landau expansion of free energy in
powers of order parameter.

From the all considered above we can see, that \emph{averaged over
a system Bethe-Salpeter amplitude} $\eta$ \emph{and}
$\eta^{\ast}$\emph{ - the amplitude of pairing }$\Delta$ and
$\Delta^{+}$\emph{have the properties, which are analogous to the
properties of a order parameter}. Thus, we have solved the basic
problem of statistical mechanics: the calculation of a partition
function (free energy) and, in particular, of a vacuum amplitude
in a system of interacting particles for case, when the
interaction causes a phase transition, that is symmetry of a
system changes at a switching of the interaction.

\section{The pairing with nonzero momentum of center of mass of a pair.}\label{nonzeromomentum}

Let fermions with momentums $\textbf{k}+\frac{\textbf{q}}{2}$ and
$-\textbf{k}+\frac{\textbf{q}}{2}$ pair up, so that the momentum
of a center of mass of a pair is $\textbf{q}$. The free
propagators, corresponding to these states are
\begin{eqnarray}\label{6.1}
&&G_{0}\left(\textbf{k}+\frac{\textbf{q}}{2},\omega\right)=\frac{1}{\omega-\varepsilon\left(\textbf{k}+\frac{\textbf{q}}{2}\right)}\equiv\frac{1}{\omega-\varepsilon_{+}}\nonumber\\
&&G_{0}\left(-\textbf{k}+\frac{\textbf{q}}{2},-\omega\right)=\frac{1}{-\omega-\varepsilon\left(-\textbf{k}+\frac{\textbf{q}}{2}\right)}\equiv\frac{1}{-\omega-\varepsilon_{-}}.
\end{eqnarray}
We can suppose that
\begin{eqnarray}\label{6.1a}
\varepsilon_{+}=\frac{1}{2m}\left(\textbf{k}+\frac{\textbf{q}}{2}\right)^{2}-\mu\approx\varepsilon+\frac{\textbf{kq}}{2m},\qquad
\varepsilon_{-}=\frac{1}{2m}\left(-\textbf{k}+\frac{\textbf{q}}{2}\right)^{2}-\mu\approx\varepsilon-\frac{\textbf{kq}}{2m}.
\end{eqnarray}
The corresponding mass operator describing an interaction of
particles with fluctuations of pairing is:
\begin{eqnarray}\label{6.2}
  (-i)\Sigma_{q} &=& -i\Delta
iG_{0}\left(-\textbf{k}+\frac{\textbf{q}}{2},-\omega\right)i\Delta^{+}=(-i)\frac{\Delta\Delta^{+}}{\omega+\varepsilon_{-}}.
\end{eqnarray}
Then can find the dressed propagator from Dyson equation:
\begin{eqnarray}\label{6.3}
G_{S}=\frac{1}{G_{0}^{-1}-\Sigma_{q}}=\frac{\omega+\varepsilon_{-}}{(\omega-E_{+})(\omega-E_{-})},
\end{eqnarray}
where the specters of quasi-particles are
\begin{eqnarray}\label{6.4}
&&E_{+}=\frac{\varepsilon_{+}-\varepsilon_{-}}{2}+\sqrt{\left(\frac{\varepsilon_{+}+\varepsilon_{-}}{2}\right)+|\Delta|^{2}}
\approx\frac{\textbf{kq}}{2m}+\sqrt{\varepsilon^{2}+|\Delta|^{2}}\nonumber\\
&&E_{-}=\frac{\varepsilon_{+}-\varepsilon_{-}}{2}-\sqrt{\left(\frac{\varepsilon_{+}+\varepsilon_{-}}{2}\right)+|\Delta|^{2}}
\approx\frac{\textbf{kq}}{2m}-\sqrt{\varepsilon^{2}+|\Delta|^{2}}.
\end{eqnarray}
We can see, that if to assume $q=0$, then the specter (\ref{6.4})
turn into the usual specter of boholons:
$E=\pm\sqrt{\varepsilon^{2}+|\Delta|^{2}}$. The critical momentum
$q_{cr}$ exists when a minimum of the specter(\ref{6.4}) touches
Fermy surface. Then for excitation of a system it is necessary
infinitely small energy. This means, that superfluidity of Fermy
gas is absent. The critical momentum is:
\begin{equation}\label{6.5}
    E_{+}(q=q_{cr},\varepsilon=0)=E_{-}(q=q_{cr},\varepsilon=0)=0\Rightarrow
q_{cr}=\frac{2}{v_{F}}|\Delta|.
\end{equation}
Similar pairing can take place in high-temperature superconductors
(cuprates) with a mirror nesting
$\varepsilon\left(\textbf{k}+\frac{\textbf{q}}{2}\right)=\varepsilon\left(-\textbf{k}+\frac{\textbf{q}}{2}\right)$
of regions of Fermy contour \citep{belya2,belya}.

\section{Free energy in a case of slow spatial inhomogeneity.}\label{spaceinhom}

In the previous sections we supposed, that amplitudes of pairing
$\Delta$ and $\Delta^{+}$ don't depend on spatial coordinates.
This takes place in interminable, homogeneous, isotropic and
isolated from external fields superconductor. However in a total
case these conditions are not realized. For example, in a
sufficiently strong magnetic field the inclusions of normal phase
can exist in volume of a superconductor. Another example - a
contact of a superconductor and a normal metal. In this case the
order parameter is suppressed in a boundary layer of a
superconductor, however it appears in boundary layer of a normal
metal.

\begin{figure}[h]
\includegraphics[width=8cm]{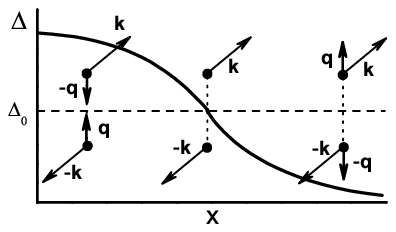}
\caption{The pairing of fermions in a spatially inhomogeneous
system. $\Delta_{0}$ is the equilibrium value of a gap in a
homogeneous system.} \label{fig6}
\end{figure}

Let's consider some region of a superconductor, where a
distribution of $\Delta$ is inhomogeneous and the amplitude of
pairing can be both smaller and larger than its equilibrium value
$\Delta_{0}$ - Fig.\ref{fig6}. When a quasi-particle propagates
along the axis $Ox$ its energy is constant
$E=\sqrt{\varepsilon(k)^{2}+|\Delta|^{2}}$, but its momentum
changes. If a pair moves into region, where
$|\Delta|<|\Delta_{0}|$, then forces appear tearing the pair. Each
element of the pair gets some increment of momentum $\textbf{q}$:
$(\textbf{k},-\textbf{k})\rightarrow
(\textbf{k}+\textbf{q},-\textbf{k}-\textbf{q})$. If the pair moves
into region, where $|\Delta|>|\Delta_{0}|$, then a forces appear
increasing a bound energy of the pair. Each element of the pair
gets some increment of momentum $\textbf{q}$ too. Since the
amplitude of pairing is function of coordinates
$\Delta(\textbf{r})$, moreover we suppose that order parameter is
real $\Delta=\Delta^{+}$ in momentum space and it has an identical
dimension in $\textbf{q}$-space and in $\textbf{r}$-space, then we
can write the Fourier-transformations:
\begin{eqnarray}\label{7.1}
  \Delta(\textbf{r}) &=& \sum_{\textbf{q}}\Delta(q)e^{i\textbf{qr}}=\frac{V}{(2\pi)^{3}}\int\Delta(q)e^{i\textbf{qr}}d^{3}q, \nonumber\\
  \Delta^{+}(\textbf{r}) &=& \sum_{\textbf{q}}\Delta(q)e^{-i\textbf{qr}}=\frac{V}{(2\pi)^{3}}\int\Delta(q)e^{-i\textbf{qr}}d^{3}q,\\
  \Delta(q) &=& \frac{1}{V}\int\Delta(\textbf{r})e^{-i\textbf{qr}}d^{3}r=\frac{1}{V}\int\Delta^{+}(\textbf{r})e^{i\textbf{qr}}d^{3}r.\nonumber
\end{eqnarray}

\begin{figure}[h]
\includegraphics[width=8cm]{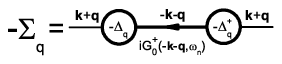}
\caption{The diagram for the mass operator $\Sigma$ describing an
interaction of a fermion with fluctuations of pairing in a
spatially inhomogeneous system.} \label{fig7}
\end{figure}

The mass operator for above-mentioned process is shown in
Fig.\ref{fig7}. In analytical representation it has a view:
\begin{equation}\label{7.2}
    -\Sigma_{q}(\textbf{k},\omega_{n})=(-\Delta_{q})iG_{0}^{+}(-\textbf{k}-\textbf{q},\omega_{n})(-\Delta^{+}_{q})=\frac{-\Delta_{q}\Delta^{+}_{q}}{i\omega_{n}+\varepsilon(\textbf{k}+\textbf{q})},
\end{equation}
where the free propagator $G_{0}$ is
\begin{equation}\label{7.3}
    G_{0}=\frac{1}{i\omega_{n}-\varepsilon(\textbf{k}+\textbf{q})}=
i\frac{i\omega_{n}+\varepsilon_{q}(k)}{(i\omega_{n})^{2}-\varepsilon^{2}_{q}(k)}.
\end{equation}
Then from Dyson equation we can obtain the dressed propagator:
\begin{eqnarray}\label{7.4}
  \frac{1}{G_{0}} &=& \frac{1}{G_{S}}-i\Sigma_{q}\Rightarrow
G_{S}=i\frac{i\omega_{n}+\varepsilon_{q}}{(i\omega_{n})^{2}-E_{q}^{2}},
\end{eqnarray}
where $E_{q}$ is the specter of quasi-particles in a
nonhomogeneous system:
\begin{equation}\label{7.5}
  E_{q}^{2}=\varepsilon_{q}^{2}+|\Delta_{q}|^{2},\qquad
\varepsilon_{q}\equiv\varepsilon(\textbf{k}+\textbf{q})\approx\varepsilon(k)+\frac{\textbf{kq}}{m},\qquad
|\textbf{k}|\simeq k_{F}.
\end{equation}
Dyson equation can be represented in a form of set of Gor'kov
equations, from where the expressions for anomalous propagators
follow:
\begin{eqnarray}\label{7.6}
\begin{array}{c}
  (i\omega_{n}-\varepsilon_{q})G+\Delta_{q} F^{+}=i \\
\\
  (i\omega_{n}+\varepsilon_{q})F^{+}+G\Delta_{q}=0\\
\end{array}\Rightarrow
\begin{array}{c}
  F^{+}(\textbf{k}+\textbf{q},\omega_{n})=\frac{-i\Delta_{q}^{+}}{(i\omega_{n})^{2}-E^{2}_{q}} \\
\\
  F(\textbf{k}+\textbf{q},\omega_{n})=(F^{+}(\textbf{k}+\textbf{q},\omega_{n}))^{+}=\frac{i\Delta_{q}}{(i\omega_{n})^{2}-E^{2}_{q}}\\
\end{array}
\end{eqnarray}
If to suppose $q=0$, then we shall have the expressions
(\ref{5.12}-\ref{5.13}).

Now let's suppose that $\Delta(\textbf{r})$ changes very slowly on
a coherence length $l(T)$ which characterizes a size of Cooper
pair. Then we can suppose
$\varepsilon(\textbf{k}+\textbf{q})=\varepsilon(\textbf{k})$ in
the specter of quasi-particles, such that
$E_{q}\approx\sqrt{\varepsilon^{2}(k)+|\Delta_{q}|^{2}}$. However
we must keep $\varepsilon(\textbf{k}+\textbf{q})$ in numerator of
the expressions (\ref{7.3}) and (\ref{7.4}) for $G$. Hence the
normal propagator has a form:
\begin{eqnarray}\label{7.7}
  G(\textbf{k}+\textbf{q},\tau)=-i\theta_{\tau}\left(g^{+}_{q}A(\textbf{k}+\textbf{q})e^{-E_{q}\tau}+g^{-}_{q}B(\textbf{k}+\textbf{q})e^{E_{q}\tau}\right)+
i\theta_{-\tau}\left(g^{-}_{q}A(\textbf{k}+\textbf{q})e^{-E_{q}\tau}+g^{+}_{q}B(\textbf{k}+\textbf{q})e^{E_{q}\tau}\right),
\end{eqnarray}
where
\begin{eqnarray}\label{7.8}
A(\textbf{k}+\textbf{q})\approx
A_{q}(k)+\frac{1}{2E_{q}}\frac{\textbf{kq}}{m}, \qquad
B(\textbf{k}+\textbf{q})\approx
B_{q}(k)-\frac{1}{2E_{q}}\frac{\textbf{kq}}{m}.
\end{eqnarray}
The anomalous propagators are
\begin{eqnarray}\label{7.9}
  F^{+}_{\alpha\beta}(\textbf{k}+\textbf{q},\tau)&=&ig_{\alpha\beta}\frac{\Delta^{+}_{q}}{2E_{q}}
\left[\left(g_{q}^{+}e^{-E_{q}\tau}-g_{q}^{-}e^{E_{q}\tau}\right)\theta_{\tau}-\left(g_{S}^{+}e^{E_{q}\tau}-g_{q}^{-}e^{-E_{q}\tau}\right)\theta_{-\tau}\right],\nonumber\\
F_{\alpha\beta}(\textbf{k}+\textbf{q},\tau)&=&ig_{\alpha\beta}\frac{\Delta_{q}}{2E_{q}}
\left[-\left(g_{q}^{+}e^{-E_{q}\tau}-g_{q}^{-}e^{E_{q}\tau}\right)\theta_{\tau}+\left(g_{q}^{+}e^{E_{q}\tau}-g_{q}^{-}e^{-E_{q}\tau}\right)\theta_{-\tau}\right].
\end{eqnarray}
We can see, that in the approximation of slowness of changes of
$\Delta(\textbf{r})$ the anomalous propagators depend on
 $q$ by means of $\Delta(q)$ only.

Kinetic energy of a system is determined by the following way:
\begin{eqnarray}\label{7.10}
  \langle
W\rangle&=&-2i\sum_{\textbf{k}}\varepsilon(\textbf{k}+\textbf{q})G(\textbf{k}+\textbf{q},\tau\rightarrow
0^{-})=2\sum_{\textbf{k}}\varepsilon(\textbf{k}+\textbf{q})\left(g^{-}_{q}A(\textbf{k}+\textbf{q})+g^{+}_{q}B(\textbf{k}+\textbf{q})\right)\nonumber\\
&=&2\sum_{\textbf{k}}\varepsilon(k)\left(g^{-}_{q}A_{q}+g^{+}_{q}B_{q}\right)+2\sum_{\textbf{k}}\left(g^{-}_{q}-g^{+}_{q}\right)\frac{1}{E_{q}}\frac{(\textbf{kq})^{2}}{m^{2}}\nonumber\\
&=&W_{n}+V\frac{\nu_{F}}{2}\int_{-\omega_{D}}^{\omega_{D}}\left[\tanh\left(\frac{\beta|\varepsilon|}{2}\right)\frac{\varepsilon^{2}}{|\varepsilon|}-
\tanh\left(\frac{\beta
E_{q}}{2}\right)\frac{\varepsilon^{2}}{E_{q}}\right]d\varepsilon\nonumber\\
&+&V\frac{\nu_{F}}{2}\frac{1}{3}v_{F}^{2}q^{2}\int_{-\omega_{D}}^{\omega_{D}}\left[\tanh\left(\frac{\beta|\varepsilon|}{2}\right)\frac{1}{|\varepsilon|}-
\tanh\left(\frac{\beta
E_{q}}{2}\right)\frac{1}{E_{q}}\right]d\varepsilon.
\end{eqnarray}
We can see, that the term, which is proportional to $q^{2}$, is
added to the kinetic energy (\ref{5.17}) (with the replacement
$E\rightarrow E_{q}=\sqrt{\varepsilon^{2}(k)+|\Delta(q)|^{2}}$).
In the approximation of slowness of changes the expressions for
entropy and vacuum amplitude coincide with the expressions
(\ref{5.19}) and (\ref{5.24}) accordingly, however it should be
written $\Delta(q)$ instead of $\Delta=\texttt{const}$. Then we
can write the free energy:
\begin{equation}\label{7.11}
    \Omega(q)=\Omega_{n}(q)+\Omega(\Delta_{q})+V\frac{\nu_{F}}{2}\frac{1}{3}v_{F}^{2}q^{2}\int_{-\omega_{D}}^{\omega_{D}}\left[\tanh\left(\frac{\beta|\varepsilon|}{2}\right)\frac{1}{|\varepsilon|}-
\tanh\left(\frac{\beta
E_{q}}{2}\right)\frac{1}{E_{q}}\right]d\varepsilon,
\end{equation}
where $\Omega(\Delta_{q})$ coincides with the expression
(\ref{5.27}), where the replacement $\Delta\rightarrow\Delta(q)$
was done.

Expanding the free energy (\ref{7.11}) in powers of $\Delta$ we
can obtain the expression:
\begin{equation}\label{7.12}
    \Omega(q)=\Omega_{n}(q)+V\left(\alpha(T)\Delta_{q}^{2}+\frac{1}{2}b\Delta_{q}^{4}+\gamma q^{2}\Delta_{q}^{2}\right),
\end{equation}
where the coefficient $\gamma$ is
\begin{eqnarray}\label{7.13}
 \gamma=\frac{\nu_{F}}{2}\frac{7\zeta(3)v_{F}^{2}}{24\pi^{2}T_{c}^{2}}=\frac{\nu_{F}}{2}l_{0}^{2},
\end{eqnarray}
where $l_{0}$ is a coherence length at $T=0$ (Pippard length). The
expansion (\ref{7.12}) has a form of Landau expansion of free
energy in powers of order parameter at the condition $ql_{0}\ll
1$. We can see, that a spatial inhomogeneity increases the free
energy of a superconductor. Hence in most cases we can be
restricted by the term $\sim q^{2}$ in the expansion, because more
fast changes of $\Delta$ increase the free energy essentially.

The full free energy of a system in a spatial inhomogeneous case
can be obtained by summation of the expression (\ref{7.11}) over
all possible $\textbf{q}$:
\begin{eqnarray}\label{7.14}
\Omega=\Omega_{n}+\sum_{\textbf{q}}\Omega_{s}(q)=\Omega_{n}+\frac{V}{(2\pi)^{3}}\int\Omega_{s}(q)d^{3}q
=\Omega_{n}+\frac{V^{2}}{(2\pi)^{3}}\int\left(\alpha(T)\Delta_{q}^{2}+\frac{1}{2}b\Delta_{q}^{4}+\gamma
q^{2}\Delta_{q}^{2}\right)d^{3}q
\end{eqnarray}
Let's pass from momentum space to real space using the expressions
(\ref{7.1}):
\begin{eqnarray}\label{7.15}
\int\Delta_{q}\Delta_{q}d^{3}q=\int\Delta_{q}\left[\frac{1}{V}\int\Delta(\textbf{r})e^{-i\textbf{qr}}d^{3}r\right]d^{3}q
=\frac{1}{V}\int\Delta(\textbf{r})\left[\int\Delta(q)e^{-i\textbf{qr}}d^{3}q\right]d^{3}r
=\frac{(2\pi)^{3}}{V^{2}}\int\Delta(\textbf{r})\Delta^{+}(\textbf{r})d^{3}r,
\end{eqnarray}
\begin{eqnarray}\label{7.16}
&&\int\textbf{q}\Delta_{q}\textbf{q}\Delta_{q}d^{3}q=\int\textbf{q}\Delta_{q}\left[\frac{1}{V}\int
e^{-i\textbf{qr}}(-i)\frac{\partial}{\partial\textbf{r}}\Delta(\textbf{r})d^{3}r\right]d^{3}q\nonumber\\
&&=\frac{-i}{V}\int\left[\int\textbf{q}\Delta(q)e^{-i\textbf{qr}}d^{3}q\right]\frac{\partial}{\partial\textbf{r}}\Delta(\textbf{r})d^{3}r
=\frac{(2\pi)^{3}}{V^{2}}\int\left[\frac{\partial}{\partial\textbf{r}}\Delta(\textbf{r})\right]\frac{\partial}{\partial\textbf{r}}\Delta^{+}(\textbf{r})d^{3}r.
\end{eqnarray}
For the term $\Delta^{4}$ and terms with more high powers the
situation is more difficult. This is because a square of a Fourier
transform is not equal to a Fourier transform of a square:
$\left(\frac{1}{V}\int\Delta(\textbf{r})e^{-i\textbf{qr}}d^{3}r\right)^{2}\neq\frac{1}{V}\int\Delta^{2}(\textbf{r})e^{-i\textbf{qr}}d^{3}r$.
Apparently this fact results to some nonlocality of a
superconductor's state in zero magnetic field described in
\citep{hook}, where a value of gap in a point is determined by a
distribution of gap in some neighborhood: $\Delta(\textbf{r})=\int
d\textbf{r}'Q(\textbf{r},\textbf{r}')\Delta'(\textbf{r})$. However
in first approximation this correlation can be neglected and we
can write the expansion of free energy in powers of
$\Delta\Delta^{+}$ in real space:
\begin{eqnarray}\label{7.17}
\Omega=\Omega_{n}+\int\left[\alpha(T)|\Delta(\textbf{r})|^{2}+\frac{b}{2}|\Delta(\textbf{r})|^{4}+\gamma\left|\frac{\partial}{\partial\textbf{r}}\Delta(\textbf{r})\right|^{2}\right]d^{3}r.
\end{eqnarray}
This expansion coincides with Ginzburg-Landau expansion in zero
magnetic field.

\section{Free energy of a superconductor in magnetic field.}\label{magnetic}

In this section we shall generalize the previous results for the
case, when a superconductor is placed in a magnetic field
$\textbf{H}(\textbf{r})=\texttt{rot}\textbf{A}(r)$. Our aim is to
obtain the functional of free energy
$\Omega\left(\Delta(\textbf{r}),\frac{\partial}{\partial\textbf{r}}\Delta(\textbf{r}),\textbf{A}(\textbf{r})\right)$,
which is correct for an arbitrary value of the relation
$\Delta(T)/T$, for an arbitrary scale of a change of
$\Delta(\textbf{r})$ in comparison with a coherent length $l(T)$,
for an arbitrary value of a magnetic penetration depth
$\lambda(T)$ in comparison with a coherent length $l_{0}$
(nonlocal electromagnetic response). Thus, the all three
restriction on Ginzburg-Landau functional, described in section
\ref{formulation}, are excluded.

Let the microscopic magnetic field exists in a superconductor with
a potential $\textbf{A}$ and an intensity $\textbf{H}$:
\begin{equation}\label{8.1}
\textbf{A}(\textbf{r})=\sum_{\textbf{q}}\textbf{a}(\textbf{q})e^{i\textbf{qr}}\Rightarrow
\textbf{H}(\textbf{r})=\texttt{rot}\textbf{A}(\textbf{r})=i\sum_{\textbf{q}}\textbf{q}\times\textbf{a}(\textbf{q})e^{i\textbf{qr}}.
\end{equation}
Then the magnetic field inducts a current:
\begin{equation}\label{8.2}
\textbf{J}(\textbf{r})=\frac{c}{4\pi}\texttt{rot}\textbf{H}(\textbf{r})
=-\frac{c}{4\pi}\sum_{\textbf{q}}\textbf{q}\times\textbf{q}\times\textbf{a}(\textbf{q})e^{i\textbf{qr}}
=-\frac{c}{4\pi}\sum_{\textbf{q}}\left(\textbf{q}(\textbf{qa})-\textbf{a}q^{2}\right)e^{i\textbf{qr}}
\equiv\sum_{\textbf{q}}\textbf{j}(\textbf{q})e^{i\textbf{qr}}.
\end{equation}
Energy of the magnetic field is
\begin{equation}\label{8.3}
    W_{f}=\frac{1}{8\pi}\int\left|\textbf{H}(\textbf{r})\right|^{2}d^{3}r=\frac{V}{8\pi}\sum_{\textbf{q}}\left(q^{2}a_{q}^{2}-(\textbf{q}\textbf{a}_{q})^{2}\right).
\end{equation}
A magnetic field affects on a superconductor essentially. In the
first place, a distribution of order parameter becomes
inhomogeneous. As it was shown in the section \ref{spaceinhom}, an
inhomogeneity leads to same growth of momentum of each element of
a pair: $\textbf{k}\rightarrow\textbf{k}+\textbf{q}$,
$-\textbf{k}\rightarrow -\textbf{k}-\textbf{q}$, moreover the
order parameter depends on the momentum
$\Delta=\Delta(\textbf{q})$. In the second place, an ordinary
momentum must be replaced by a canonical momentum:
$\textbf{k}+\textbf{q}\rightarrow\textbf{k}+\textbf{q}-\frac{e}{c}\textbf{a}_{q}$,
moreover the order parameter depends on the momentum
$\Delta=\Delta(\textbf{q}-\frac{e}{c}\textbf{a}_{q})$.

\begin{figure}[h]
\includegraphics[width=16cm]{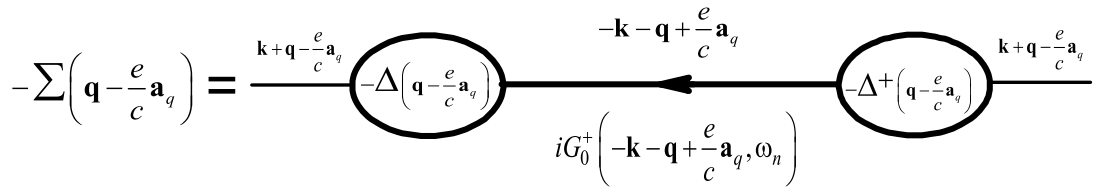}
\caption{The diagram for the mass operator $\Sigma$ describing the
interaction of a charged fermion with fluctuations of pairing in a
spatially inhomogeneous system situated in magnetic field with a
potential $\textbf{a}(\textbf{q})$.} \label{fig8}
\end{figure}

The mass operator for a process of interaction of a fermion (with
charge $e$) with a fluctuation of pairing is shown in
Fig.\ref{fig8}. In analytical representation this diagram has the
form:
\begin{eqnarray}\label{8.4}
-\Sigma\left(\textbf{k}+\textbf{q}-\frac{e}{c}\textbf{a}_{q},\omega_{n}\right)&=&
-\Delta\left(\textbf{q}-\frac{e}{c}\textbf{a}_{q}\right)iG_{0}^{+}\left(-\textbf{k}-\textbf{q}+\frac{e}{c}\textbf{a}_{q},\omega_{n}\right)\left(-\Delta^{+}\left(\textbf{q}-\frac{e}{c}\textbf{a}_{q}\right)\right)\nonumber\\
&=&\frac{-\Delta\left(\textbf{q}-\frac{e}{c}\textbf{a}_{q}\right)\Delta^{+}\left(\textbf{q}-\frac{e}{c}\textbf{a}_{q}\right)}{i\omega_{n}+\varepsilon\left(\textbf{k}+\textbf{q}-\frac{e}{c}\textbf{a}_{q}\right)},
\end{eqnarray}
where the free propagator $G_{0}$ is
\begin{equation}\label{8.5}
    G_{0}=\frac{1}{i\omega_{n}-\varepsilon(\textbf{k}+\textbf{q}-\frac{e}{c}\textbf{a}_{q})}=
i\frac{i\omega_{n}+\varepsilon\left(\textbf{k}+\textbf{q}-\frac{e}{c}\textbf{a}_{q}\right)}{(i\omega_{n})^{2}-\varepsilon^{2}\left(\textbf{k}+\textbf{q}-\frac{e}{c}\textbf{a}_{q}\right)}.
\end{equation}
Then from Dyson equation we can obtain the dressed propagator:
\begin{eqnarray}\label{8.6}
  \frac{1}{G_{0}} &=& \frac{1}{G_{S}}-i\Sigma_{q}\Rightarrow
G_{S}=i\frac{i\omega_{n}+\varepsilon\left(\textbf{k}+\textbf{q}-\frac{e}{c}\textbf{a}_{q}\right)}{(i\omega_{n})^{2}-E^{2}\left(\textbf{k}+\textbf{q}-\frac{e}{c}\textbf{a}_{q}\right)},
\end{eqnarray}
where $E$ is the specter of quasi-particles in a inhomogeneous
system situated in magnetic field:
\begin{eqnarray}\label{8.7}
&&E^{2}\left(\textbf{k}+\textbf{q}-\frac{e}{c}\textbf{a}_{q}\right)=\varepsilon^{2}\left(\textbf{k}+\textbf{q}-\frac{e}{c}\textbf{a}_{q}\right)+|\Delta\left(\textbf{q}-\frac{e}{c}\textbf{a}_{q}\right)|^{2}\equiv E_{\textbf{q},\textbf{a}}^{2}\nonumber\\
&&\varepsilon\left(\textbf{k}+\textbf{q}-\frac{e}{c}\textbf{a}_{q}\right)\approx\varepsilon(k)+\frac{\textbf{k}\left(\textbf{q}-\frac{e}{c}\textbf{a}_{q}\right)}{m}\equiv
\varepsilon_{\textbf{q},\textbf{a}}, \qquad |\textbf{k}|\simeq
k_{F},
\end{eqnarray}
where we have introduced the notations $E_{\textbf{q},\textbf{a}}$
and $\varepsilon_{\textbf{q},\textbf{a}}$ for convenience. Then
Dyson equation can be represented in the form of Gor'kov
equations. From these equations the expressions for anomalous
propagators follow:
\begin{eqnarray}\label{8.8}
\begin{array}{c}
  (i\omega_{n}-\varepsilon_{\textbf{q},\textbf{a}})G+\Delta_{\textbf{q},\textbf{a}} F^{+}=i \\
\\
  (i\omega_{n}+\varepsilon_{\textbf{q},\textbf{a}})F^{+}+G\Delta_{\textbf{q},\textbf{a}}=0\\
\end{array}\Rightarrow
\begin{array}{c}
  F^{+}(\textbf{k}+\textbf{q}-\frac{e}{c}\textbf{a}_{q},\omega_{n})=\frac{-i\Delta_{\textbf{q},\textbf{a}}^{+}}{(i\omega_{n})^{2}-E^{2}_{\textbf{q},\textbf{a}}} \\
\\
  F(\textbf{k}+\textbf{q}-\frac{e}{c}\textbf{a}_{q},\omega_{n})=(F^{+}(\textbf{k}+\textbf{q}-\frac{e}{c}\textbf{a}_{q},\omega_{n}))^{+}=\frac{i\Delta_{\textbf{q},\textbf{a}}}{(i\omega_{n})^{2}-E^{2}_{\textbf{q},\textbf{a}}}\\
\end{array}
\end{eqnarray}
If to suppose $q=0$ è $a=0$, then we shall have the expressions
(\ref{5.12}-\ref{5.13}).

In the space $(\textbf{k},t)$ the normal propagator has the form:
\begin{eqnarray}\label{8.9}
G(\textbf{k}+\textbf{q}-\frac{e}{c}\textbf{a}_{q},\tau)&=&-i\theta_{\tau}\left(g^{+}_{\textbf{q},\textbf{a}}A_{\textbf{q},\textbf{a}}e^{-E_{\textbf{q},\textbf{a}}\tau}+g^{-}_{\textbf{q},\textbf{a}}B_{\textbf{q},\textbf{a}}e^{E_{\textbf{q},\textbf{a}}\tau}\right)\nonumber\\
&&+i\theta_{-\tau}\left(g^{-}_{\textbf{q},\textbf{a}}A_{\textbf{q},\textbf{a}}e^{-E_{\textbf{q},\textbf{a}}\tau}+g^{+}_{q}B_{\textbf{q},\textbf{a}}e^{E_{\textbf{q},\textbf{a}}\tau}\right),
\end{eqnarray}
where
\begin{eqnarray}\label{8.10}
A\left(\textbf{k}+\textbf{q}-\frac{e}{c}\textbf{a}_{q}\right)=\frac{1}{2}\left(1+\frac{\varepsilon_{\textbf{q},\textbf{a}}}{E_{\textbf{q},\textbf{a}}}\right),\qquad
B\left(\textbf{k}+\textbf{q}-\frac{e}{c}\textbf{a}_{q}\right)=\frac{1}{2}\left(1-\frac{\varepsilon_{\textbf{q},\textbf{a}}}{E_{\textbf{q},\textbf{a}}}\right).
\end{eqnarray}
The anomalous propagators are
\begin{eqnarray}\label{8.11}
  F^{+}_{\alpha\beta}(\textbf{k}+\textbf{q}-\frac{e}{c}\textbf{a}_{q},\tau)&=&ig_{\alpha\beta}\frac{\Delta^{+}_{\textbf{q},\textbf{a}}}{2E_{\textbf{q},\textbf{a}}}
\left[\left(g_{\textbf{q},\textbf{a}}^{+}e^{-E_{\textbf{q},\textbf{a}}\tau}-g_{\textbf{q},\textbf{a}}^{-}e^{E_{\textbf{q},\textbf{a}}\tau}\right)\theta_{\tau}-\left(g_{\textbf{q},\textbf{a}}^{+}e^{E_{\textbf{q},\textbf{a}}\tau}-g_{\textbf{q},\textbf{a}}^{-}e^{-E_{\textbf{q},\textbf{a}}\tau}\right)\theta_{-\tau}\right],\nonumber\\
F_{\alpha\beta}(\textbf{k}+\textbf{q}-\frac{e}{c}\textbf{a}_{q},\tau)&=&ig_{\alpha\beta}\frac{\Delta_{\textbf{q},\textbf{a}}}{2E_{\textbf{q},\textbf{a}}}
\left[-\left(g_{\textbf{q},\textbf{a}}^{+}e^{-E_{\textbf{q},\textbf{a}}\tau}-g_{\textbf{q},\textbf{a}}^{-}e^{E_{\textbf{q},\textbf{a}}\tau}\right)\theta_{\tau}+\left(g_{\textbf{q},\textbf{a}}^{+}e^{E_{\textbf{q},\textbf{a}}\tau}-g_{\textbf{q},\textbf{a}}^{-}e^{-E_{\textbf{q},\textbf{a}}\tau}\right)\theta_{-\tau}\right].
\end{eqnarray}
where $g_{\textbf{q},\textbf{a}}^{+}$ and
$g_{\textbf{q},\textbf{a}}^{-}$ are statistical multipliers:
\begin{eqnarray}\label{8.12}
g^{-}\left(\textbf{k}+\textbf{q}-\frac{e}{c}\textbf{a}_{q}\right)=\frac{1}{e^{\beta
E_{\textbf{q},\textbf{a}}}+1},\qquad
g^{+}\left(\textbf{k}+\textbf{q}-\frac{e}{c}\textbf{a}_{q}\right)=\frac{1}{e^{-\beta
E_{\textbf{q},\textbf{a}}}+1}.
\end{eqnarray}
We can see, that the normal and anomalous propagators have a
complicated dependence on vector $\textbf{q}$, amplitudes of
pairing $\Delta,\Delta^{+}$ and magnetic field
$\textbf{a}(\textbf{q})$.

Kinetic energy of a system is determined in the following way:
\begin{eqnarray}\label{8.13}
  \langle
W\rangle&=&-2i\sum_{\textbf{k}}\varepsilon\left(\textbf{k}+\textbf{q}-\frac{e}{c}\textbf{a}_{q}\right)G\left(\textbf{k}+\textbf{q}-\frac{e}{c}\textbf{a}_{q},\tau\rightarrow
0^{-}\right)=2\sum_{\textbf{k}}\varepsilon_{\textbf{q},\textbf{a}}\left(g^{-}_{\textbf{q},\textbf{a}}A_{\textbf{q},\textbf{a}}+g^{+}_{\textbf{q},\textbf{a}}B_{\textbf{q},\textbf{a}}\right)\nonumber\\
&=&\sum_{\textbf{k}}\varepsilon_{\textbf{q},\textbf{a}}\left(1-\frac{\varepsilon_{\textbf{q},\textbf{a}}}{E_{\textbf{q},\textbf{a}}}\tanh\frac{\beta
E_{\textbf{q},\textbf{a}}}{2}\right)\nonumber\\
&=&W_{n}+\sum_{\textbf{k},\left(|\varepsilon(k)|<\omega_{D},|\textbf{k}|\simeq
k_{F}\right)}\varepsilon_{\textbf{q},\textbf{a}}\left(\frac{\varepsilon_{\textbf{q},\textbf{a}}}{|\varepsilon_{\textbf{q},\textbf{a}}|}\tanh\frac{\beta
|\varepsilon_{\textbf{q},\textbf{a}}|}{2}-\frac{\varepsilon_{\textbf{q},\textbf{a}}}{E_{\textbf{q},\textbf{a}}}\tanh\frac{\beta
E_{\textbf{q},\textbf{a}}}{2}\right)\equiv W_{n}+W_{S}.
\end{eqnarray}
We can see, that the kinetic energy depends on the vectors
$\textbf{q}$ and $\textbf{a}(\textbf{q})$ in a complicated way. If
we suppose $a=0$ and $q$ is small, then we shall obtain the
expression (\ref{7.10}).

Free energy of a superconductor is the sum of the following terms:
\begin{equation}\label{8.14}
    \Omega=\Omega_{n}+W_{S}-\frac{1}{\beta}S_{S}+\Omega_{\lambda}+W_{\texttt{field}}(\textbf{a}),
\end{equation}
where $W_{S}$ is the kinetic energy of fermions of a system in
superconductive phase, $S_{S}$ is the entropy of boholons,
$\Omega_{\lambda}=-\frac{1}{\beta}\ln R(\beta)$ is the energy
corresponding to an interaction, $W_{\texttt{field}}(\textbf{a})$
is the energy of the magnetic field (\ref{8.3}). The expressions
for $S$ and $\Omega_{\lambda}$ are obtained from the expressions
(\ref{5.19}) and (\ref{5.24}) (without transition to integration
over $\varepsilon$ only) by the replacement
$\Delta\rightarrow\Delta_{\textbf{q},\textbf{a}}$, $E\rightarrow
E_{\textbf{q},\textbf{a}}$,
$\varepsilon\rightarrow\varepsilon_{\textbf{q},\textbf{a}}$,
$f\rightarrow f_{\textbf{q},\textbf{a}}$. Hence the free energy of
a superconductor for given $\textbf{q}$ is
\begin{eqnarray}\label{8.15}
  \Omega(\textbf{q},\textbf{a}_{q})&=&\Omega_{n}(\textbf{q},\textbf{a}_{q})+\sum_{\textbf{k},\left(|\varepsilon(k)|<\omega_{D},|\textbf{k}|\simeq
k_{F}\right)}\varepsilon_{\textbf{q},\textbf{a}}\left(\frac{\varepsilon_{\textbf{q},\textbf{a}}}{|\varepsilon_{\textbf{q},\textbf{a}}|}\tanh\frac{\beta
|\varepsilon_{\textbf{q},\textbf{a}}|}{2}-\frac{\varepsilon_{\textbf{q},\textbf{a}}}{E_{\textbf{q},\textbf{a}}}\tanh\frac{\beta
E_{\textbf{q},\textbf{a}}}{2}\right)\nonumber\\
&+&\sum_{\textbf{k},\left(|\varepsilon(k)|<\omega_{D},|\textbf{k}|\simeq
k_{F}\right)}\frac{2}{\beta}\left[f_{\textbf{q},\textbf{a}}^{S}\ln
f_{\textbf{q},\textbf{a}}^{S}+(1-f_{\textbf{q},\textbf{a}}^{S})\ln
(1-f_{\textbf{q},\textbf{a}}^{S})-f_{\textbf{q},\textbf{a}}^{0}\ln
f_{\textbf{q},\textbf{a}}^{0}-(1-f_{\textbf{q},\textbf{a}}^{0})\ln
(1-f_{\textbf{q},\textbf{a}}^{0})\right]\nonumber\\
&+&\frac{\lambda}{V}\sum_{\textbf{k},\left(|\varepsilon(k)|<\omega_{D},|\textbf{k}|\simeq
k_{F}\right)}\frac{\Delta_{\textbf{q},\textbf{a}}}{2E_{\textbf{q},\textbf{a}}}\tanh\frac{\beta
E_{\textbf{q},\textbf{a}}}{2}
\sum_{\textbf{k},\left(|\varepsilon(k)|<\omega_{D},|\textbf{k}|\simeq
k_{F}\right)}\frac{\Delta_{\textbf{q},\textbf{a}}}{2E_{\textbf{q},\textbf{a}}}\tanh\frac{\beta
E_{\textbf{q},\textbf{a}}}{2}\nonumber\\
&+&\frac{V}{8\pi}\left(q^{2}a_{q}^{2}-(\textbf{q}\textbf{a}_{q})^{2}\right)
\equiv\Omega_{n}(\textbf{q},\textbf{a}_{q})+\Omega_{S}(\textbf{q},\textbf{a}_{q})+w_{\texttt{field}}(\textbf{q},\textbf{a}_{q}),
\end{eqnarray}
where
\begin{equation}\label{8.16}
    f_{\textbf{q},\textbf{a}}^{S}(k)=\frac{1}{e^{\beta
E_{\textbf{q},\textbf{a}}}+1},\qquad
f_{\textbf{q},\textbf{a}}^{0}(k)=\frac{1}{e^{\beta
|\varepsilon_{\textbf{q},\textbf{a}}|}+1}
\end{equation}
are occupation numbers of states by boholons, moreover with help
of the term $f_{\textbf{q},\textbf{a}}^{0}$ a normal part of
entropy is separated, such that $\Omega_{S}(\Delta=0)=0$. The full
free energy is sum of the expression (\ref{8.15}) over all
possible $\textbf{q}$:
\begin{eqnarray}\label{8.17}
\Omega=\Omega_{n}+\sum_{\textbf{q}}\left\{\Omega_{S}(\textbf{q},\textbf{a}_{q})+w_{\texttt{field}}(\textbf{q},\textbf{a}_{q})\right\}.
\end{eqnarray}
\emph{Unlike Ginzburg-Landau functional the obtained functional of
free energy} (\ref{8.17}) \emph{is correct for an arbitrary value
of the relation} $\Delta(T)/T$, \emph{for an arbitrary scale of a
change of $\Delta(\textbf{r})$ in comparison with a coherent
length} $l(T)$, \emph{for an arbitrary value of a magnetic
penetration depth $\lambda(T)$ in comparison with a coherent
length} $l_{0}$ - \emph{it describes a nonlocal response to
magnetic field}. However the obtained expression is complicated
for analyze.  For its simplification let's suppose that
$\Delta(\textbf{r})$ changes in space slowly. Then it is necessary
to expand the expression (\ref{8.15}) in degrees of
$\textbf{q}-\frac{e}{c}\textbf{a}_{q}$ keeping terms which are
proportional to the vector in second degree only. Then supposing
$E_{\textbf{q},\textbf{a}}\approx\sqrt{\varepsilon^{2}(k)+\Delta_{\textbf{q},\textbf{a}}^{2}}$
we have:
\begin{eqnarray}\label{8.18}
\Omega(\textbf{q},\textbf{a}_{q})=&&\Omega_{n}(\textbf{q},\textbf{a}_{q})+V\frac{\nu_{F}}{2}\int_{-\omega_{D}}^{\omega_{D}}\left[\tanh\left(\frac{\beta|\varepsilon|}{2}\right)\frac{\varepsilon^{2}}{|\varepsilon|}-
\tanh\left(\frac{\beta
E}{2}\right)\frac{\varepsilon^{2}}{E}\right]d\varepsilon\nonumber\\
&&+\frac{2V}{\beta}\frac{\nu_{F}}{2}\int_{-\omega_{D}}^{\omega_{D}}\left[f_{S}\ln
f_{S}+(1-f_{S})\ln (1-f_{S})-f_{0}\ln f_{0}-(1-f_{0})\ln
(1-f_{0})\right]d\varepsilon\nonumber\\
&&+V\frac{\nu_{F}}{2}g\int_{-\omega_{D}}^{\omega_{D}}\tanh\left(\frac{\beta
E}{2}\right)\frac{\Delta}{2E}d\varepsilon\int_{-\omega_{D}}^{\omega_{D}}\tanh\left(\frac{\beta
E}{2}\right)\frac{\Delta}{2E}d\varepsilon\nonumber\\
&&+V\frac{\nu_{F}}{2}\frac{1}{3}v_{F}^{2}\left(\textbf{q}-\frac{e}{c}\textbf{a}_{q}\right)^{2}\int_{-\omega_{D}}^{\omega_{D}}\left[\tanh\left(\frac{\beta|\varepsilon|}{2}\right)\frac{1}{|\varepsilon|}-
\tanh\left(\frac{\beta
E}{2}\right)\frac{1}{E_{q}}\right]d\varepsilon+\frac{V}{8\pi}\left(q^{2}a_{q}^{2}-(\textbf{q}\textbf{a}_{q})^{2}\right),
\end{eqnarray}
where $g=\lambda\frac{\nu_{F}}{2}$ is the effective interaction
constant, and the gap is
$\Delta=\Delta\left(\textbf{q}-\frac{e}{c}\textbf{a}_{q}\right)$.

Let's consider a high-temperature limit of the free energy:
$\Delta\beta_{c}\ll 1$ at $T\rightarrow T_{c}$. This means, that
the expression (\ref{8.18}) can be expended in degrees of
$\Delta_{\textbf{q},\textbf{a}}$:
\begin{equation}\label{8.19}
\Omega=\Omega_{n}+V\sum_{\textbf{q}}\left(\alpha(T)\Delta^{2}_{\textbf{q},\textbf{a}}+\frac{1}{2}b\Delta^{4}_{\textbf{q},\textbf{a}}+\frac{1}{3}d\Delta^{6}_{\textbf{q},\textbf{a}}+\gamma\left(\textbf{q}-\frac{e}{c}\textbf{a}_{q}\right)^{2}\Delta_{\textbf{q},\textbf{a}}^{2}\right)
+\frac{V}{8\pi}\sum_{\textbf{q}}\left(q^{2}a_{q}^{2}-(\textbf{q}\textbf{a}_{q})^{2}\right),
\end{equation}
where the coefficients $\alpha(T),b,d$ are determined by the
formulas (\ref{5.33}), and the coefficient $\gamma$ is determined
by the formula (\ref{7.13}). The expansion (\ref{8.19}) has a form
of Ginzburg-Landau expansion of free energy in degrees of order
parameter. Observed configuration of the order parameter
$\Delta_{\textbf{q},\textbf{a}}$ and the magnetic field
$\textbf{a}(\textbf{q})$ minimizes the free energy:
\begin{eqnarray}
  \frac{\delta\Omega}{\delta\Delta}=0 &\Rightarrow& \alpha(T)\Delta_{\textbf{q},\textbf{a}}+b\Delta^{3}_{\textbf{q},\textbf{a}}+d\Delta^{5}_{\textbf{q},\textbf{a}}+\gamma\left(\textbf{q}-\frac{e}{c}\textbf{a}_{q}\right)^{2}\Delta_{\textbf{q},\textbf{a}}=0 \label{8.20}\\
  \frac{\delta\Omega}{\delta\textbf{a}}=0 &\Rightarrow&
\textbf{j}(\textbf{q})=2e\gamma\textbf{q}\Delta^{2}_{\textbf{q},\textbf{a}}-2\gamma\frac{e^{2}}{c}\Delta^{2}_{\textbf{q},\textbf{a}}\textbf{a}(\textbf{q})\label{8.21},
\end{eqnarray}
where $\textbf{j}(\textbf{q})$ is Fourier component of a current:
\begin{equation}\label{8.22}
\textbf{j}(\textbf{q})=-\frac{c}{4\pi}\textbf{q}\times\textbf{q}\times\textbf{a}(\textbf{q})
=-\frac{c}{4\pi}\left(\textbf{q}(\textbf{qa})-\textbf{a}q^{2}\right).
\end{equation}

\begin{figure}[tbp]
  \includegraphics[width=8cm]{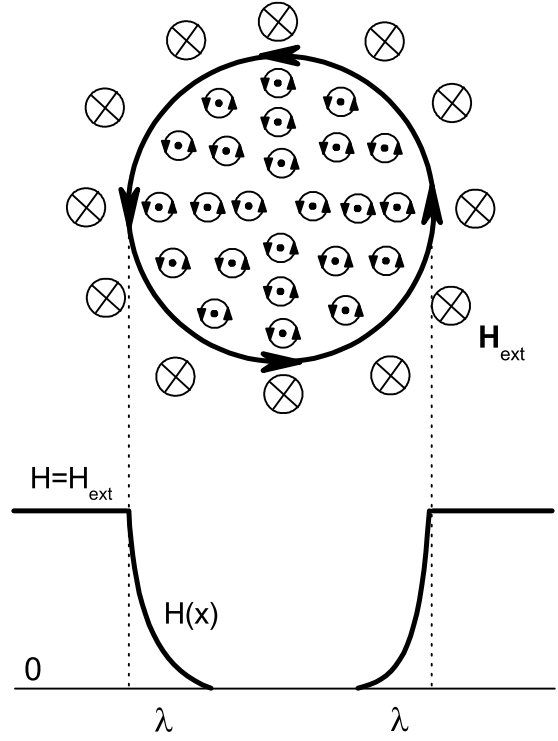}
  \caption{The screening of the external magnetic field $H_{ext}$, impressed in parallel to
a superconductive cylinder, by inducted closed currents. The
inducted currents are directed so as to compensate the external
field. The resultant current goes around a lateral area of
cylinder, but strives to zero inside of the cylinder. The area
where the resultant current is not equal to zero is surface layer
with thickness $\sim\lambda$. The magnetic field penetrates in a
superconductor in the depth $\sim\lambda$ too.}\label{fig9}
\end{figure}

These equations take more simple form in a transverse gauge
$\textbf{q}\cdot \textbf{a}(\textbf{q})=0$. This gauge gives a
condition of closure of a current (special case of conservation of
charge):
\begin{equation}\label{8.22a}
\textbf{q}\cdot\textbf{a}(\textbf{q})=0\Rightarrow
\textbf{j}(\textbf{q})=-\frac{c}{4\pi}\textbf{a}(\textbf{q})q^{2}\Rightarrow\textbf{q}\cdot\textbf{j}(\textbf{q})=0\Leftrightarrow
\texttt{div}\textbf{J}(\textbf{r})=0.
\end{equation}
The closed currents (\ref{8.22a}) screen a magnetic field in a
superconductor - Fig.\ref{fig9}. The currents is analogy to
molecular currents of Ampere, their resulting gives rise to
observed magnetic effects.
In transverse gauge the functional of free energy has a form:
\begin{equation}\label{8.23}
\Omega=\Omega_{n}+V\sum_{\textbf{q}}\left(\alpha(T)\Delta^{2}_{\textbf{q},\textbf{a}}+\frac{1}{2}b\Delta^{4}_{\textbf{q},\textbf{a}}+\frac{1}{3}d\Delta^{6}_{\textbf{q},\textbf{a}}+\gamma\left(\textbf{q}^{2}+\frac{e^{2}}{c^{2}}\textbf{a}_{q}^{2}\right)\Delta_{\textbf{q},\textbf{a}}^{2}\right)
+\frac{V}{8\pi}\sum_{\textbf{q}}q^{2}a_{q}^{2}.
\end{equation}
The equations of extremals are
\begin{eqnarray}
&&\alpha(T)\Delta_{\textbf{q},\textbf{a}}+b\Delta^{3}_{\textbf{q},\textbf{a}}+d\Delta^{5}_{\textbf{q},\textbf{a}}+\gamma\left(\textbf{q}^{2}+\frac{e^{2}}{c^{2}}\textbf{a}_{q}^{2}\right)\Delta_{\textbf{q},\textbf{a}}=0 \label{8.24}\\
&&\textbf{j}(\textbf{q})=-2\gamma\frac{e^{2}}{c}\Delta^{2}_{\textbf{q},\textbf{a}}\textbf{a}(\textbf{q})\label{8.25}.
\end{eqnarray}
From (\ref{8.25}) one can see, that the value
$Q=-2\gamma\frac{e^{2}}{c}\Delta^{2}_{\textbf{q},\textbf{a}}$ is a
Fourier transform of a kernel in the integral law of a magnetic
response (Pippard law). The order parameter is function of
$\textbf{q}$ and $\textbf{a}(\textbf{q})$:
\begin{equation}\label{8.26}
\Delta^{2}_{\textbf{q},\textbf{a}}(T)=\frac{|\alpha(T)|}{b}\left(1-\frac{\gamma}{|\alpha(T)|}\left(\textbf{q}^{2}+\frac{e}{c}\textbf{a}_{q}^{2}\right)\right)
\equiv\frac{|\alpha(T)|}{b}\left(1-l^{2}(T)\left(\textbf{q}^{2}+\frac{e}{c}\textbf{a}_{q}^{2}\right)\right),
\end{equation}
where smallness of $1/q$ in comparison with the coherent length
$l(T)$: $ql(T)\ll 1$ is supposed, and we assumed that the
coefficient $d=0$ for simplification. From the formula
(\ref{8.26}) one can see, that the kernel of the magnetic response
$Q$ is function of magnetic field. Hence the electrodynamics of a
superconductor is nonlinear. If to suppose $\Delta=\texttt{const}$
at given temperature, then we shall obtain London equation:
\begin{equation}\label{8.27}
\textbf{j}(\textbf{q})=-2\gamma\frac{e^{2}}{c}\Delta^{2}(T)\textbf{a}(\textbf{q})\equiv-\frac{c}{4\pi\lambda^{2}(T)}\textbf{a}(\textbf{q})
\Rightarrow\lambda^{2}(T)=\frac{c^{2}}{8\pi
e^{2}}\frac{b}{|\alpha(T)|\gamma},
\end{equation}
where $\lambda^{2}(T)$ is the magnetic penetration depth in a
superconductor. It is necessary to note, that
$\lambda\propto\frac{m}{e^{2}n_{S}}=\frac{2m}{(2e)^{2}n_{S}/2}$.
This means that the exchange of mass, charge and concentration of
superconductive electrons to corresponding values of Cooper pairs
doesn't change the observed values.

However it is necessary to note, that in a high-temperature limit
a gap (and a kernel $Q$) depends on vector $\textbf{q}$ and field
$\textbf{a}(\textbf{q})$ strongly. From the formula (\ref{8.26})
one can see, that the gap decreases at an increase of $q$, hence a
magnetic penetration depth $\lambda$ increases. Moreover with a
rise of temperature this dependence becomes stronger (at $T=T_{C}$
we have $\lambda=\infty$). Besides at temperature $T=T_{C}$ the
critical magnetic field is zero $H_{C}=0$. This means, that in the
limit $T\rightarrow T_{C}$ any magnetic field
$\textbf{H}(\textbf{q})$ can not be considered as weak. Therefore
it suppresses order parameter essentially and penetrates in a
superconductor deeply (in macroscopic distant even). For example,
the penetration of magnetic field along a core of Abrikosov vortex
in a type II superconductor. Such structure exists in infinitely
weak magnetic field at $T\rightarrow T_{C}$.

For research of the nonlocal characteristics of the functional of
free energy (\ref{8.15},\ref{8.17}) let's consider a
low-temperature limit $\Delta\beta\gg 1$ at $T\rightarrow 0$. A
value of gap is close to the value at zero temperature
$\Delta(T)\leq\Delta_{0}$. Moreover, magnetic field is weak, such
that it changes a value of gap lightly, that is the magnetic field
is much smaller than critical field $H\ll H_{C}$. Either as above,
we assume that a change of a gap in space is slow. Starting from
aforesaid and using the expansion (\ref{5.29}) we obtain the free
energy:
\begin{equation}\label{8.29}
\Omega=\Omega_{n}+V\sum_{\textbf{q}}\left(\alpha_{0}(T)+b_{0}(T)\Delta_{\textbf{q},\textbf{a}}+d_{0}\Delta^{2}_{\textbf{q},\textbf{a}}+\gamma\left(\textbf{q}-\frac{e}{c}\textbf{a}_{q}\right)^{2}\Delta_{\textbf{q},\textbf{a}}^{2}\right)
+\frac{V}{8\pi}\sum_{\textbf{q}}\left(q^{2}a_{q}^{2}-(\textbf{q}\textbf{a}_{q})^{2}\right),
\end{equation}
where coefficients $\alpha_{0}(T),b_{0}(T),d_{0}$ are determined
by the formulas (\ref{5.30a}), and coefficient $\gamma$ is
determined by the formula (\ref{7.13}). The observed
configurations of order parameter $\Delta_{\textbf{q},\textbf{a}}$
and magnetic field $\textbf{a}(\textbf{q})$ minimized free energy:
\begin{eqnarray}
  \frac{\delta\Omega}{\delta\Delta}=0 &\Rightarrow& b_{0}(T)+2d_{0}\Delta_{\textbf{q},\textbf{a}}+\gamma\left(\textbf{q}-\frac{e}{c}\textbf{a}_{q}\right)^{2}\Delta_{\textbf{q},\textbf{a}}=0 \label{8.30}\\
  \frac{\delta\Omega}{\delta\textbf{a}}=0 &\Rightarrow&
\textbf{j}(\textbf{q})=2e\gamma\textbf{q}\Delta^{2}_{\textbf{q},\textbf{a}}-2\gamma\frac{e^{2}}{c}\Delta^{2}_{\textbf{q},\textbf{a}}\textbf{a}(\textbf{q})\label{8.31}.
\end{eqnarray}
In the transverse gauge $\textbf{q}\cdot \textbf{a}(\textbf{q})=0$
the functional of free energy and the equations for the extremals
have a form:
\begin{eqnarray}
&&\Omega=\Omega_{n}+V\sum_{\textbf{q}}\left(\alpha_{0}(T)+b_{0}(T)\Delta_{\textbf{q},\textbf{a}}+d_{0}\Delta^{2}_{\textbf{q},\textbf{a}}+\gamma\left(\textbf{q}^{2}+\frac{e^{2}}{c^{2}}\textbf{a}_{q}^{2}\right)\Delta_{\textbf{q},\textbf{a}}^{2}\right)
+\frac{V}{8\pi}\sum_{\textbf{q}}q^{2}a_{q}^{2}\label{8.32}\\
&& b_{0}(T)+2d_{0}\Delta_{\textbf{q},\textbf{a}}+2\gamma\left(\textbf{q}^{2}+\frac{e^{2}}{c^{2}}\textbf{a}_{q}^{2}\right)\Delta_{\textbf{q},\textbf{a}}=0 \label{8.33}\\
&&\textbf{j}(\textbf{q})=-2\gamma\frac{e^{2}}{c}\Delta^{2}_{\textbf{q},\textbf{a}}\textbf{a}(\textbf{q})\label{8.34}.
\end{eqnarray}
If in the equation (\ref{8.34}) to assume $\Delta=\texttt{const}$
at given temperature, then we shall have London equation again:
\begin{equation}\label{8.35}
\textbf{j}(\textbf{q})=-2\gamma\frac{e^{2}}{c}\Delta^{2}(T)\textbf{a}(\textbf{q})\equiv-\frac{c}{4\pi\lambda^{2}(T)}\textbf{a}(\textbf{q})
\Rightarrow\lambda^{2}(T)=\frac{c^{2}}{2\pi
e^{2}}\frac{d_{0}^{2}}{b_{0}^{2}(T)\gamma},
\end{equation}
The set of equations (\ref{8.33},\ref{8.34}) allows to generalize
London equation. From the equation (\ref{8.33}) we can find value
of a gap:
\begin{equation}\label{8.36}
    \Delta_{\textbf{q},\textbf{a}}(T)=\frac{-b_{0}(T)}{2d_{0}+2\gamma\left(\textbf{q}^{2}+\frac{e^{2}}{c^{2}}\textbf{a}_{q}^{2}\right)}
\end{equation}
Then the equation for current has a form:
\begin{equation}\label{8.37}
\textbf{j}(\textbf{q})=-2\gamma\frac{e^{2}}{c}\frac{b_{0}^{2}(T)}{\left(2d_{0}+2\gamma\left(\textbf{q}^{2}+\frac{e^{2}}{c^{2}}\textbf{a}_{q}^{2}\right)\right)^{2}}\textbf{a}(\textbf{q})
\equiv Q(\textbf{q},\textbf{a})\textbf{a}(\textbf{q})
\end{equation}
This equation is the nonlocal and nonlinear generalization of
London equation in a long wavelength limit $q\rightarrow 0$,
because the kernel $Q(\textbf{q},\textbf{a})$ is function of $q$
and magnetic field $\textbf{a}(\textbf{q})$. However the equation
(\ref{8.37}) is correct when a magnetic field is much weaker than
the critical field $H\ll H_{C}$.

Let's neglect by the nonlinearity, that is we suppose that the
kernel $Q$ is function of $q$ only. Then we have:
\begin{equation}\label{8.38}
\textbf{j}(\textbf{q})=-2\gamma\frac{e^{2}}{c}\frac{b_{0}^{2}(T)}{\left(2d_{0}+2\gamma\textbf{q}^{2}\right)^{2}}\textbf{a}(\textbf{q})
=-2\gamma\frac{e^{2}}{c}\frac{\Delta_{0}^{2}(T)}{\left(1+\frac{\gamma}{d_{0}}\textbf{q}^{2}\right)^{2}}\textbf{a}(\textbf{q})
\approx-2\gamma\frac{e^{2}}{c}\Delta_{0}^{2}(T)\left(1-l_{0}^{2}\textbf{q}^{2}\right)\textbf{a}(\textbf{q})
\equiv Q(q)\textbf{a}(\textbf{q}),
\end{equation}
where we took into account a slowness of changes of a gap in
space: $l_{0}q\ll 1$.
$2\frac{\gamma}{d_{0}}=2\frac{l_{0}^{2}\nu_{F}/2}{\nu_{F}}=l^{2}_{0}$
is a coherent length at temperature $T=0$. Thus \emph{we have the
nonlocal kernel} $Q(q)$, \emph{where radius of a nonlocality is
equal to the coherent length} $l_{0}$. \emph{This result
corresponds to nonlocal Pippard electrodynamics} (long wavelength
limit). \emph{This fact proves nonlocality of the obtained
functional of free energy of a superconductor} (\ref{8.15},
\ref{8.17}). For generalization in case of a large value $l_{0}q$
it is necessary to expand the free energy (\ref{8.15}) in degrees
of $q$. Then we can obtain a short wavelength limit of $Q$: $Q\sim
1/ql_{0}$. Starting from correctness of the asymptotics
(\ref{8.19}) and (\ref{8.29}) of the functional (\ref{8.15},
\ref{8.17}) we can make a conclusion about its correctness for
description of a superconductive phase.

\section{Conclusion.}\label{conclusion}

In this paper on the example of superconductivity we described the
type II phase transition on a microscopic level, namely starting
from first principles. This means, that the method of calculation
of a free energy $\Omega(T,N/V)$) has been developed in a range of
temperatures, which includes a point of pase transition, without
introducing any artificial parameters of type of order parameter
and sourses of ordering, but starting from microscopic parameters
of Hamiltonian only. Moreover, the theorems about connection of a
vacuum amplitude with thermodynamics potentials are realized.

Microscopic picture of a phase transition lies in the following.
At switching of attraction between particles of a Fermy system the
instability relatively formation of bound states of two fermions
rises. The given states are characterized by Bethe-Solpiter
amplitudes - amplitudes of pairing, which can be found exactly in
the case of an isolated pair. However, in consequence of
statistical correlations between pairs the amplitudes are
determined by dynamics of all particles of a system. Their
observing value is result of averaging over the system. Thus, a
collective (condensate) of pairs exists. A particle propagating
through a system interacts with fluctuations of pairing. As a
result of such interaction a dispersion law of quasi-particles is
changed and anomalous propagators appear. This means, that a
spontaneous symmetry breakdown takes place. After consideration of
interaction of particles with fluctuations of pairing all
characteristics of a system must be calculated over the new vacuum
with broken symmetry. So, calculation of a vacuum amplitude over
new ground state gives a possibility to use the theorem about
connection of a vacuum amplitude with a ground state energy (with
free energy at nonzero temperature). As a result, the free energy
is function of amplitudes $\Omega=\Omega(\Delta\Delta^{+})$, and
their observed value minimizes the free energy. Analyze of the
obtained functional of free energy shows, that the amplitude of
pairing plays a part of an order parameter. Namely, the order
parameter is Bethe-Solpiter amplitude averaged over a system due
to statistical and dynamical correlations. Since in Nambu-Gor'kov
formalism (the method of anomalous propagators) any phase
transition can be described \citep{matt2}, then our method can be
generalized to the rest transitions (ferromagnetism and
antiferromagnetism, waves of charge and spin density,
ferroelectricity and so on).

The functional of a superconductor's free energy (\ref{8.15},
\ref{8.17}) has been obtained in this paper using the developed
method of microscopic description of phase transitions and
generalizing its in the cases of spatial inhomogeneity and
presence of magnetic field. The functional generalizes
Ginzburg-Landau functional for cases of arbitrary temperatures,
arbitrary spatial inhomogeneities and a nonlocality of a magnetic
response. The equations of superconductor's state are extremals of
the functional obtained by variation over the gap $\Delta$ and the
magnetic field $\textbf{a}$. The equations determined equilibrium
configurations of a gap and a magnetic field at given conditions.

\appendix

\section{The method of uncoupling of correlations and Dyson equation.}\label{appendixA}

As it was shown in \citep{migdal} for propagators $G$
(one-particle), $K_{2}$ (two-particle), $K_{3}$ (tree-particle)
... $K_{n}$ (n-particle) the set of coupling equations can be
written. The set of equations is analogous to BBGKY hierarchy for
a s-particle probability density function. For example, for $G$
and $K_{2}$ these equations have a view:
\begin{eqnarray}
  G(1,1') &=& G_{0}(1,1')+i\int G_{0}(1,2)U(2,3)K_{2}(2,3;1',3^{+})dx_{2}dx_{3}\label{A1} \\
  K_{2}(1,2;1',2') &=& G(2,2')G_{0}(1,1')-G(2,1')G_{0}(1,2')+i\int
G_{0}(1,3)U(3,4)K_{3}(3,2,4;1',2',4^{+})dx_{3}dx_{4}\label{A2},
\end{eqnarray}
where $3^{+}\equiv(\xi_{3},t_{3}+0)$, $U(2,3)\equiv
V(\xi_{2},\xi_{3})\delta(t_{2}-t_{3})$, $dx\equiv d\xi dt$. The
cross term $G(2,2')G_{0}(1,1')-G(2,1')G_{0}(1,2')$ appeared as
result of calculation of Fermy symmetry of particles (it would be
$"+"$ for bosons). The equation for a three-particle propagator
$K_{3}$ will be is determined by four-particle propagator $K_{4}$
and so on: $K_{n}=f(K_{n+1})$.

\begin{figure}[h]
\includegraphics[width=7cm]{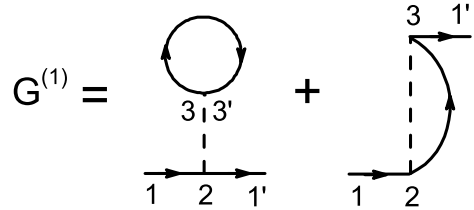}
\caption{The correction of first order $G_{1}$ obtained by
uncoupling of correlations in the equation (\ref{A1}).}
\label{fig10}
\end{figure}

It is obviously that this set of equations can not be solved.
However, in most cases for description of a system it is enough to
know functions $G$ and $K_{2}$ (less). Then the method of
uncoupling of correlations is used. The function $K_{2}$ in a zero
approximation is
\begin{equation}\label{A3}
    K_{2}^{(0)}=G_{0}(1,1')G_{0}(2,2')-G_{0}(1,2')G_{0}(2,1').
\end{equation}
We can see, that in an absence of interaction the two-particle
propagator is represented in a multiplicative form by the
one-particle free propagators $G_{0}$. Statistical correlation
exists only in the course of Pauli exclusion principle. Then in
first approximation let's use the free propagator $K_{2}^{(0)}$
instead of the dressed propagators $K_{2}$ in the formula
(\ref{A1}). Hence the correction for the dressed one-particle
propagator $G$ is
\begin{eqnarray}\label{A4}
  G^{(1)}(1,1') &=& i\int G_{0}(1,2)U(2,3)G_{0}(3,3^{+})G_{0}(2,1')dx_{2}dx_{3}-i\int
G_{0}(1,2)U(2,3)G_{0}(3,1')G_{0}(2,3)dx_{2}dx_{3}.
\end{eqnarray}
The correction (\ref{A4}) is represented graphically in
Fig.\ref{fig10}. Integration is carried over coordinates of the
internal lines. The first term corresponds to direct Hartree
interaction, the second term corresponds to exchange Fock
interaction.
In order to obtain the next approximation for $G$ it is necessary
to find $K_{2}^{(1)}$. In a symbolic representation we have the
equations:
\begin{eqnarray}\label{A5}
&&K_{2}^{(1)}=G^{(1)}G_{0}-G_{0}G^{(1)}+iG_{0}UK_{3}^{(0)}\nonumber\\
&&G^{(2)}=iG_{0}UK_{2}^{(1)}.
\end{eqnarray}

\begin{figure}[h]
\includegraphics[width=8cm]{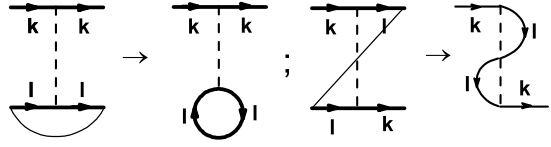}
\caption{The graphical method of uncoupling of correlations. In
the diagram for the first order correction $K_{1}$ of two-particle
propagator we connect two lines taking into account conservation
of momentum and spin. As a result, we have the first order
correction $G_{1}$ for one-particle propagator.} \label{fig11}
\end{figure}

The procedure of uncoupling of correlations can be represented in
another way. Let a correction for two-particle propagator
$K^{(1)}$ is determined by the matrix elements: $V_{klkl}$ is a
direct interaction and $V_{kllk}$ is an exchange interaction. A
two-particle propagator has two entering momentums (represented by
lines with corresponding indexes) and two outgoing momentums. An
one-particle propagator has one entering momentum and one outgoing
momentum. The procedure of uncoupling of correlations consist in
the fact, that we connect two lines in $K$ taking into account
conservation of momentum and spin: Fig.\ref{fig11}. The connection
means integration over intermediate momentums and energy
parameters (in a momentum-energy representation $\xi,t\rightarrow
\textbf{k},\omega$) in the formula (\ref{A4}). As a result, we
have the same diagrams for $G$ as in Fig.\ref{fig10}.

The procedure can be generalized to higher corrections (with two
and more lines of interaction). In general case the rules of
diagram technique are:
\begin{enumerate}
\item The multiplier $iG(\textbf{k},\omega)$ is associated with
each bold line, the multiplier $iG_{0}(\textbf{k},\omega)$ is
associated with each thin line \item The multiplier $-iV_{klmn}$
is associated with each dashed line of interaction between
particles, and the multiplier $-iV_{kl}$ is associated with line
of interaction of a particle with an external field. \item The
multiplier $-1$ associated with each fermion loop and we make
summation over all possible spin configurations. \item Momentum
$\textbf{k}+\textbf{l}=\textbf{m}+\textbf{n}$, energy parameter
and spin are reserved in every vertex. \item Integration is made
over each intermediate momentum and summation is made over each
intermediate energy parameter:
$\sum_{\textbf{k}}\rightarrow\frac{1}{V}\int\frac{d^{3}k}{(2\pi)^{3}}$
($V$ is volume of a system) and $\int\frac{d\omega}{2\pi}$.
\end{enumerate}

The diagrams of the type Fig.\ref{fig10} are summarized with help
of the mass operator $\Sigma$ - any diagram without external
lines. Hence a dressed propagator $G$ can be found from Dyson
equation:
\begin{eqnarray}\label{A6}
iG=iG_{0}+iG_{0}(-i)\Sigma iG\Rightarrow
G=\frac{1}{G_{0}^{-1}-\Sigma}
\end{eqnarray}

The mass operator $\Sigma$ has the sense of a mean field of all
particles of a system acting on a test (marked) article. In this
fact the sense of the procedure of uncoupling of correlations is:
interaction and propagation of all particle of a system is reduced
to propagation of each particle in the mean field of all rest
particles.

Other approach exists (more widely represented in literature) for
obtaining of the diagram expansion for $G$. In this approach a
one-particle propagator is determined as
\begin{equation}\label{A7}
G(\textbf{k}_{2},\textbf{k}_{1},t_{2}-t_{1})=\lim_{
\begin{array}{c}
  T_{1}\rightarrow -\infty(1-i\delta) \\
  T_{2}\rightarrow +\infty(1-i\delta) \\
\end{array}}
\frac{-i\langle\Phi_{0}|T[\widetilde{U}(T_{2},T_{1})\widehat{C}_{\textbf{k}_{2}}(t_{2})\widehat{C}_{\textbf{k}_{2}}^{+}(t_{2})]|\Phi_{0}\rangle}
{\langle\Phi_{0}|\widetilde{U}(T_{2},T_{1})|\Phi_{0}\rangle},
\end{equation}
where $\widehat{C}^{+}(t),\widehat{C}(t)$ is creation and
annihilation operators in \emph{interaction representation},
$\widetilde{U}$ is evolution operator in interaction
representation, $\Phi_{0}$ is ground state of a system of
\emph{noninteracting} fermions. This definition of a propagator is
equivalent to the definition
\begin{eqnarray}\label{A8}
  G(\textbf{k},t_{2}-t_{1})=-i\langle\Psi_{0}|T[C_{\textbf{k},\sigma}(t_{2})C_{\textbf{k},\sigma}^{+}(t_{1})]|\Psi_{0}\rangle,
\end{eqnarray}
where $\widehat{C}^{+}(t),\widehat{C}(t)$ is creation and
annihilation operators in Heisenberg representation, $\Psi_{0}$ is
ground state of a system of \emph{interacting} fermions, and
expansion of $G$ in series of perturbation theory is possible if
the condition of adiabaticity is realized (\ref{1.9}):
\begin{equation}\label{A9}
    \langle\Phi_{0}|\Psi_{0}\rangle\neq0.
\end{equation}

In the method of uncoupling of correlations we didn't use the
condition (\ref{A9}) and Wick theorem unlike the standard
formulation of the perturbation theory based on (\ref{A7}). Thus
\emph{the advantage of stated above method of uncoupling of
correlations consists in that we can formulate a perturbation
theory without using of the adiabatic hypothesis} (\ref{A9}).

\section{The method of uncoupling of correlations for a vacuum amplitude in the case of normal processes.}\label{appendixB}

\begin{figure}[h]
\includegraphics[width=8cm]{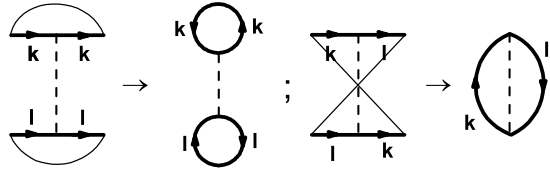}
\caption{The procedure of uncoupling of correlations in vacuum
amplitude for a correction of first order. In the average of
matrix element of interaction operator $V\langle
C^{+}C^{+}CC\rangle$ we connect the lines taking into account
conservation of momentum and spin. As a result, we obtain the
correction of first order $R_{1}$ for a vacuum amplitude of a view
$V\langle C^{+}C\rangle\langle C^{+}C\rangle$.} \label{fig12}
\end{figure}

Let us consider the processes of direct and exchange interaction
of first order, which is described by the matrix elements
$V_{klkl}$ and $V_{lkkl}$, moreover the interaction doesn't act to
spins of particles. Contribution to vacuum amplitude of such
processes is (let's suppose $t_{2}>t_{1}$ for definiteness):
\begin{eqnarray}\label{B1}
&&R(t)=1+\frac{1}{1!}\int_{0}^{t}dt_{1}\sum_{\alpha,\beta}\sum_{\textbf{k},\textbf{l}}\left(-\frac{i}{2}V_{klkl}\right)
\langle\Phi_{0}|C_{\textbf{l},\beta}^{+}(t_{1})C_{\textbf{k},\alpha}^{+}(t_{1})C_{\textbf{k},\alpha}(t_{1})C_{\textbf{l},\beta}(t_{1})|\Phi_{0}\rangle
\nonumber\\
&&+\frac{1}{1!}\int_{0}^{t}dt_{1}\sum_{\alpha}\sum_{\textbf{k},\textbf{l}}\left(-\frac{i}{2}V_{lkkl}\right)
\langle\Phi_{0}|C_{\textbf{k},\alpha}^{+}(t_{1})C_{\textbf{l},\alpha}^{+}(t_{1})C_{\textbf{k},\alpha}(t_{1})C_{\textbf{l},\alpha}(t_{1})|\Phi_{0}\rangle
\nonumber\\
&&+\frac{1}{2!}\int_{0}^{t}dt_{2}\int_{0}^{t}dt_{1}\sum_{\alpha,\beta}\sum_{\textbf{k},\textbf{l}}\left(-\frac{i}{2}V_{klkl}\right)
\sum_{\alpha',\beta'}\sum_{\textbf{k}',\textbf{l}'}\left(-\frac{i}{2}V_{k'l'k'l'}\right)\nonumber\\
&&\times\langle\Phi_{0}|C_{\textbf{l}',\beta'}^{+}(t_{2})C_{\textbf{k}',\alpha'}^{+}(t_{2})C_{\textbf{k}',\alpha'}(t_{2})C_{\textbf{l}',\beta'}(t_{2})C_{\textbf{l},\beta}^{+}(t_{1})C_{\textbf{k},\alpha}^{+}(t_{1})C_{\textbf{k},\alpha}(t_{1})C_{\textbf{l},\beta}(t_{1})|\Phi_{0}\rangle
\nonumber\\
&&+\frac{1}{2!}\int_{0}^{t}dt_{2}\int_{0}^{t}dt_{1}\sum_{\alpha}\sum_{\textbf{k},\textbf{l}}\left(-\frac{i}{2}V_{lkkl}\right)
\sum_{\alpha'}\sum_{\textbf{k}',\textbf{l}'}\left(-\frac{i}{2}V_{l'k'k'l'}\right)\nonumber\\
&&\times\langle\Phi_{0}|C_{\textbf{k}',\alpha'}^{+}(t_{2})C_{\textbf{l}',\alpha'}^{+}(t_{2})C_{\textbf{k}',\alpha'}(t_{2})C_{\textbf{l}',\alpha'}(t_{2})C_{\textbf{k},\alpha}^{+}(t_{1})C_{\textbf{l},\alpha}^{+}(t_{1})C_{\textbf{k},\alpha}(t_{1})C_{\textbf{l},\alpha}(t_{1})|\Phi_{0}\rangle\nonumber\\
&&+\frac{1}{2!}2\int_{0}^{t}dt_{2}\int_{0}^{t}dt_{1}\sum_{\alpha,\beta}\sum_{\textbf{k},\textbf{l}}\left(-\frac{i}{2}V_{klkl}\right)
\sum_{\alpha'}\sum_{\textbf{k}',\textbf{l}'}\left(-\frac{i}{2}V_{l'k'k'l'}\right)\nonumber\\
&&\times\langle\Phi_{0}|C_{\textbf{k}',\alpha'}^{+}(t_{2})C_{\textbf{l}',\alpha'}^{+}(t_{2})C_{\textbf{k}',\alpha'}(t_{2})C_{\textbf{l}',\alpha'}(t_{2})C_{\textbf{l},\beta}^{+}(t_{1})C_{\textbf{k},\alpha}^{+}(t_{1})C_{\textbf{k},\alpha}(t_{1})C_{\textbf{l},\beta}(t_{1})|\Phi_{0}\rangle\nonumber\\
&&+...
\end{eqnarray}
For approximate calculation $R(t)$ we shall use the method of
uncoupling of correlations, which lies in the fact that an average
of four creation and annihilation operators is represented by a
product of averages of pairs of the operators: $\langle
C^{+}C^{+}CC\rangle\rightarrow\langle C^{+}C\rangle\langle
C^{+}C\rangle$. The averages correspond to propagators of
particles with initial and final states corresponding to the
matrix element of interaction $V_{klmn}$ taking into account
conservation of momentum and spin. It is achieved by preliminary
transposition of the operators before the uncoupling taking into
account Fermy commutation. For Hartree and Fock processes the
procedure corresponds to the diagrams in Fig.\ref{fig12}. In each
process of scattering we connect incoming and outgoing lines
taking into account the laws of conservation. The obtained diagram
must not have free ends. Then we have the amplitude of transition
"vacuum-vacuum". Analytically it will be so:
\begin{eqnarray}\label{B2}
&&R(t)\approx
1+(-1)^{2}\frac{1}{1!}\int_{0}^{t}dt_{1}\sum_{\alpha,\beta}\sum_{\textbf{k},\textbf{l}}\left(-\frac{i}{2}V_{klkl}\right)
\langle\Phi_{0}|C_{\textbf{l},\beta}^{+}(t_{1})C_{\textbf{l},\beta}(t_{1})|\Phi_{0}\rangle\langle\Phi_{0}|C_{\textbf{k},\alpha}^{+}(t_{1})C_{\textbf{k},\alpha}(t_{1})|\Phi_{0}\rangle
\nonumber\\
&&+(-1)\frac{1}{1!}\int_{0}^{t}dt_{1}\sum_{\alpha}\sum_{\textbf{k},\textbf{l}}\left(-\frac{i}{2}V_{lkkl}\right)
\langle\Phi_{0}|C_{\textbf{k},\alpha}^{+}(t_{1})C_{\textbf{k},\alpha}(t_{1})|\Phi_{0}\rangle\langle\Phi_{0}|C_{\textbf{l},\alpha}^{+}(t_{1})C_{\textbf{l},\alpha}(t_{1})|\Phi_{0}\rangle
\nonumber\\
&&+(-1)^{4}\frac{1}{2!}\int_{0}^{t}dt_{2}\int_{0}^{t}dt_{1}\sum_{\alpha,\beta}\sum_{\textbf{k},\textbf{l}}\left(-\frac{i}{2}V_{klkl}\right)
\sum_{\alpha',\beta'}\sum_{\textbf{k}',\textbf{l}'}\left(-\frac{i}{2}V_{k'l'k'l'}\right)\nonumber\\
&&\times\langle\Phi_{0}|C_{\textbf{l},\beta}^{+}(t_{1})C_{\textbf{l},\beta}(t_{1})|\Phi_{0}\rangle\langle\Phi_{0}|C_{\textbf{k},\alpha}^{+}(t_{1})C_{\textbf{k},\alpha}(t_{1})|\Phi_{0}\rangle
\langle\Phi_{0}|C_{\textbf{l}',\beta'}^{+}(t_{2})C_{\textbf{l}',\beta'}(t_{2})|\Phi_{0}\rangle\langle\Phi_{0}|C_{\textbf{k}',\alpha'}^{+}(t_{2})C_{\textbf{k}',\alpha'}(t_{2})|\Phi_{0}\rangle
\nonumber\\
&&+(-1)^{2}\frac{1}{2!}\int_{0}^{t}dt_{1}\int_{0}^{t}dt_{2}\sum_{\alpha}\sum_{\textbf{k},\textbf{l}}\left(-\frac{i}{2}V_{lkkl}\right)\sum_{\alpha'}\sum_{\textbf{k}',\textbf{l}'}\left(-\frac{i}{2}V_{l'k'k'l'}\right)\nonumber\\
&&\times\langle\Phi_{0}|C_{\textbf{k},\alpha}^{+}(t_{1})C_{\textbf{k},\alpha}(t_{1})|\Phi_{0}\rangle\langle\Phi_{0}|C_{\textbf{l},\alpha}^{+}(t_{1})C_{\textbf{l},\alpha}(t_{1})|\Phi_{0}\rangle
\langle\Phi_{0}|C_{\textbf{k}',\alpha'}^{+}(t_{2})C_{\textbf{k}',\alpha'}(t_{2})|\Phi_{0}\rangle\langle\Phi_{0}|C_{\textbf{l}',\alpha'}^{+}(t_{2})C_{\textbf{l}',\alpha'}(t_{2})|\Phi_{0}\rangle
\nonumber\\
&&+\frac{1}{2!}(-1)^{2}(-1)2\int_{0}^{t}dt_{2}\int_{0}^{t}dt_{1}\sum_{\alpha,\beta}\sum_{\textbf{k},\textbf{l}}\left(-\frac{i}{2}V_{klkl}\right)\sum_{\alpha'}\sum_{\textbf{k}',\textbf{l}'}\left(-\frac{i}{2}V_{l'k'k'l'}\right)\nonumber\\
&&\times\langle\Phi_{0}|C_{\textbf{l},\beta}^{+}(t_{1})C_{\textbf{l},\beta}(t_{1})|\Phi_{0}\rangle\langle\Phi_{0}|C_{\textbf{k},\alpha}^{+}(t_{1})C_{\textbf{k},\alpha}(t_{1})|\Phi_{0}\rangle
\langle\Phi_{0}|C_{\textbf{k}',\alpha'}^{+}(t_{2})C_{\textbf{k}',\alpha'}(t_{2})|\Phi_{0}\rangle\langle\Phi_{0}|C_{\textbf{l}',\alpha'}^{+}(t_{2})C_{\textbf{l}',\alpha'}(t_{2})|\Phi_{0}\rangle\nonumber\\
&&+\ldots=1+(R_{1}^{Hartree}+R_{1}^{Fock})+\frac{1}{2!}(R_{1}^{Hartree}+R_{1}^{Fock})^{2}+\ldots=\exp(R_{1}^{Hartree}+R_{1}^{Fock})
\end{eqnarray}

The uncoupling lets to write expression for $R_{1}(t)$ via normal
propagators:
\begin{eqnarray}
R_{1}^{Hartree}&=&(2s+1)^{2}\int_{0}^{t}dt\sum_{\textbf{k},\textbf{l}}\left(-\frac{i}{2}V_{klkl}\right)iG_{0}(\textbf{l},t-t)iG_{0}(\textbf{k},t-t)\nonumber\\
&=&(-2i)\sum_{\textbf{k},\textbf{l}}V_{klkl}B_{0}(l)B_{0}(k)t\\
 R_{1}^{Fock}&=&(2s+1)(-1)\int_{0}^{t}dt\sum_{\textbf{k},\textbf{l}}\left(-\frac{i}{2}V_{lkkl}\right)iG_{0}(\textbf{l},t-t)iG_{0}(\textbf{k},t-t)\nonumber\\
&=&i\sum_{\textbf{k},\textbf{l}}V_{lkkl}B_{0}(l)B_{0}(k)t.
\end{eqnarray}
The multiplier $(2s+1)$ is result of summation over spin states
(number of spin configurations), $s=1/2$. We can see, that the
method of uncoupling of correlations lets to calculate vacuum
amplitude simply, selecting contributions of processes of each
type. The method can be generalized to the processes of higher
order. We can see, that the proposed method of obtaining of
diagram expansion for a vacuum amplitude by uncoupling of
correlations doesn't demand  of an use of Wick theorem and the
adiabatic hypothesis.

\end{document}